\documentstyle[12pt]{article}
\def\x{\stackrel{\otimes}{,}}

\advance \topmargin by -\headheight
\advance \topmargin by -\headsep
     
\evensidemargin \oddsidemargin
\begin{document}
\newcommand{\sect}[1]{\setcounter{equation}{0}\section{#1}}
\renewcommand{\theequation}{\thesection.\arabic{equation}}

\topmargin -.6in

\def\rf#1{(\ref{eq:#1})}
\def\lab#1{\label{eq:#1}} 
\def\br{\begin{eqnarray}}
\def\er{\end{eqnarray}}
\def\be{\begin{equation}}
\def\ee{\end{equation}}
\def\0{\nonumber}
\def\lb{\lbrack}
\def\rb{\rbrack}
\def\({\left(}
\def\){\right)}
\def\v{\vert}
\def\bv{\bigm\vert}
\def\lskip{\vskip\baselineskip\vskip-\parskip\noindent}
\relax
\newcommand{\nit}{\noindent}
\newcommand{\ct}[1]{\cite{#1}}
\newcommand{\bi}[1]{\bibitem{#1}}
\def\a{\alpha}
\def\b{\beta}
\def\ca{{\cal A}}
\def\cm{{\cal M}}
\def\cn{{\cal N}}
\def\cf{{\cal F}}
\def\d{\delta}
\def\D{\Delta}
\def\eps{\epsilon}
\def\g{\gamma}
\def\G{\Gamma}
\def\grad{\nabla}
\def\h{ {1\over 2}  }
\def\hc{\hat{c}}
\def\hd{\hat{d}}
\def\hg{\hat{g}}
\def\hp{ {+{1\over 2}}  }
\def\hm{ {-{1\over 2}}  }
\def\k{\kappa}
\def\l{\lambda}
\def\L{\Lambda}
\def\lg{\langle}
\def\m{\mu}
\def\n{\nu}
\def\o{\over}
\def\om{\omega}
\def\O{\Omega}
\def\p{\phi}
\def\pa{\partial}
\def\pr{\prime}
\def\ra{\rightarrow}
\def\rh{\rho}
\def\rg{\rangle}
\def\s{\sigma}
\def\t{\tau}
\def\th{\theta}
\def\ti{\tilde}
\def\wti{\widetilde}
\def\inte{\int dx }
\def\xb{\bar{x}}
\def\yb{\bar{y}}

\def\tr{\mathop{\rm tr}}
\def\Tr{\mathop{\rm Tr}}
\def\partder#1#2{{\partial #1\over\partial #2}}
\def\ds{{\cal D}_s}
\def\wtwo{{\wti W}_2}
\def\lie{{\cal G}}
\def\alie{{\widehat \lie}}
\def\dlie{{\cal G}^{\ast}}
\def\elie{{\widetilde \lie}}
\def\edlie{{\elie}^{\ast}}
\def\hlie{{\cal H}}
\def\wlie{{\widetilde \lie}}

\def\rlx{\relax\leavevmode}
\def\inbar{\vrule height1.5ex width.4pt depth0pt}
\def\IZ{\rlx\hbox{\sf Z\kern-.4em Z}}
\def\IR{\rlx\hbox{\rm I\kern-.18em R}}
\def\IC{\rlx\hbox{\,$\inbar\kern-.3em{\rm C}$}}
\def\one{\hbox{{1}\kern-.25em\hbox{l}}}

\def\PRL#1#2#3{{\sl Phys. Rev. Lett.} {\bf#1} (#2) #3}
\def\NPB#1#2#3{{\sl Nucl. Phys.} {\bf B#1} (#2) #3}
\def\NPBFS#1#2#3#4{{\sl Nucl. Phys.} {\bf B#2} [FS#1] (#3) #4}
\def\CMP#1#2#3{{\sl Commun. Math. Phys.} {\bf #1} (#2) #3}
\def\PRD#1#2#3{{\sl Phys. Rev.} {\bf D#1} (#2) #3}
\def\PLA#1#2#3{{\sl Phys. Lett.} {\bf #1A} (#2) #3}
\def\PLB#1#2#3{{\sl Phys. Lett.} {\bf #1B} (#2) #3}
\def\JMP#1#2#3{{\sl J. Math. Phys.} {\bf #1} (#2) #3}
\def\PTP#1#2#3{{\sl Prog. Theor. Phys.} {\bf #1} (#2) #3}
\def\SPTP#1#2#3{{\sl Suppl. Prog. Theor. Phys.} {\bf #1} (#2) #3}
\def\AoP#1#2#3{{\sl Ann. of Phys.} {\bf #1} (#2) #3}
\def\PNAS#1#2#3{{\sl Proc. Natl. Acad. Sci. USA} {\bf #1} (#2) #3}
\def\RMP#1#2#3{{\sl Rev. Mod. Phys.} {\bf #1} (#2) #3}
\def\PR#1#2#3{{\sl Phys. Reports} {\bf #1} (#2) #3}
\def\AoM#1#2#3{{\sl Ann. of Math.} {\bf #1} (#2) #3}
\def\UMN#1#2#3{{\sl Usp. Mat. Nauk} {\bf #1} (#2) #3}
\def\FAP#1#2#3{{\sl Funkt. Anal. Prilozheniya} {\bf #1} (#2) #3}
\def\FAaIA#1#2#3{{\sl Functional Analysis and Its Application} {\bf #1} (#2)
#3}
\def\BAMS#1#2#3{{\sl Bull. Am. Math. Soc.} {\bf #1} (#2) #3}
\def\TAMS#1#2#3{{\sl Trans. Am. Math. Soc.} {\bf #1} (#2) #3}
\def\InvM#1#2#3{{\sl Invent. Math.} {\bf #1} (#2) #3}
\def\LMP#1#2#3{{\sl Letters in Math. Phys.} {\bf #1} (#2) #3}
\def\IJMPA#1#2#3{{\sl Int. J. Mod. Phys.} {\bf A#1} (#2) #3}
\def\AdM#1#2#3{{\sl Advances in Math.} {\bf #1} (#2) #3}
\def\RMaP#1#2#3{{\sl Reports on Math. Phys.} {\bf #1} (#2) #3}
\def\IJM#1#2#3{{\sl Ill. J. Math.} {\bf #1} (#2) #3}
\def\APP#1#2#3{{\sl Acta Phys. Polon.} {\bf #1} (#2) #3}
\def\TMP#1#2#3{{\sl Theor. Mat. Phys.} {\bf #1} (#2) #3}
\def\JPA#1#2#3{{\sl J. Physics} {\bf A#1} (#2) #3}
\def\JSM#1#2#3{{\sl J. Soviet Math.} {\bf #1} (#2) #3}
\def\MPLA#1#2#3{{\sl Mod. Phys. Lett.} {\bf A#1} (#2) #3}
\def\JETP#1#2#3{{\sl Sov. Phys. JETP} {\bf #1} (#2) #3}
\def\JETPL#1#2#3{{\sl  Sov. Phys. JETP Lett.} {\bf #1} (#2) #3}
\def\PHSA#1#2#3{{\sl Physica} {\bf A#1} (#2) #3}
\def\PHSD#1#2#3{{\sl Physica} {\bf D#1} (#2) #3}

\begin{titlepage}
\vspace*{-2 cm}
\noindent

\begin{flushright}
IFT--P.018/97\\ 
CBPF--NF--019/98 
\end{flushright}

\vskip 1 cm
\begin{center}
{\Large\bf Parafermionic Reductions of WZW Model } \vglue 1  true cm
{ J.F. Gomes}$^{\dagger}$,
 { G.M. Sotkov}\footnote{On leave of absence from the Institute for Nuclear
 Research and Nuclear Energy, Bulgarian Academy of Sciences, 1784, Sofia,
 Bulgaria}$^*$ and { A.H. Zimerman}$^{\dagger}$\\

\vspace{1 cm}

$^{\dagger}${\footnotesize Instituto de F\'\i sica Te\'orica - IFT/UNESP\\
Rua Pamplona 145\\
01405-900, S\~ao Paulo - SP, Brazil}\\
jfg@axp.ift.unesp.br, zimerman@axp.ift.unesp.br\\

\vspace{1 cm} 

$^*${\footnotesize Centro Brasileiro de Pesquisas F\'\i sicas-CBPF\\
Rua Xavier Sigaud 150\\
22290-180, Rio de Janeiro, RJ}\\
sotkov@cbpfsu1.cat.cbpf.br\\
\medskip
\end{center}

\normalsize
\vskip 0.2cm

\begin{center}
{\large {\bf ABSTRACT}}\\
\end{center}

\noindent
{\footnotesize 
We investigate a class of conformal Non-Abelian-Toda models representing a
 noncompact
$SL(2,R)/U(1)$ parafermionions (PF) interacting with a specific abelian Toda
theories and having a global $U(1)$ symmetry.  A systematic derivation of the
conserved currents, their algebras and the exact solution of these models is
presented.   An important property of this class of models is the affine
$SL(2,R)_q$ algebra spanned by charges of the chiral and antichiral nonlocal
currents and the $U(1)$ charge.  The classical (Poisson Brackets) algebras of
symmetries $V{G_n}$ of these models appears to be of mixed PF-$W{G_n}$ type. 
They contain together with the local quadratic terms specific for the
$W_n$-algebras the nonlocal terms similar to the ones of the classical
PF-algebra.  The renormalization of the spins of the nonlocal currents is the
main new feature of the quantum $V{A_n}$-algebras.  The quantum  
 $V{A_2}$-algebra   and its degenerate representations are studied in detail.
    } 
\vglue 1 true cm

\end{titlepage}

\sect{INTRODUCTION}

The identification of 2-dimensional critical phenomena 
as conformal minimal models of the (extended) Virasoro algebra \cite{BPZ} 
provide powerful algebraic tools for calculation of the critical exponents.
Within this framework the long standing problem of classification of the  
universality classes in two dimensions is reduced to the problem of 
exhausting all the extensions of the Virasoro algebra.  The first part of the 
list of extensions of the Virasoro algebra contains the well known Lie 
algebraic lower spin extensions ($s\leq 2$) including the conformal 
current algebra ($s=1$), the $N=1,2,3,4 $ super Virasoro algebras 
($s\leq 3/2$ ), etc.  An important step in completing this list was
 proposed 
by Zamolodchikov and Fateev \cite{Z}, \cite{ZF1} and \cite{ZF2}.  They 
observe that for describing the critical behavior of a large class of 
statistical mechanical systems one has to consider {\it two new types of  
non-Lie algebraic extensions}.  The first one include together with the 
stress tensor ($s=2$) a new set of local higher spin ($s=3,4, \cdots ,N$) 
currents which close an associative algebra of quadratic relations known  
as $W_N$-algebra \cite{F-L}.  The second one represents a {\it nonlocal 
extension} of the Virasoro algebra with a set of fractional spins ($s = 
{k(N-k)\over N}, \;\;\; k=1,2,\cdots ,N-1$) nonlocal currents--the 
$Z_N$-parafermionic algebra (PF) (see ref. \cite{Ge} for a generalized PF 
algebras). 

 An universal method for deriving all these algebras as well 
as   the (classical ) Lagrangean of models with such symmetries consists  
in considering the {\it gauged $G/H$- WZW models} and their algebras of 
symmetries \cite{Pol1},\cite{Al-Shat}, \cite{Orai}.
For example the $W_n$-algebra appears from the $SL(n,R)$ current algebra  
by gauging the nilpotent subalgebras $N^{\pm}(n)$ \cite{Pol1},\cite{Orai}. 
 The 
parafermionic algebras arises when the Cartan subalgebra of $SU(n)$ is 
gauged away \cite{ZF1}, \cite{Ge}.  The natural question to ask is 
{\it whether gauging another subalgebras}, say, of mixed type $U(1)\oplus 
N^+(n-1)\oplus N^-(n-1)$ ( i.e. a part of the nilpotent subalgebras 
$N^{\pm}(n)$ with only one Cartan generator) one produces a {\it new type of 
extensions} of Virasoro algebra different from $W$ and PF algebras.   
This question was raised in a slightly different form by Gervais and 
Saveliev \cite{GS}, studying the symmetries of the (classical )$B_n$ 
nonabelian Toda theories (NA).  An explicit form of such (classical) 
nonlocal and non--Lie algebra (called $V$ algebra), unifying together the 
nonlocal PF-currents and the local $WB_{n-1}$currents was constructed in  
ref \cite {Bila} for the case of $B_2$-NA- Toda model.

In our recent paper \cite {Gosz} we have found the classical and quantum 
algebras of symmetries of the first few models of the   family of $A_n$ -NA-Toda 
theories. The present paper is devoted to the systematic construction of the 
(classical) $VG^{(j,1)}_{n}$-algebras ($G_n = A_n, B_n, C_n \;\; or D_n$) 
of mixed $PFA_1-WG_{n-1}$ type and their quantization. 
 They arise as algebras of the conserved 
currents of the simplest family of $G^{(j,1)}_n-$NA-Toda theories 
representing a noncompact $SL(2,R)/U(1)$ PF's interacting with 
$G_{n-j}\otimes G_{j-1}$ abelian Toda model ($G_0 =1\;\;j=1,\cdots n$).  
This family of NA-Toda theories  can be obtained as a 
specific (parafermionic) reduction of the $G_n$-WZW model imposing a set  
of constraints similar to these ones leading to standard $G_n$ abelian 
Toda model, but with one of the constraints $J_{-\alpha_j} = \mu_j $ (and 
$\bar J_{\alpha_j} = \bar \mu_j$) removed ($j = 1,\cdots ,n $ fixed),i.e. 
we {\it leave unconstrained} one of the simple negative (positive) root 
current.  This changes the group of the residual gauge transformations 
and contrary to the abelian Toda case the transformation responsible for  
the vanishing of the Cartan subalgebra current $J_{\lambda_jH}$ (j-fixed) 
does not exist. We therefore  require the {\it additional constraint}  
$J_{\lambda_jH} = \bar J_{\lambda_jH} = 0$.  The main consequence of this 
{\it parafermionic constraint} is that two of the chiral (antichiral) conserved 
current  $V^+_j = J_{\alpha_j}$ and $ V^-_j = J_{-\alpha_1 -\cdots 
\alpha_n}$ of the reduced model ($G^{(j,1)}_n$ -NA-Toda) appears to be 
{\it nonlocal (PF-type) currents}.  The simplest representative of this family 
of NA-Toda models $G^{(1,1)}_n$ is given by the lagrangean
\br
{\cal L}^{(1)}_{n} &=& -{k\over {2\pi }}\{ \tilde \eta_{ik}g^{\mu \nu}\partial 
_{\mu}\phi_{i}\partial _{\nu}\phi_k + {e^{k_{12}\phi_1}\over \Delta}g^{\mu 
\nu}\partial _\mu \psi \partial_ {\nu}\chi  \nonumber \\ 
&- &({2\o k})^2 \sum_{i=1}^{n-1}{2\over 
{\alpha_i^2}}e^{-\tilde k_{ij}\phi_j} - {e^{k_{12}\phi_1}\over 
{2\Delta}}k_{12}\epsilon_{\mu \nu }\partial _{\mu}\phi_1(\psi \partial 
_{\nu}\chi - \chi \partial _{\nu}\psi ) \} 
\label{lagr11}
\er
where $\Delta = 1+ {1\over 2{\cal K}_{11}}e^{k_{12}\phi_1} \psi \chi $,
($k_{12}= -1 $ for $A_n, B_n, C_n, D_n$ except for $B_2$ where $k_{12} =-2$;
${\cal K}_{11}={n\over n+1}$   for $A_n$ and ${\cal K}_{11}=1$ for $B_n$).
The corresponding equations of motion of
(\ref{lagr11}) can be also derived by a slight modification of the
Leznov-Saveliev method starting with an appropriate grading operator $Q=
\sum_{i=2}^{n} 2 {{\lambda_i H}\o {\a_i^2}}$, (see Sect. 2 and app. A).

  Apart from 
the two {\it nonlocal chiral currents} $V^{\pm}$ ( of spin $s^{\pm} = 
{(n+1)\over 2}$ for $A_n$, $s^{\pm} = n$ for $B_n$ and $C_n$, $s^{\pm} = 
n-1$ for $D_n$) each  of these models exhibit other $n-1$ 
{\it local chiral }
(antichiral) currents $ 
W_{n-l+2},\quad l=2,\cdots ,n$ of spins $s_l = n-l+2$ for $A_n$;
 of spins 
 ($s_i= 2,4, \cdots 2(n-i+1)$ for $B_n$ and $C_n$ and 
 of spins $s_l= 2,4, \cdots 2(n-2),n$ for $D_n$.The {\it nonchiral} $U(1)$ current 
\be
J_{\mu} = -{k\over 4\pi}{(\psi \partial _{\mu}\chi - \chi \partial 
_{\mu}\psi -k_{12} \psi \chi \partial _{\mu}\phi_1 )\over \Delta }e^{k_{12}\phi_1}
\label{u1}  
\ee
generating {\it global} $U(1)$ gauge transformations $\phi_i^{\prime} = 
\phi_i$, $ \psi^{\prime} = \psi e^{\alpha} $ and $ \chi^{\prime} = \chi 
e^{-\alpha}$ ( $\alpha $ constant) completes the list of conserved  
currents for the  $G_n^{(1,1)}$ model.

One of the {\it basic results} presented in this paper is the explicit form of 
the Poisson brackets (PB's) algebra $VA^{(1,1)}_{n}$ of the conserved 
currents of $A^{(1,1)}_n$ -NA Toda theory.  This new algebra appears to be a
{\it natural unification} of the main features of $W$- and PF-algebras.  As it is
shown in Sect. 3, the    PB's of the local 
currents are quite similar to the classical $WA_{n-1}$ algebra but 
including new terms in the quadratic part - the product $V^+V^-$ and its  
derivatives.  The local and nonlocal currents obey PB's with quadratic 
terms in the form $V^{\pm}W_k$ ( and its derivatives).  Finally the 
PB-algebra of the nonlocal currents contains specific 
{\it nonlocal quadratic} 
terms (see sect 3):
\br
&&\{ V^{\pm} (\sigma),   V^{\pm} (\sigma^{\pr})\}
 = - {{n+1} \o nk^2 } \epsilon (\sigma - \sigma^{\pr})V^{\pm}
(\sigma)V^{\pm}(\sigma^{\pr})
\nonumber \\
&&\{ V^{+} (\sigma),   V^{-} (\sigma^{\pr}) \}
  =  {{n+1} \o nk^2 } \epsilon (\sigma - \sigma^{\pr})V^{+}
(\sigma)V^{-}(\sigma^{\pr})+ ({k\o 2})^{n-1} \pa_{\sigma^{\pr}}^{n} \d (\sigma -
\sigma^{\pr})   \nonumber \\
 &&\hskip 3cm  -   \sum_{s=0}^{n-2}({k\o 2})^{s-1} W_{n-s}(\sigma ^{\pr})
  \pa ^{s}_{\sigma^{\pr}}
\d (\sigma - \sigma^{\pr})
\label{VV}
\er
where $\epsilon (\sigma ) = {\rm sign} \;(\sigma )$ .  Our 
{\it main observation} is that 
the {\it nonlocal terms }in (\ref{VV})
 ( those with $\epsilon (\sigma )$) are of {\it PF-type}. 
  One obstacle of such identification  is the {\it discrepancy} between 
the {\it fractal spins} of the PF currents and the (half) {\it integer spins}
 of the 
nonlocal currents $V^{\pm}$.  The precise statement is that the 
{\it semi-classical limit} $N \rightarrow \infty$ of the operator product 
expansion (OPE) algebra of certain $W$--reduced $G_n$ parafermions 
$\Psi_{\alpha}$ of spins $s_{\alpha} = {{n+1}\over 2} - {\alpha^2\over  
2N}$ coincides with the PB's -algebra (\ref{VV}) of $V^{\pm}$, 
($s^{\pm}={{n+1}\over 2}$).  The $Z_N$-parafermions  provide the 
 simplest example ($G=A_1$, $n=1$) of 
quantization of $V^{\pm}$ and their classical algebra (\ref{VV}).  We 
identify the quantized $V^{\pm}$ with the PF currents $\Psi_1$ and $ 
\Psi_1^{\dagger}$ of spins $ s^{\pm}= 1-{1\over N}$ namely, $ V^+ = 
{{1\over \sqrt {N}}}\Psi_1  $, $ V^- ={{1\over \sqrt {N}}}\Psi_1^{\dagger}  
$.  This way we impose the OPE's of $V^{\pm}$ to be of the form ( see 
sect. 3 of ref. \cite{ZF1}):
\br
V^{\pm}(1)V^{\pm}(2) &=& {{N-1}\over (N)^{3/2}}(z_{12})^{\Delta_2^{\pm} -2 
\Delta_1^{\pm}} \( V_2^{\pm}(2) + O(z_{12})\) \nonumber \\
V^{-}(1)V^{+}(2) &=& {{1}\over N}(z_{12})^{ -2 
\Delta_1^{\mp}} \( I + {2\Delta_1 \over c}T(2)z_{12}^2 +O(z_{12}^3)\)
\label{opevv} 
\er
where $\Delta^{\pm}_2 = 2 - {4\over N}$ , $\Delta_1^{\pm} = 1 - {1\over 
N}$, $ c= 2{{N-1}\over {N+2}}$, $N=2,3,\cdots $.  
Next we define the classical PB's as certain limit of the OPE's 
(\ref{opevv}):
\be
\{ V^a(1),V^b(2) \} = lim_{N\rightarrow \infty }(-{iN\over 
2\pi})(V^a(1)V^b(2) - V^b(2)V^a(1))
\label{1.5}
\ee
where $a,b = \pm 1$.  The last step is to verify that the $lim_{N 
\rightarrow \infty }$ of (\ref{opevv}) indeed reproduces the PB's 
algebra (\ref{VV}) with $n=1$.  One can also derive the OPE's (\ref{opevv}) and the
renormalization of the spins of $V^{\pm}$, $\Delta^{quant} = \Delta^{class} -
{1\o N}$ following the procedure of the quantum hamiltonian reduction
\cite{Ber-Oo,ZF1}. 
Starting with the bosonized form of the $SL(2)$ current algebra (see for
instance \cite{Ger}) and
imposing the constraints $J_3 = 0$, one obtain the free field
 representation of
the PF-currents.  The OPE's (\ref{opevv}) as well as the
 new (anomalous) dimensions of $V^{\pm}$appears as 
 a simple consequence of this
 construction.

The purpose of this discussion of the parafermionic properties of the 
nonlocal currents $V^{\pm}$ is to point out that 
their {\it quantization requires deep 
changes in the classical algebraic structure} (\ref{VV}) namely, $i)$ 
renormalization of the bare spins $s_{cl}^{\pm} = {{n+1}\o 2}$ to $s^{\pm}_{q}
= {{n+1}\o 2}(1- {{1\o {2k+n+1}}})$; $ii)$Breaking the global 
$U(1)$ symmetry to some discrete group; $iii)$ The quantum counterpart of 
the PB's  of the charges $Q^{\pm}_{m+{{(1+l)(n+1)}\over {2(2k+n+1)}}}\;\;\; 
(m\in Z, 
l=1,\cdots ,2k+n)$ of $V^{\pm}$ are the so called {\it PF commutators} (an 
infinite sum of bilinears of the charges) (see sect.4 of ref 
\cite{ZF1}).  The quantization of the local currents $W_p$ is similar to  
the one of the classical $WG_{n-1}$-algebras.  It consists in the 
familiar substitution $i\hbar \{ \;\; \}_{PB} = [ \;\; ]_{comm}$ 
followed by certain 
changes in the structure constants and of the central charge.
No spin renormalization and PF type commutators are required in this case. 
All these new features of the quantum $VA_n$-algebras  
  we  shall demonstrate in Sect. 9 on the example of the
quantization  of the $VA_2^{(1,1)} \equiv V_3^{(1,1)}$.   The quantum
counterparts of the PB's algebra (\ref{VV}) appears to be the ``parafermionic
commutation relations '' (\ref{9.14}) and (\ref{9.15}).  The method we are using
allows us to find also the {\it anomalous dimensions} of the ``completely degenerate ''
representations of the quantum $V_3^{(1,1)}$-algebra.

The two chiral algebras $VG_n^{(1,1)}$ and $\bar VG_n^{(1,1)}$  
together with the nonchiral $U(1)$ current (\ref{u1}) of charge $Q_0 =  
\int J_0 d\sigma $
\br
\{Q_0, V^{\pm}(\sigma ), \} &= &\pm V^{\pm}(\sigma ), \quad \quad   
\{Q_0, W_p(\sigma ), \} = 0\nonumber \\
\{Q_0, \bar{V}^{\pm}(\sigma ), \} &= &\pm \bar V^{\pm}(\sigma ), \quad \quad   
\{Q_0, \bar W_p(\sigma ), \} = 0
\label{qw}
\er
$p=2,3,\cdots ,n$ {\it do not exhaust} all the symmetries of the 
$G^{(1,1)}_n$ -NA Toda model (\ref{lagr11}).  It turns out that certain 
charges of the {\it chiral nonlocal} currents , $Q^+ =\int V^+ d\sigma $ and $ 
Q^- = \int \sigma^{n-1}V^- d\sigma $ have nonvanishing equal time PB's  
with the {\it antichiral nonlocal} charges $\bar Q^- =\int \bar V^-d\sigma $ 
and $\bar Q^+ = \int \sigma^{n-1}\bar V^+ d\sigma $.  They are linked by 
the {\it topological charge} 
\be
H_0 = {k\over 2\pi}\int_{-\infty }^{\infty } \partial \varphi d\sigma = -Q_0 
+ {k\over 2\pi }{\cal K}_{11}k_{12}\int_{-\infty}^{\infty} \partial \phi_1 
d\sigma 
\ee
in the following algebra
  \be
\{Q^{\pm},\bar Q^{\mp} \} = \pm {k\pi \over 2}\int_{-\infty}^{\infty} 
d\sigma \partial e^{\pm {1\o {\cal K}_{11}}\varphi } , \quad \quad \{Q^{\pm}, \bar 
Q^{\pm} \} = 0
\label{qalg}
\ee
Simple redefinitions of the nonlocal charges (see Sect. 4) allows to 
rewrite the algebra (\ref{qw}) and (\ref{qalg}) in the standard form of 
the {\it affine q-deformed $SL(2,R)$ PB'S algebra} of level zero 
 \cite{Bern-Leclair,Drinf}.  
Note that the deformation parameter
 $q=e^{-{2\pi \over k}{1\o {\cal K}_{11}}}$  is a function of the {\it  
 classical (bare) coupling constant}.  It is 
important to mention that the $SL(2,R)_q$ PB's algebra appears in the 
NA-Toda theories (\ref{lagr11}) as {\it algebra of the (Noether ) symmetries 
of their equations of motion} and leaves invariant the NA-Toda 
hamiltonian.  One has to distinguish this classical  $q$-deformed
${\hat SL}(2,R)_q$ PB algebra  
(generated by the nonlocal conserved currents) from the {\it Poisson-Lie  
group}  $G_n(r)$ of the monodromy matrices $M \in G_n$ satisfying together 
with the standard group multiplication laws the Sklyanin PB's algebra \cite{Skly},
\be
\{ M \x M \} = {-2\pi \over k}[r, M\otimes M]
\ee
where $r$ is the classical r-matrix.  The latter appears in 
$G_n$ WZW models \cite{Al-Shat}, \cite{Bab}, \cite{F-L1}
 and the abelian $G_n$-Toda theories (as well as in a 
large class of integrable models \cite{Fad}) as the right hand 
transformation that leaves invariant the  Poisson structure of the
corresponding models.  Some preliminary results concerning 
the  $G_n(r) $-algebra  generated by the classical monodromy matrices of  
 $G_n^{(1,1)}$ -NA-Toda models are given in Sect. 4.

The Poisson-Lie groups are known to be the classical analog of the 
quantum group $U_q(G_n)$ encoded in the quantum exchange algebra 
\cite{Gomez-Siera},\cite{Al-Shat},\cite{F-L2},\cite{Bab-Bon}. 
 One might wonder 
which is the {\it quantum counterpart of the classical $\hat SL(2,R)_q$  PB's 
algebra}.  Partial answer to this question is given by  the 
simplest example of $n=1$, $N=2$ (the $Z_2$ PF, i.e. critical Ising model ).
 The  
quantum nonlocal charges $Q^+$ and $\bar Q^-$ coincide with the Ramond 
sector of zero modes $\psi_0$ and $\bar \psi_0$ of the (chiral) Ising 
fermions.  It turns out that their {\it commutator} is proportional to the 
fermion parity operator $\Gamma $ 
\be
[\psi_0, \bar \psi_0 ]= i\alpha \Gamma, \quad \Gamma \psi_0 \Gamma^{-1} = 
-\psi_0, \quad \Gamma \bar \psi_0 \Gamma^{-1} = - \bar \psi_0
\ee
and $\Gamma ^2=1$.  This algebra appears to be the quantum analog of 
(\ref{qalg}) for this particular case.  It is important to note that the  
{\it nonvanishing} commutator of the left and right fermionic zero modes is not in 
contradiction with the {\it holomorphic factorization} of the critical Ising 
model.  What is crucial for this factorization is that the {\it
anticommutator} 
$[\psi_m,\bar \psi_n]_+ = 0 $, $n,m \in Z$ indeed {\it vanishes}.

To make the discussion of the $SL(2,R)_q$ symmetries of the $G_n^{(1,1)}$ 
NA-Toda models complete, we have to demonstrate that classical solutions  
with nontrivial topological charge $H_0 \neq 0$ (i.e. $\varphi (\infty,0) 
\neq \varphi(-\infty ,0)$) do exist.  We derive in Section 5 the general 
solution of (\ref{lagr11}) in a simple and explicit form, appropriate for  
the analysis of these asymptotics.  Our construction is the NA-Toda 
analog of the Gervais-Bilal's \cite{Ger-Bil} solution of the abelian Toda 
models.  It is based on the {\it important observation} that the fields 
$\psi,\;\; \chi$ and $\phi_i$, $i=1,\cdots ,n-1$ of the NA-Toda theory  
(\ref{lagr11}) can be realized in terms of the corresponding abelian Toda 
fields $\varphi_A$, $A= 1,\cdots n$ and the chiral nonlocal currents 
$V^+$ and $\bar V^-$ considered as independent variables.  The origin of  
this transformation of the solutions of the $G_n^{(1,1)}$ NA-Toda  into 
those of the $G_n$ abelian Toda (and vice-versa) is in the fact that 
both can be realized as gauged $G_n/H_i$-WZW models with $H_1^{\pm} = 
U(1)\otimes N^{\pm}(n-1)$, $H_2^{\pm} = N^{\pm}(n)$.  Therefore the 
transformation we have found can be identified as (field dependent) 
gauge transformations $h(V^+)\otimes \bar h(\bar V^-) \in G_n\otimes 
G_n$ that maps $G/H_1$-WZW  into $G/H_2$-WZW, $g_1 =\bar hg_2 h$, $g_i  
\in G_n/H_i$.  This provides us with a powerful method for explicit      
  construction of these transformations.  Consider a set of constraints,  
gauge fixing conditions and remaining currents which define the reduction 
of $G_n$ WZW model to $G_n^{(1,1)}$  NA-Toda. For $G_n = A_n$, 
\be
J^{NA} = V^+E_{-\a_1} + \sum_{i=2}^{n}E_{-\a_i} + V^-E_{\a_1+\a_2+\cdots 
+\a_n} + \sum_{i=2}^{n} W^{NA}_{n-i+2}E_{\a_i+\a_{i+1}+\cdots +\a_n}
\label{naw}
\ee
(similar for $\bar J^{NA}$) and those leading to the abelian Toda are
\be
J^{A} =  \sum_{i=1}^{n}E_{-\a_i} +  
 \sum_{i=1}^{n} W^{A}_{n-i+2}E_{\a_i+\a_{i+1}+\cdots +\a_n}
\label{aw}
\ee
The transformations $h(V^+)$, $\bar h(\bar V^-)$ that maps (\ref{naw}) 
into (\ref{aw}) satisfy the following system of first order differential  
equations 
\be
(J^{A} + {k\over 2}\partial )h^{-1} = h^{-1} J^{NA} \quad \quad
(\bar J^{NA} - {k\over 2}\bar \partial )\bar h^{-1} = \bar h^{-1} 
\bar J^{A} 
\label{transf}
\ee
We present in Sect. 6 the explicit form of the solutions of eqns. 
(\ref{transf}) in the $A_n$ case i.e. $h, \;\; \bar h \in SL(n+1,R)$.  

The fact that one can {\it connect all the coset models}
 obtained from a given 
$G_n$-WZW model (i.e. all the hamiltonian reductions of $G_n$-WZW) by 
{\it specific (current dependent) $G_n$ gauge transformations} leads to 
important consequences concerning the symmetry structure of the 
$G_n^{(1,1)}$-NA-Toda models.  As one can see from eqns. 
(\ref{naw}-\ref{transf}) the transformation $h(V^+)$ gives as a byproduct 
the explicit constructions of the currents $W^{A}_p$, $p=2,\cdots n+1$ in 
terms of the conserved currents $V^{\pm}$ and $W^{NA}_i,$ $i=2,\cdots , 
n$ of the $G_n^{(1,1)}$-NA-Toda model.  We further verify (using the 
$VA_n^{(1,1)}$ PB's algebra only) that these $W^A_p$ indeed does close 
the $W_{n+1}$ algebra.  Thus $h(V^+)$ maps the $VA_n^{(1,1)} = 
V_{n+1}^{(1,1)}$ algebra into the $W_{n+1}$-one.  This shows that $W_{n+1}$, 
 which leaves in the  
{\it universal enveloping} of $V_{n+1}^{(1,1)}$ appears as an
 {\it algebra of 
symmetries} of the $A_n^{(1,1)}$-NA-Toda theories as well.

The gauge transformation between different set of constraints imposed on  
$G_{n}$-WZW currents play an important role in the description of the 
symmetries of a larger class of $G_n^{(j,1)}$-NA-Toda models ($J_{-\a_j} 
= V^+_{(j)}$, $J_{\lambda_jH} =0,\;\;\; j= 1,\cdots ,n$ arbitrary 
fixed).  Again as in the $j=1$ case we find a transformation $h(V_j^+)$ ( 
as solution of eqn. (\ref{transf}))which maps them into $G_n$-abelian 
Toda theory.  The new phenomena occurs when we consider 
the transformation $H(j_1,j_2) = 
h(V^+_{j_1})h(V^+_{j_2})^{-1}$ between $G_n^{(j_1,1)}$ and 
$G_n^{(j_2,1)}$-NA Toda models  ($j_1 \neq j_2$).  Both contain equal 
number of independent fields, both are $W_{n+1}$-invariant and the 
transformation $H(j_1,j_2) $ realizes the map
\be
V^{(j_1,1)}_{n+1} \longrightarrow W_{n+1} \longrightarrow V_{n+1}^{(j_2,1)}
\ee
that mixes their algebras of symmetries.  However their lagrangeans, 
their symmetry algebras $V_{n+1}^{(j_1,i)}$ and $V_{n+1}^{(j_2,1)}$ as 
well as the spins of their conserved currents are quite different.  
Nevertheless, as we claim in Sect. 7, they are classically equivalent 
models related by complicated nonlocal change of field variables: $ 
g(j_1) = \bar H(j_1,j_2)g(j_2)H(j_1,j_2) $.  This  is the reason why we 
are mainly considering the $j=1$ model, all the rest $j\neq 1$ 
$G_n^{(j,1)}$-NA-Toda models being equivalent to it.

The $G_n^{(j,1)}$-NA-Toda models we are studying in this paper are the 
{\it nearest neighbours}
 of the abelian $G_n$ Toda theories.  They are defined 
by the set of constraints and gauge fixings conditions (\ref{naw}) and 
(\ref{aw}).  The only difference with the abelian  $G_n$-Toda is that the 
constraint $ J_{-\a_j}^{(A)} $ is removed ($ J_{-\a_j}^{(NA)} = V_j^+(z)$) and 
one new PF-type constraint  has to be imposed $J_{\lambda_jH}=0$.  These 
modifications of the abelian Toda  constraints reflects on the properties 
of the remaining currents: $W_{n+1}$ splits in two nonlocal currents 
$V^+$ and $ V^-$, the new nonlocal algebra $VG_n^{(1,1)}$ replaces the 
$WG_n$ algebra and finally the chiral and antichiral nonlocal charges 
$Q^{\pm}$ and $ \bar Q^{\mp}$ together with the topological charge $H_0$
generate the $SL(2,R)_q$.   One could wonder {\it how general} is 
this way of describing the NA-Toda models.  Say, abandoning more abelian  
Toda constraints   
$J_{-\a_i}=1 \;\;\;i=1, \cdots ,l\;\;\;l\leq n$, i.e. $J_{-\a_i}=V^+_i$ and 
requiring $J_{\lambda_iH} = 0 \;\;\;i=1,\cdots ,l$ are we getting {\it new 
NA-Toda models}?  The answer to this question is indeed positive:  This 
set of constraints and gauge fixing conditions 
\be
J^{NA(l)} = \sum_{i=1}^{l} V^+_iE_{-\a_i} + \sum_{i=l+1}^n E_{-\a_i} + 
\sum_{i=1}^l
V^-_iE_{\a_i+\cdots +\a_{n}} + \sum_{i=l+1}^nW_{n-i+2}E_{\a_i+\cdots  
+\a_n}
\ee
(for $G_n = A_n$)
defines a family of conformal invariant $G_n^{([j]_l,l)}$-NA-Toda models  
($[j]_l = [j_1,j_2, \cdots ,j_l]$, labels the positions of the 
PF-constraints).  Their properties are quite similar to the simplest 
$G_n^{(1,1)}$-NA-Toda model.  They have $2l$ chiral nonlocal currents 
$V^{\pm}_i$, ($i=1,\cdots ,l$) and (n-l) chiral local ones $W_{n-i+2},  
\;\;\; i=l+1, \cdots ,n)$.  The complete discussion of the symmetries 
of $A_n^{([j]_l,l)}$-NA-Toda models, their general solutions, their 
relation with the abelian $A_n$-Toda, etc will be  presented in our 
forthcoming paper \cite{GSSZ2}.

Although the $VG_n^{([j]_l,l)}$-algebras do not exhaust all the 
parafermionic extensions of the $WG_n$-algebras, they make the picture of 
the extended Virasoro algebras more complete.  The application of the quantum
$VG_n$-algebras and their minimal conformal models is not restricted to the
problem of classification of the universality classes in two dimensions only. 
As it is well known \cite{LS} certain 2-d NA-Toda theories naturally appears
in the construction of cylindrically  symmetric instantons solutions of self
dual Yang-Mills theories in four dimensions.  This is a strong indication that
the quantum $A_n$-NA-Toda models (and their integrable off-critical
pertubations ) provide powerful tools for the nonperturbative quantization of
instantons and merons.

\sect{NA-Toda's as gauged WZW models}

The $G_{n}^{(1,1)}$-NA-Toda theories we are going to study are originally
defined as an integrable system of field equations  given by the zero
curvature condition:
\begin{eqnarray}
[\partial - {\cal A},\bar{\partial} - \bar{{\cal A}}]=0,
\label{2.1}
\end{eqnarray}
for the specific Lax connections~\footnote{all algebraic notations and definitions
used here are collected in App. A}:
\br
{\cal A}&=&\frac{\psi \partial \chi}{2{\cal K}_{11}\Delta}e^{k_{12}\phi_1} 
\lambda_{1}\cdot H+\sum_{j=1}^{n-1}\partial
\phi_{j}{\a_{j+1}\cdot H\o {\a_{j+1}^2}}
+\frac{\partial
\chi}{\Delta}e^{k_{12}{1\o 2}\phi_1}E_{\alpha_{1}}+
({2\o k})\sum_{i=1}^{n-1}
e^{-{1\o 2}\tilde k_{ij}\phi_{j}} E_{\alpha_{i+1}},
\nonumber
\\
\bar{{\cal A}}&=&-\frac{\chi \bar{\partial}\psi}{2{\cal K}_{11}\Delta}
e^{k_{12}\phi } 
\lambda_{1}\cdot H-\sum_{j=1}^{n-1}\bar{\partial}
\phi_{j}{\a_{j+1}\cdot H\o {\a_{j+1}^2}}
-\frac{\bar{\partial}
\psi}{\Delta}e^{k_{12}{1\o 2}\phi_1}E_{-\alpha_{1}}- ({2\o k})
\sum_{i=1}^{n-1}
e^{-{1\o 2}\tilde k_{ij}\phi_{j}} E_{-\alpha_{i+1}}, \nonumber\\
&&
\label{2.2}
\end{eqnarray}
where $H_{i}$, $E_{\pm{\alpha}}$ denote the generators of the
$G_{n}$-algebra; $\pm \alpha_{i}$ are its simple roots, ${\alpha }$-an
arbitrary root; $k_{ij}$ and ${\cal K}_{ij}$-the Cartan matrix and its inverse,
respectively, $\tilde k_{ij}$,$\tilde {\cal K}_{ij}$  and $\tilde \a_j$ are the
corresponding  matrices and roots for $G_{n-1}$,
 $\lambda_{i}= {\cal K}_{ij} \a_j$ is the $i^{th}$
 fundamental weight.

The problem we address here is {\it to find an action which reproduces the
equations of motion for the fields $\psi (z,\bar{z})$, $\chi (z,\bar{z})$,
$\phi_i (z,\bar{z})$ $(i=1,2,...,n-1)$ } encoded on eqns. (\ref{2.1}),
\br
\pa \bar \pa \phi_l = ({2\o k})^2e^{-\tilde k_{ls}\phi_s} -
\tilde {\cal K}_{1l}({\tilde \a^2_l \o 2})
{{\pa \chi \bar \pa \psi} \o \Delta^2}e^{k_{12}\phi_1},
\nonumber 
\er
\be
\bar \pa ({\pa \chi \o \Delta }e^{k_{12}\phi_1}) = -{{\bar \pa \psi \pa \chi }\o
{2{\cal K}_{11}\Delta^2}}\chi e^{2k_{12}\phi_1}, \quad \quad 
 \pa ({\bar \pa \psi \o \Delta }e^{k_{12}\phi_1}) = -{{\bar \pa \psi \pa \chi }\o
{2{\cal K}_{11}\Delta^2}}\psi e^{2k_{12}\phi_1} 
\label{2.3}
\ee
where we have used ${\tilde {\cal K}} _{1i} {\cal K}_{11} =
{\cal K}_{1\,i+1}$. As is well
known, the $G_{n}$-abelian Toda \cite{Orai} and the $A_{1}$-NA-Toda theories
\cite{Witten1} are equivalent to specific gauged WZW models: $G_{n}/N_{L}\otimes
N_{R}$ and $SL(2,R)/U(1)$ respectively. This fact suggests to look for a  
subgroup
$H\subset G_{n}$ such that the corresponding $G_{n}/H$-WZW model provides an
action for the $G_{n}^{(1,1)}$-NA-Toda theories. According to the Hamiltonian
reduction recipe \cite{Al-Shat,Orai,Ber-Oo} 
the first step consists in imposing a specific
set of constraints on the WZW-currents:
\begin{eqnarray}
J_{-\alpha_{i}}&=&\bar{J}_{\alpha_{i}} = 1, \quad i=2, \cdots ,n
\nonumber
\\
J_{-[\alpha]}&=&\bar{J}_{[\alpha]} = 0, \quad \a \;\; {\rm non -\; simple \;\; root }
\nonumber
\\
J_{\lambda_{1}\cdot H}&=&\bar {J}_{\lambda_{1}\cdot H} = 0 
\label{2.4}
\end{eqnarray}
on the WZW currents
\begin{eqnarray}
J&=&({k\o 2})g^{-1}\partial g=\sum_{all\; roots}J_{\left\{ \alpha \right\} }
E_{\left\{ \alpha \right\} }+\sum_{i=2}^{n}J_{i}{{\a_i \cdot H}\o {\a_i^2}} +
J_{\lambda_1H} \lambda_1 \cdot H,
\nonumber \\
\bar{J}&=&-({k\o 2})\bar{\partial}gg^{-1}=\sum_{all\; roots}
\bar{J}_{\left\{ \alpha \right\} }
E_{\left\{ \alpha \right\} }
+\sum_{i=2}^{n}\bar{J}_{i}{{\a_i \cdot H}\o {\a_i^2}} + 
\bar J_{\lambda_1H} \lambda_1 \cdot H,
\nonumber
\end{eqnarray}
  An important characteristic of the constraints (\ref{2.4}) is the group of
{\it residual gauge transformations} $H_+^{L} \otimes H_-^{R} \in G_n^{L} \otimes
 G_n^{R} $ that leaves them unchanged:
\br
H_+^{L} &=& \{ g_+^{L}(z) \in G_n^{L}:   g_+^{L}(z)= exp {[{1\o
{2{\cal K}_{11}}}\lambda_1 H \omega_0(z)]} 
exp {[\sum_{[\a]_1}\omega_{[\a]_1}(z)E_{[\a ]_1}]}\} \nonumber \\
H_-^{R} &=& \{ g_-^{R}(\bar z) \in G_n^{R}:   g_-^{R}(\bar z)= 
exp {[\sum_{-[\a]_1}\bar \omega_{-[\a]_1}(\bar z)E_{-[\a]_1}]}exp {[{1\o
{2{\cal K}_{11}}}\lambda_1 H \bar \omega_0(\bar z)]} \} 
\nonumber
\er
where $[\a ]_1$ denotes all positive roots except $\a_1$.  This symmetry allows
us to  remove the remaining irrelevant (fields) degrees of freedom by choosing
an appropriate gauge.  Note that eqs. (\ref{2.4}) can be also considered as
specific gauge fixing conditions for the `` constraints'' subgroup 
 $H_-^{c} \otimes H_+^{c} \in G_n^{L} \otimes G_n^{R}$
\br
H_+^{c} &=& \{ g_+^{c}(\bar z) \in G_n^{R}:   g_+^{c}(\bar z)=exp {[{1\o
{2{\cal K}_{11}}}\lambda_1 H \bar \omega_0(\bar z)]} exp {[\sum_{[\a]_1}
 \bar \omega_{[\a]_1}( \bar z)E_{[\a]_1}]}\} \nonumber \\
H_-^{c} &=& \{ g_-^{c}( z) \in G_n^{L}:   g_-^{c}( z)= 
exp {\sum_{-[\a]_1} \omega_{-[\a]_1}( z)E_{-[\a]_1}}exp {[{1\o
{2{\cal K}_{11}}}\lambda_1 H  \omega_0( z)]} \}
\nonumber 
\er
We can the further combine $H_+^{L}, H_-^{R}$ and $H_{\pm}^{c}$ in two
{\it noncommuting } and {\it nonchiral }  subgroups $H_{\pm} \in G_n$:
\br
&&\hskip-0.7cm H_+ = \{ g_+=g_+^{L}(z)g_+^{c}(\bar z)  \in G_n:  
 g_+(z, \bar z)= exp {[{1\o
{2{\cal K}_{11}}}\lambda_1 H R]} 
exp {[\sum_{[\a]_1}\psi_{[\a]_1}E_{[\a]_1}]} \}\nonumber \\
&&\hskip-0.7cm H_- = \{ g_-=g_-^{c}( z)g_-^{R}(\bar z)  \in G_n:  
 g_-(z, \bar z)=  
exp {[\sum_{[\a]_1}\chi_{[\a]_1}E_{-[\a]_1}]} exp {[{1\o
{2{\cal K}_{11}}}\lambda_1 H R]}\}     
\label{2.5a}
\er
where $\chi_{[\a_1]}(z, \bar z)$, $\psi_{[\a_1]}(z, \bar z)$ and 
$R(z, \bar z)$ are arbitrary functions of $z= \tau + \sigma$ and $\bar z= 
\tau - \sigma$.  In this way, all the information concerning the 
``nonphysical'' (gauge)
degrees of freedom (that should be removed by Hamiltonian reduction procedure)
is encoded in the subgroups $H_{\pm} \in G_n$ and in the following two constant
matrices $\epsilon^{\pm} = \sum_{i=2}^{n} ({{\a_i^2}\o k})E_{\pm \a_i}$
(which indicate
the currents that are constrained to one).  The remaining ``physical'' degrees
of freedom belong to the factor group $H_0^{f} = H_0 /H_0^{0} $, ($H_0^{0} = \{
g_0^{0} \in G_n: g_0^{0} = exp {[{1\o {2{\cal K}_{11}}}\lambda_1 H R]}, [g_0^{0},
\epsilon^{\pm} ] =0 \}$):
\be
H_0^{f} = \{ g_0^{f} \in G_n: g_0^{f}(z, \bar z) = exp{[\chi E_{-\a_1}]}
exp{[\sum_{i=1}^{n-1}{{{2\a_{i+1}H} \o {\a_{i+1}^2}}}\phi_i ]} exp{[\psi E_{\a_1}
]}
\label{2.5b}
\ee
This decomposition of $G_n$ into $H_{\pm}$ and $H_0^{f}$ provides us with a
specific parametrization for each $g(z, \bar z) \in G_n: g(z, \bar z)=g_-
g_0^{f} g_+ $.  The latter is the {\it crucial ingredient} in the derivation of the
NA-Toda eqns. (\ref{2.3}) from the original $G_n$-WZW equations $\bar \pa J =  
\pa \bar J =0$ by imposing the constraints (\ref{2.4}) and the corresponding
gauge fixing conditions.  The proof is quite similar to the one of the abelian
Toda case \cite{Orai}(see also \cite{LAF}).  We leave the details to appendix
A.

The natural splitting of the original  $G_n$ valued WZW fields $g(z,\bar
z)$into {\it irrelevant} $g_{\pm} \in H_{\pm}$ and {\it physical} ones $g_0^{f} \in
H_0^{f}$ induced by  the constraints (\ref{2.4}) is an indication that the
$G_n^{(1,1)}$-NA Toda models (\ref{2.3}) can be described as gauged 
${H_-}\setminus {G_n}/{H_+}$
($\equiv G_n/H$)-WZW models. 
 Given the  subgroups $H_{\pm} \in G_n$
and the constant matrices $\epsilon^{\pm} \in {\cal H}_{\pm}$.  The standard
procedure \cite{Orai}, \cite{Ga-Ku} to construct the corresponding action
consists in the introduction of auxiliary gauge fields $A(z, \bar z) \in {\cal
H}_{-}$  and $\bar A(z, \bar z) \in {\cal
H}_{+}$:
\be
A=h_{-}^{-1}\partial h_{-},\quad \quad 
\bar{A}=\bar{\partial}h_{+}h_{+}^{-1},\quad h_{\pm} \in H_{\pm}
\label{2.6}
 \ee
interacting in an $H_{\pm}$ invariant way with the WZW field $g\in G_n$:
\be
\hskip-0.3cm S(g,A,\bar{A})=S(g)-\frac{k}{2\pi}\int dzd\bar{z} 
 \Tr\left\{A(\bar{\partial}gg^{-1}-\epsilon_{+})+\bar{A}(g^{-1}\partial
g-\epsilon_{-})+Ag\bar{A}g^{-1}+A_{0}\bar{A}_{0}\right\}
\label{2.7}
\ee
where $S(g)$ is the $G_n$-WZW action:
\begin{eqnarray}
S(g)=-\frac{k}{4\pi}\int dzd\bar{z} \Tr(g^{-1}\partial
gg^{-1}\bar{\partial}g)-\frac{k}{12\pi}\int_{B}d^{3}x\epsilon_{ijk}
\Tr(g^{-1}\partial_{i}gg^{-1}\partial_{j}gg^{-1}\partial_{k}g)
\nonumber
\end{eqnarray}
and $A_{0}=h_{0}^{-1}\partial h_{0}$
($\bar{A}_{0}=\bar{\partial}h_{0}h_{0}^{-1}$) ($h_0 \in H_0^0$) 
is the diagonal part of
$A$ and $\bar{A}$.   Since 
 $H_{\pm}$ (\ref{2.5a}) are
by construction semidirect products of nilpotent 
$N_{\pm}^{(1)}$ and diagonal
$H_0^0$ subgroups, the gauge fields $A \in {\cal H}_-$ and 
 $\bar A \in {\cal H}_+$ covariantly split into nilpotent
  $A_- (\bar A_+)$ and
 diagonal $A_0 (\bar A_0)$ parts $
A=A_{-}+A_{0},\quad  
\bar{A}=\bar{A}_{0}+\bar{A}_{+} $.

 The structure of the $A (\bar A)$ dependent terms in (\ref{2.7}) represents a
 mixture of the familiar (vector) $U(1) \equiv H^0_0$ gauged WZW model(of
 PF-type) and the nilpotent $N^{(1)}_{\pm}$ gauged WZW, similar to the one that
 gives the abelian Toda model \cite{Orai}.  The specific combination of terms
 that appears in (\ref{2.7})is fixed by the requirement of $H_{\pm}$-invariance
 of the action $S(g,A,\bar{A})$, i.e. under the following transformations
 \br
 g^{\pr} &=&  \a_{-} g \a_{+},
  \quad A_0^{\pr} = A_0 + \a_0^{-1} \pa \a_0, \quad
 \bar  A_0^{\pr} =\bar  A_0 + \bar \pa \a_0 \a_0^{-1} \nonumber \\
 A^{\pr} &=& \a_-^{-1}A\a_- + \a_-^{-1} \pa \a_-, \quad 
 \quad \bar  A^{\pr} =\a_+\bar A \a_+^{-1} + \bar \pa \a_+ \a_+^{-1}
 \label{2.8}
 \er
 where $\a_{\pm}(z, \bar z) \in H_{\pm}$ and $\a_0 \in H_0^0$.

What remains  to be shown is  that the action (\ref{2.7}) indeed describes
the $G_{n}^{(1,1)}$-NA-Toda theories. The way we are going to prove it
consists in deriving an effective action (\ref{lagr11}) of
 the NA-Toda model by
integrating out the auxiliary fields $A$, $\bar{A}$ in (\ref{2.7}). In order
to perform the (matrix) functional integral of Gauss type in $A$ and
$\bar{A}$, we first simplify (\ref{2.7}) by gauge fixing the 
$H_{\pm}$-symmetries.
We choose $\alpha_{\pm}=g_{\pm}^{-1}$, therefore
$g^{\prime}=
 \alpha_{-}g\alpha_{+}=\alpha_{-}g_{-}g_{0}^f g_{+}\alpha_{+}=g_{0}^f$,
i.e., $
S(g,A,\bar{A})=S(g_{0}^f,A^{\prime},\bar{A}^{\prime})$.
Equally well, one can consider this as a change of the variables
$A,\bar{A}\rightarrow A^{\prime},\bar{A}^{\prime}$ in the functional
integral. Taking into account the specific form of 
$g_{0}^f\in H_{0}^f$ (see eqn.
(\ref{2.5b})) and that $A^{\prime}\in {\cal H}_{-}$, $\bar{A}^{\prime}\in
{\cal H}_{+}$ we verify that the following trace identities hold:
\br
&&{\rm Tr}A^{\prime}\bar{\partial}g_{0}^f{(g_{0}^f)}^{-1}=
{\rm Tr}A_{0}^{\prime}\bar{\partial}g_{0}^f{(g_{0}^f)}^{-1},
\quad \quad
{\rm Tr}\bar{A}^{\prime}{(g_{0}^f)}^{-1}\partial g_{0}^f=
{\rm Tr}\bar{A}_{0}^{\prime}{(g_{0}^f)}^{-1}\partial g_{0}^f
\nonumber
\\
&&{\rm Tr}A^{\prime}g_{0}^f\bar{A}^{\prime}{(g_{0}^f)}^{-1}=
{\rm Tr}A_{0}^{\prime}g_{0}^f\bar{A}^{\prime}{(g_{0}^f)}^{-1}+
{\rm Tr}A_{-}^{\prime}g_{0}^f\bar{A}_{+}^{\prime}{(g_{0}^f)}^{-1}.
\nonumber
\er
As a consequence, the functional integral over $A$, $\bar{A}$
 splits in a
product of two integrals: the first one lying in the diagonal subalgebra
${\cal H}_{0}$ of ${\cal H}_{0}^f$ :
\br
&&\hskip2cm Z_{0}=\int DA_{0}^{\prime}D\bar{A}_{0}^{\prime}
\exp \left\{ -G_{0}(A_{0}^{\prime},\bar{A}_{0}^{\prime})\right\} ,\nonumber\\
&&\hskip-2cm G_{0}=-\frac{k}{2\pi}\Tr\int dzd\bar{z}
[A_{0}^{\prime}g_{0}^f\bar{A}_{0}^{\prime}{(g_{0}^f)}^{-1}
+\bar{A}_{0}^{\prime}({(g_{0}^f)}^{-1}\partial
g_{0}^f) + 
A_{0}^{\prime}(\bar{\partial}g_{0}^f{(g_{0}^f)}^{-1})+A_{0}^{\pr} 
\bar A_{0}^{\prime}],
\label{2.9}
\er
 and the second one, lying in the nilpotent subalgebras ${\cal N}_{\pm}^{(1)}$:
\br
&&Z_{+-}=\int DA^{\prime}D\bar{A}_{+}^{\prime}
\exp \left\{ -G_{+-}(A_{-}^{\prime},\bar{A}_{+}^{\prime})\right\} ,\nonumber\\
&&G_{+-}=-\frac{k}{2\pi}{\rm Tr}\int dzd\bar{z}
[A_{-}^{\prime}g_{0}^f\bar{A}_{+}^{\prime}{(g_{0}^f)}^{-1}
-\bar{A}_{+}^{\prime}\epsilon_{-}-A_{-}^{\prime}\epsilon_{+}].
\label{2.10}
\er
According to the definition of ${\cal H}_{0}$, an arbitrary 
$A_{0}^{\prime}\in
{\cal H}_{0}^0$ can be parametrized by only one function
$a_{0}^{\prime}(z,\bar{z})$:
\be
A_{0}^{\prime}=\frac{1}{2{\cal K}_{11}}\lambda_{1}\cdot H a_{0}^{\prime},
\quad \quad 
\bar{A}_{0}^{\prime}=\frac{1}{2{\cal K}_{11}}\lambda_{1}\cdot H 
\bar{a}_{0}^{\prime}.
\label{2.11}
\ee
We first simplify $G_0(a_0^{\pr}, \bar a_0^{\pr})$ taking into account the
explicit form (\ref{2.5b}), (\ref{2.11}) of $g_0^f$, and $A^{\pr}_0, 
\bar A^{\pr}_0 $ and the basic trace formulas for the  $G_n$-generators
\br
\Tr (H_iH_j )= \d_{ij}; \quad \Tr (H_i E_{\a}) = 0; \quad \Tr (E_{\a}E_{\b}) = {2\o
{{\a}^2}}\d_{\a+\b,0}
\nonumber
\er
The result is 
\br
G_0(a_0^{\pr},\bar a_0^{\pr} ) = -{k\o {2\pi }}\int
\frac{dzd\bar{z}}{2{\cal K}_{11}}
\left\{ a_{0}^{\prime}\bar{a}_{0}^{\prime}\Delta
-e^{ k_{12}\phi_{1}} (a_{0}^{\prime}\chi \bar{\partial}\psi
+\bar{a}_{0}^{\prime}\psi \partial \chi )\right\} ,
\nonumber
\er
where $\Delta=1+\frac{1}{2{\cal K}_{11}}e^{ k_{12}\phi_{1}} \psi \chi $.
Thus the matrix integral (\ref{2.9}) reduces to simple functional Gauss integral
for the  scalars $a_0^{\pr},\bar a_0^{\pr}$:
\be
Z_0 = {{2{\cal K}_{11}}\o k\Delta }e^{-S_0^{eff}}, \quad 
S_{0}^{{ eff}}=\frac{k}{2\pi}\int dzd\bar{z}e^{
2k_{12}\phi_{1}} \frac{ \psi \chi \bar{\partial}\psi \partial
\chi}{2{\cal K}_{11}\Delta}
\nonumber 
\ee
The calculation of $S(g_0^{f})$ can be easely performed applying the
Polyakov-Wiegman decomposition formula  for each of the multipliers in
$g_0^f$ (see (\ref{2.5b})):
\be
S(g_{0}^f)=-\frac{k}{4\pi}\int dzd\bar{z}\left\{ \tilde \eta_{ik}\partial
\phi_{i}\bar{\partial}\phi_{k}+2\exp \left\{ k_{12}\phi_{1}\right\} \partial \chi
\bar{\partial}\psi \right\}
\label{2.13}
\ee
where $\tilde \eta_{ik} = 4 {{\a_i \a_j}\o {\a_i^2 \a_j^2}}$ is the
Killing-Cartan form for $G_{n-1}$ (obtained by deleting the first point of the
Dynkin diagram of $G_n$).
In order to take the integral (\ref{2.10}) we have to rewrite $G_{+-}$
separating the exact  square term
\be
G_{+-} =-{k\o {2\pi }}\int dz d\bar z tr\left\{ (A^{\pr}_{-} - g_0^f \epsilon
_{-}{(g_0^f)}^{-1})g_0^f (\bar A^{\pr}_{+} - {(g_0^f)}^{-1} \epsilon
_{+}{g_0^f}){(g_0^f)}^{-1} - \epsilon_+ g_0^f \epsilon_- {(g_0^f)}^{-1} \right\}
\nonumber 
\ee
Its contribution to the effective action yields
\be
S_{+-}^{{ eff}}
=\frac{k}{2\pi}\int dzd\bar{z}
{ tr}[\epsilon_{+}{g_{0}^f}\epsilon_{-}{g_{0}^f}^{-1}]
=\frac{k}{2\pi}\int dzd\bar{z}({2\o k})^2
\sum_{i=1}^{n-1}\frac{2}{\alpha_{i+1}^{2}}\exp \left\{
-\tilde k_{ij}\phi_{j}\right\}
\label{2.14}
\ee
  Combining together
$S_0^{eff}$, $S(g_0^f)$ and $ S_{+-}^{eff}$ we find that the effective classical
action  (all the determinant factors from the Gauss integration and changes of
variables neglected) for the $H_{\pm}$ gaugeed $G_n$-WZW model has the form
\be
S_{G/H}^{{ eff}}=
-\frac{k}{2\pi}\int dzd\bar{z}\left\{ \frac{1}{2}\tilde \eta_{ik}\partial
\phi_{i}\bar{\partial}\phi_{k}+e^{ k_{12}\phi_{1}} 
\frac{\bar{\partial}\psi \partial \chi}{\Delta}
-({2\o k})^2\sum_{i=1}^{n-1}\frac{2}{\alpha_{i+1}^{2}}
e^{ -\tilde k_{ij}\phi_{j}} \right\} .
\label{2.15}
\ee

Finally, comparing the equations of motion derived from (\ref{2.15}) with the
NA-Toda ones (\ref{2.3}), we realize that they do coincide. Therefore,
the $S_{G/H}^{{\bf eff}}$ is the action for the $G_{n}^{(1,1)}$-NA-Toda
theories we were looking for. This completes our proof that the class of
NA-Toda models we are considering are equivalent to the gauged
${H_{-}}\setminus {G_{n}}/{H_{+}}$-WZW models.

Note that the second term in (\ref{2.15}) contains both symmetric and
antisymmetric parts:
\begin{eqnarray}
\frac{e^{ k_{12}\phi_{1}}} {\Delta}\bar{\partial}\psi \partial \chi
=\frac{ e^{k_{12}\phi_{1}} }
{\Delta}(g^{\mu \nu}\partial_{\mu}\psi\partial_{\nu}\chi
+\epsilon_{\mu \nu}\partial_{\mu}\psi \partial_{\nu}\chi ),
\nonumber
\end{eqnarray}
where $g_{\mu \nu}$ is the 2-D metric of signature $ g_{\mu
\nu}= diag (1,-1)$. For $n=1$ ($G_{n}\equiv A_{1}$, $\phi_{1}$ is zero) the
antisymmetric term is a total derivative:
\begin{eqnarray}
\epsilon_{\mu \nu}\frac{\partial_{\mu}\psi \partial_{\nu}\chi}{1+\psi \chi}
=\frac{1}{2}\epsilon_{\mu \nu}\partial_{\mu}
\left( \ln \left\{ 1+\psi \chi \right\}
\partial_{\nu}\ln{\frac{\chi}{\psi}}\right),
\nonumber
\end{eqnarray}
and it can be neglected. This $A_{1}$-NA-Toda model is known to describe the
2-D black hole solution for (2-D) string theory \cite{Witten1}. The
$G_{n}^{(1,1)}$-NA-Toda model ($G_{n}=B_{n}$ or $D_{n}$) can be used in the
description of specific (n+1)-dimensional black string theories \cite{GS},
 with $\hskip1cm$ n-1-flat and
2-non flat directions ($g^{\mu
\nu}G_{ab}(X)\partial_{\mu}X^{a}\partial_{\nu}X^{b}$, $X^{a}=(\psi ,\chi
,\phi_{i})$), containing axions ($\epsilon_{\mu
\nu}B_{ab}(X)\partial_{\mu}X^{a}\partial_{\nu}X^{b}$) and tachions
($\exp \left\{ -k_{ij}\phi_{j}\right\} $), as well.
For this geometric interpretation of the NA-Toda models,
it is convenient to rewrite the antisymmetric term in
(\ref{2.15}) as:
\begin{eqnarray}
\frac{e^{k_{12}\phi_{1}} }{\Delta}\epsilon_{\mu
\nu}\partial_{\mu}\psi
\partial_{\nu}\chi =
-\frac{1}{2}\epsilon_{\mu \nu}k_{12}\partial_{\mu}\phi_{1}(\psi
\partial_{\nu}\chi -\chi \partial_{\nu}\psi
)\frac{\exp \left\{ k_{12}\phi_{1}\right\} }{\Delta}+{\cal K}_{11}\epsilon_{\mu
\nu}\partial_{\mu}\left(
\ln \left\{ \Delta \right\} \partial_{\nu}\ln{\frac{\chi}{\psi}}\right) .
\nonumber
\end{eqnarray}
Discarding the total derivative term, the action (\ref{2.15}) takes its final
form:
\begin{eqnarray}
S_{G/H}^{{ eff}}&=& S_{G_{n}^{(1,1)}}^{NA}
=-\frac{k}{2\pi}\int d^{2}z \{ \tilde \eta_{ik}g^{\mu 
\nu}\partial_{\mu}\phi_{i}\partial_{\nu}\phi_{k}
+\frac{e^{ k_{12}\phi_{1}} }{\Delta }g^{\mu \nu}\partial_{\mu}\psi
\partial_{\nu}\chi  
\nonumber
\\
&-&
({2\o k})^2\sum_{i=1}^{n-1}\frac{2}{\alpha_{i+1}^{2}}e^{
-\tilde k_{ij}\phi_{i}}  
-
\frac{1}{2}\epsilon_{\mu \nu}k_{12}\partial_{\mu}\phi_{1}(\psi \partial_{\nu}\chi
-\chi \partial_{\nu}\psi )\frac{e^{ k_{12}\phi_{1}}}{\Delta} \} .
\label{2.17}
\end{eqnarray}

  Our definition of the $G_n^{(1,1)}$-NA Toda models (\ref{2.15}) is based on
  an appropriate set of constraints (\ref{2.4}) on the $G_n$-WZW currents and
  their residual gauge symmetries.  As we have shown this data can be
  transformed into a (complete) set of subgroups $H_{\pm}$ , 
  the factor group $H_0^f$ of $G_n$
  and the matrices $\epsilon_{\pm}$.   This allows us to represent these
  NA-Toda models as ${H_-}\setminus {G_n}/H_+$-gauged WZW models.  One might 
  wonder whether
  they can be derived following the original Leznov-Saveliev (LS) 
   approach \cite{LS} to
  the NA-Toda theories and vice-versa, i.e. given a model from the
   LS-scheme, could
  one write it as gauged WZW model?  As it is shown in 
  appendix A starting from
  our NA-Toda data, $H_{\pm}, H_0^f$ (and $H_0$), $\epsilon _{\pm}$ one can
  construct an unique  grading operator $Q= \sum_{i=2}^{n}{{ 2\lambda_i }\o
  {\a_i^2}}H$ such that $[Q, {\cal H}_{\pm} ] =\pm \eta {\cal H}_{\pm}$, 
$[Q, { H}_{0} ] = 0$, $[{\cal H}_0^0 , \epsilon_{\pm}] = 0$, $H_0^f =
H_0/{H_0^0}  $ and finally to derive  eqns (\ref{2.3})  following the
LS-approach.   There exists however a {\it difference} between the LS-NA-Toda
models and our $G_n^{(1,1)}$-models.  It consists in the {\it following}:  given a
grading operator ( from the Kac classification table) 
$Q= \sum_{i=1}^{n}{2s_i \lambda_i H\o {\a_i^2}}$.  To construct a model of the LS-type one
has to find an appropriate (linear) combinations $\epsilon_{\pm}$ of the $G_n$
step operators of $Q$-grade $\pm 1$, such that $Q$ and $\epsilon_{\pm}$ closes
an $SL(2,R)$ algebra.   The $\epsilon_{\pm}$ of the $G_n^{(1,1)}$- models
closes an $SL(2,R)$ algebra not with $Q$ but with $Q^{\pr} = Q + {\cal H}_0^0$,
(${\cal{H}}_0^0 = \b \lambda_1 H, [\epsilon_{\pm} , {\cal H}_0^0 ]  = 0$), i.e. 
$[\eps_+, \eps_- ] = = Q + {\cal H}_0^0 $.  The equivalence of the
$G_n^{(1,1)}$-NA Toda models with certain 
$ {H_-}\setminus {G_n}/H_+ (\eps_{\pm}) $-WZW
coset models raises the question: What is the gauged WZW model describing the
NA-Toda theories of the LS-type?  One has to repeat literally the construction
presented in this section using the corresponding (LS) $\eps_{\pm}$'s and  
as $H_{\pm}$ -- the positive (negative ) $Q$-grade subgroups of $G_n$. 

 The main
advantage of the description  of the NA-Toda models as $G/H$-WZW models is that
it provides a powerful tool for the construction of conserved currents, for the
derivation of their $V_n$-algebras (see Sect 3) as well as for their
quantization (see Sect 9).

\sect{Conserved Currents and  $V^{(1,1)}_{n+1}$-algebras}

We start our analysis of the symmetries of the 
$G^{(1,1)}_n$-NA-Toda theories
(\ref{2.17}) with the construction of the improved 
(classical ) stress-tensor
$T_{\mu \nu}$ and the global $U(1)$ current $J_{\mu}$.  
Since the action
(\ref{2.17}) is manifestly translation, Lorentz and 
dilation invariant, the
corresponding $T_{\mu \nu}$ is conserved, symmetric and 
traceless and its two
nonvanishing (chiral ) components $T(z)$ and $\bar T(\bar 
z)$ are given by 
\be
T(z) = {1\o 2}\eta_{ik} \pa \phi_i \pa \phi_k +
\sum {2\o {\a_i^2}}\pa ^2\phi_i + 
{{\pa \chi \pa \psi }\o \Delta}e^{k_{12}\phi_1} + 
\gamma \pa ({{\psi \pa \chi }\o \Delta }e^{k_{12}\phi_1})
\label{3.1}
\ee
where $\gamma = \sum_{i=1}^{n-1}{\tilde {\cal K}}_{1,i} $ (i.e. $\gamma_{A_n} =
{{n-1}\o 2}, \gamma_{B_n} = n-1, \gamma_{C_n} = n-{3\o 2}$).
 $\bar T$ have the same form with $\pa ,\psi , \chi \longrightarrow 
\bar \pa , \chi ,\psi $.   Thus our 
NA-Toda  models 
(\ref{2.17}) are indeed conformally invariant.  Another 
evident symmetry of 
the action (\ref{2.17}) is under global $U(1)$ gauge 
transformations:
$\phi^{\prime}_i = \phi_i$,  $\psi^{\prime} = \psi 
e^{-\alpha }$ 
and $ \chi ^{\prime} = \chi e^{\alpha }$
($\alpha $ is a constant).  The corresponding ({\it nonchiral }) 
$U(1)$ current derived from (\ref{2.17}) is of the form 
\be
J_{\mu} = -{({k\o {4\pi }})}{{e^{k_{12}\phi_1}\o \Delta}}(\psi \pa _{\mu} \chi 
- \chi \pa _{\mu} \psi - 
k_{12}\psi \chi \pa _{\mu} \phi_1)
\label{3.2}
\ee
and its conservation reads $ \pa \bar J + \bar \pa J =0$, 
where $J = {1\o 2}(J_0 - J_1)$ and 
$\bar J = {1\o 2}(J_0 + J_1)$.

Similarly to the abelian Toda case \cite{Ger-Bil} 
 one might expect 
that the NA-Toda theories obey a larger set of  
symmetries generated by 
certain higher spin conserved currents.  This is indeed 
the case, however 
it is rather difficult to derive them from the effective 
action 
(\ref{2.17}).  A powerful and systematic method of 
exhausting all the
 symmetries
of the models (\ref{2.17}) is based on their equivalence 
with specific 
$H$-reduced $G_n$-WZW models as we have shown in Sect.2. 
Since the equations of motion of WZW theory ($\bar \pa 
J_{\alpha} = \pa \bar J_{\alpha} = 0 $) reduced by the 
constraints (\ref{2.4}) do coincide with the NA-Toda 
field equations  (\ref{2.3}), the 
remaining unconstrained  WZW currents  appears as the 
conserved currents for the 
$G^{(1,1)}_n$-NA-Toda theories as well.  The recipe for 
deriving the explicit form of these currents in terms of 
fields $\chi$, $\psi$ and $\phi_i$  consist in the 
following; i) choose an independent set of remaining 
currents by gauge fixing the residual gauge symmetries of 
(\ref{2.4}); ii) realize all WZW currents in the 
g-parametrization, $g=g_- g_0^f g_+$ (see Sect. 2)
 and next solve the 
constraints (and gauge fixings conditions) against the 
physical fields $\chi$, $\psi$ and $\phi_i$; iii) 
substitute these solutions in the remaining currents.

Let us first find an independent set of remaining 
currents and calculate their (improved) spins. As we have shown 
in Sect.2 the  constraints (\ref{2.4}) remains invariant 
under the group $H_+^{L}\otimes H_-^{R}$.
This allows us to choose a specific DS-type gauge 
\cite{Orai} such that the only independent nonvanishing 
remaining current are:
\be
J_{-\a_1} = V^+, \quad J_{\a_1 + \a_{2} + \cdots \a_n} = 
V^- \quad 
J_{\a_n + \a_{n-1} + \cdots \a_k} = W^{A}_{n-k+2}
\label{3.3}
\ee
for $A_n$ and 
\br
J_{-\a_1}& = &V^+, \quad \quad J_{\a_1 + 2\a_{2} + \cdots 2\a_n} 
= V^- \quad \nonumber \\
J_{\a_k + 2\a_{k+1} + \cdots 2\a_n} &=& W^{B}_{2(n-k+1)}, \;\; 
k=2,3,\cdots ,n-1, \quad J_{\a_n} = W_2
\label{3.4}
\er
for $B_n$ and $C_n$.

For each chiral sector, the gauge fixing condition for the 
simple root constraints $J_{-\a_i} =1$, ($i \neq 1$) 
requires $J_{{2\lambda_iH \o \a_i^2}} = 0$.  In order to 
make these constraints consistent with the conformal 
invariance (including Lorentz), we have to improve the 
WZW-stress tensor $T_{WZW}$ by 
\be
\tilde T = T_{WZW} + \sum_{i=2}^{n} \pa J_{{2\lambda_iH 
\o \a_i^2}}
\label{3.6}
\ee
such that the spin of $J_{-\a_i}$ with respect to $\tilde T$ is zero.
  Since $J_{-\a_1}$ is unconstrained we can add a 
term proportional to $\pa J_{\lambda_1H}$ i.e.
\be
T = \tilde T + X\pa J_{\lambda_1H}
\label{3.7}
\ee
The condition that fixes $X$ comes from the consistency of the conformal
transformation of $J_{\lambda_1H}$ generated by $T$, 
\br
\{T(\sigma), J_{\lambda_1H}(\sigma^{\prime }) \} =
 J_{\lambda_1H}(\sigma^{\prime } )
\d^{\prime }(\sigma - \sigma^{\prime }) + 
\pa _{\sigma^{\prime }}J_{\lambda_1H}(\sigma^{\prime } )\d
(\sigma - \sigma^{\prime }) + (X{\cal K}_{11} + 
\sum_{i=2}^{n}{\cal K}_{1i})\d^{\pr
\pr}(\sigma - \sigma^{\prime })
\nonumber
\er
with the constraint $J_{\lambda_1H} = 0$, i.e.
\be
X = - {1 \o {{\cal K}_{11}}}\sum_{i=2}^{n}{\cal K}_{1i}
\ee
Then the improved spins of the remaining currents $J_{\a}$ ( $\a$ being one of
the roots appearing in (\ref{3.3}), (\ref{3.4}), ) is given by,
\be
s(\a ) = 1 + X\lambda_1 \a + \sum_{i=2}^{n} 2 {\lambda_i \a \o {\a_i^2}}
\label{3.9}
\ee
For the $A_n^{(1,1)}$-NA-Toda models we have  $ X = -{(n-1) \o 2}$ and eqns.
(\ref{3.7}) gives
\be
s^- = s(-\a ) = {{n+1}\o 2} \quad \quad s^+ = s(\a_1 + \a_2 + \cdots \a_n ) =
{{n+1}\o 2}
\ee
In the $B_n$ case we find $X=1-n$ and rescaling that $\lambda_1 = \a_1 +
\a_2 + \cdots + \a_n $ we obtain
\br
s^- & = & s(-\a_1 ) = n \quad \quad s^+ = s(\a_1 +2\a_2 + 2\a_3 + \cdots +
 2\a_n ) = n\nonumber \\
s_2 &= & s(\a_n ) = 2, \quad \quad s_k = s(\a_k +2\a_{k+1} + \cdots + 2\a_n ) =
2(n-k+1) \nonumber
\er
The same is true for $C_n$-NA-Toda theories. 

   We have to note that the PF-type
constraint $J_{\lambda_1H} = \bar J_{\lambda_1H} = 0$ results in a system of
differential equations for the field $R$, 
\be 
\pa R = { e^{k_{12}\phi_1} \psi \pa \chi \o \Delta },  \quad \quad 
\bar \pa R = { e^{k_{12}\phi_1} \chi \bar \pa \psi \o \Delta }
\label{3.11}
\ee
Therefore the elimination of $R$ by solving (\ref{3.11}) introduce
certain nonlocal terms (of the type $e^{\a R}$) in the part of the remaining currents. 
To  construct the conserved currents for
$G^{(1,1)}_n$-NA-Toda for generic n is rather cumbersome task.  There exist
however few exceptions when we can easely perform all the calculations for
arbitrary $G_n$.  These are the two (simple root) nonlocal currents $V^+ =
J_{-\a_1}$ and $\bar V^{-} = \bar J_{\a_1}$,
\be
V^+(z) = {k\o 2}\, { e^{k_{12}\phi_1 +{1\o {2{\cal K}_{11}}}R} \pa \chi \o \Delta },
\quad \quad 
\bar V^-(\bar z) = {k\o 2}\, 
{ e^{k_{12}\phi_1 +{1\o {2{\cal K}_{11}}}R} \bar \pa \psi \o \Delta}
\label{3.12}
\ee
and the chiral components of the stress tensor $T(z) = J_{\a_n}$, and $\bar
T(\bar z) = \bar J_{-\a_n}$ which indeed 
coincides with the improved stress tensor
(\ref{3.1}), derived directly from (\ref{2.17}).  As in the case of the
abelian Toda theories \cite{Ger-Bil}, \cite{Orai} the higher spin currents
($s \geq 3$) have quite complicated form and the knowledge of their explicit
form is not necessary in the derivation of the complete algebra of
symmetries.
  Nevertheless we present here few examples
for the $A_n^{(1,1)}$-NA-Toda ($n=1,2$) mainly concerning the rest of the
nonlocal currents $V^-(z)$, $\bar V^+(\bar z)$.  Their explicit form happens
to be important for the derivation of the $SL(2,R)_q$ algebra of symmetries
in Sect.5.  For $n=1$ the remaining nonlocal currents of spin 1 have a
form even simpler than (\ref{3.12}),
\be
V_{n=1}^- = ({k\o 2})e^{-R}\pa \psi, \quad \quad \bar V_{n=1}^+ = ({k\o
2}) e^{-R}\bar \pa \chi
\label{3.13}
\ee
The full set of conserved currents in the $A_2$ case contains together with
(\ref{3.1}), (\ref{3.2}) and (\ref{3.12}) two extra nonlocal currents of spin
$3/2$,
\be
V^-_{n=2} = {({k\o 2})^2 }e^{-{3\o 4}R} \( \pa ^2\psi + {1\o 16} \psi (\pa R )^2 - \psi (\pa \phi_1 )^2 - \psi \pa ^2 \phi_1 - {1\o 4} \psi \pa
^2R - {1\o 2}\pa \psi \pa R \)
\label{3.14}
\ee
and $\bar V^+_{n=2} = V^-_{n=2}(\psi \rightarrow \chi, \pa
\rightarrow \bar \pa )$.

We now come to the {\it main problem }addressed in this paper: 
{\it to derive the
complete algebra of the symmetries of the $G_n^{(1,1)}$-NA-Toda theories
given by the action (\ref{2.17})}.  As we have shown above this algebra is
generated by the $n+1$ chiral $V^{\pm}, T, W_p^{(G)}$ and $n+1$ antichiral 
$\bar V^{\pm}, \bar T, \bar W_p^{(G)}$  conserved currents (\ref{3.3}),
(\ref{3.4}), etc together with the nonchiral $U(1)$ current ($J(z,\bar z),
\bar J(z, \bar z)$).  Given the explicit form of the conserved currents, the
standard method for deriving their algebra consists in realizing them in terms
of the fields $\chi$, $\psi$ and $\phi_i$, their conjugate momenta
(obtained from (\ref{2.17})) and their space derivatives.  From the canonical
PB's one can, in {\it principle}, calculate the algebra we seek. 
 This method of
calculating is known to be difficult and cumbersome, even for the simple
cases $n=1,2,3$ where all necessary ingredients are at hand.  Fortunately a
short cut exists and is given by a simple procedure proposed by Polyakov
\cite{Pol1} for deriving the $W_3$ algebra from the constraint $SL(3,R)$
current algebra transformations.  This method does not require any knowledge
of the explicit form of the currents.  We are going to demonstrate now how it
works in our case of parafermionic $H$-reduction of the $G_n$-WZW model
described in Sect.2.  The starting point are the $G_n$ current algebra
infinitesimal transformations 
\be
\d J = [\epsilon , J] - {k\o 2}\pa \epsilon
\label{3.15} 
\ee
with $J = \sum _{all\; roots}J_{\{ \a \} }E_{\{ \a \} } + \sum J_i H_i$ and the
same decomposition for the chiral parameters $\epsilon (z) = 
\sum _{all\; roots}\epsilon _{\{ \a \} }E_{\{ \a \} } + 
\sum \epsilon _i H_i$.  We next substitute the constraints  (\ref{2.4})
 and the gauge
fixing condition ( in the DS gauge ) in (\ref{3.15}) and further require that
the remaining gauge transformation (generated by $V^{\pm}$, $T$ and
$W_p^{(G)}$ )to leave (\ref{2.4}) invariant.  As a result, (\ref{3.15})
gives rise to a system of $n+1$ linear algebraic equations and $n^2 - 2$
first order differential equations for the redundant $\epsilon $'s we have to
solve in terms of the independent parameters $\epsilon ^{\pm}$, $\epsilon $,
$ \eta _p$ and the currents $V^{\pm}$, $T$, $W_p^{(G)}$.  The remaining $n+1$
equations of (\ref{3.15}) represent the effective transformation laws of  
$V^{\pm}$, $T$, $W_p^{(G)}$ we are looking for.  It becomes evident from the
explicit form of eqs (\ref{3.15}) that our system of differential equations, 
is
diagonal on the derivatives and that the first equation (the coefficient
 of $H_1$)
splits from the others.  Its integration is straightforward for generic
$G_n$,
\be
\epsilon _1 = {1 \o {2k^2}}\int \epsilon (\sigma - \sigma ^{\pr})(\eps ^+ V^- -
\eps ^- V^+)(\sigma ^{\pr}) d \sigma ^{\pr}
\label{3.16}
\ee
It gives rise to the only nonlocal quadratic terms in the $VG_n^{(1,1)}$
algebra.  The integration of the rest is reasonably simple but the explicit
form of the recursive relations to be solved depends heavely on the $G_n$
algebra in consideration.  This is why we choose the simplest case of PF
reduction of the $A_n$-- current algebra in order to demonstrate our method in
solving the system of algebraic and differential equations (eqn. (\ref{3.15})
reduced by the constraints (\ref{2.4}) in the DS gauge ).  It is convenient
to write eqn. (\ref{3.15}) in the Cartan- Weyl 
basis and realize the $A_n$ generators
$E_{\a}$ , $H_i$
in terms of $(n+1)\times (n+1)$ matrices $(E_{ij})_{kl} = \d_{ik} \d_{jl}$:
$H_i = E_{ii}$, $E_{\a} = E_{ij}, \;\; (i<j)$, $E_{-\a} = E_{ij}, \;\; (i>j)$,
$i,j =1,\cdots n+1 $ and $\sum_{i=1}^{n+1} H_i =0 $.  The constraints
(\ref{2.4}) takes the form 
\be
J_{i,i-1} =1, \;\; i=3,\cdots n+1 \quad \quad J_{ij} = 0, \;\; i>j \neq i-1,\quad\quad J_{11}=0
\label{3.17}
\ee
the remaining currents are then given as 
\be
J_{2,1} = V^+, \;\;\; \;\; J_{1,n+1} = V^- ,\;\;\;\;\; J_{p,n+1} =
W_{n-p+2}, \;\; \;\;p=2, \cdots ,n 
\label{3.18}
\ee
and all other elements $J_{ii}\;\;, i=2,\cdots ,n+1$, $J_{kl},\;( k<l
\neq n+1)$ are zero in the DS--gauge.  Apart from the equations for the
transformations of currents (\ref{3.18}) the rest of the system (\ref{3.15})
can be written in the following matrix form 
\be
- {k\o 2}\pa \eps _{ik} = J_{ij}\eps _{jk} - \eps _{ij}J_{jk}
\label{3.19}
\ee
(the equation for $(ik) = \{ (21),(p,n+1), (1,n+1) \} $ excluded ).  We next
choose the independent parameters to be $\eps ^- = \eps _{12}, \; \eps ^+ =
\eps _{n+1,1} $ for the transformations generated by the nonlocal currents
$V^{\pm}$, $\eps = \eps _{n+1,n} $ for conformal transformation $(T= W_2 )$
 and
 $\eta_{n-p+2} = \eps _{n+1,p} $
the transformation generated by the
highest spin currents $W_{n-p+2}, \;\; p=2, \cdots n-1 $.  The problem is to
solve eqs. (\ref{3.19}) for all others $\eps _{ik}$ in terms of $\eps
^{\pm}$, $\eps $ and $\eta_p$ and $V^{\pm}$, $T$ and $W_{p}$.  We first
derive the {\it conformal transformations} $\d _{\eps}V^{\pm}|_{\eta_p = \eps^{\pm}
= 0}$, etc , setting $\eta_p = \eps ^{\pm} = 0 $ in eqn. (\ref{3.19}).  In
this case all $\eps _{ik}$'s for $i >k$ satisfy  simple algebraic
equations and their solution is 
\be
\eps _{i,i-1} =\eps \quad i=3,\cdots , n+1 , \quad \quad  \eps _{21} = \eps V^+,
\quad \quad \eps _{ik} = 0 \;\; i>k \neq i-1
\label{3.20}
\ee
The diagonal elements $\eps _{ii},\;\; i=1,\cdots n+1$ with $\sum_{i=1}^{n+1}
\eps _{ii} = 0$ satisfy the following system of differential  recursive
relations
\br
\eps _{ii} - \eps _{i-1,i-1} = {k\o 2}\pa \eps, \quad i=3, \cdots , n
\nonumber \\
-\eps_{n,n} - \sum_{l=1}^{n} \eps _{l,l} = {k\o 2} \pa \eps, \quad \eps
_{11}=0 
\nonumber 
\er
The solution is 
\be
\eps _{ss} = {(2s-n-3)\o 2} ({k\o 2}) \pa \eps, \quad s=2, \cdots n \quad \eps
_{11} = 0
\label{3.21}
\ee
We next consider the equations for the upper triangular part of $\eps _{ik}$
and find that all elements of the first row, $\eps _{1,s}$ vanish except the
last one, 
\be
\eps_{1,s} = 0, \quad  s=2,\cdots ,n; \quad \eps_{1,n+1} = \eps V^-
\label{3.22}
\ee
The recursive relations for $\eps _{l,l+1}$, $l=1, \cdots ,n$ are of the form
\br
\eps _{l,l+1} - \eps _{l-1,l} = {k\o 2}\pa \eps_{ll} + \eps T \d_{l,n}
\nonumber 
\er
and can be easely solved by taking
\be
\eps _{l,l+1} ={k\o 2}\sum_{s=2}^{l} \pa \eps _{ss} + \eps T \d _{l,n} =
{(l-1)\o 2}(l-1-n) ({k\o 2}\pa )^2 \eps + \eps T \d_{l,n}
\label{3.23}
\ee
Similarly for $\eps _{l,l+2}$ we get for $l=2,\cdots , n-1$
\br
\eps_{l,l+2} = {k\o 2}\sum_{s=2}^{l}\pa \eps_{s,s+1} + \eps W_3 \d_{l,n-1} =
({k\o 2})^2 \sum_{s_1=2}^{l} \sum_{s_2=2}^{s_1}\pa ^{2}\eps _{s_2,s_2} + \eps
W_3 \d_{l,n-1}
\nonumber
\er
and for generic $\eps_{l,l+m}, \;\; l=2, \cdots n-m+1;\;\;m=1,\cdots , n-1$,
\be
\eps_{l,l+m} = {k\o 2} \sum_{s=2}^{l} \pa \eps _{s,s+m-1}+ \eps W_{m+1}
\d_{l,n-m+1}
\label{3.24}
\ee
Taking into account (\ref{3.21}) and the well known multiple sum formula,
\be
\sum_{s_1 =1}^{l-1} \sum_{s_2 =1}^{s_1} \cdots \sum_{s_{m-1} =1}^{s_{m-2}}
 s_{m-1}
= {(l+m-2)\o {m! (l-2)!}} = \left( \begin{array}{c} {l+m-2} \\ {m} \end{array}
\right)
\label{3.25}
\ee
we  derive the explicit form for $\eps _{l,l+m}$,
\be
\eps_{l,l+m} = \eps W_{m+1} \d_{l,n-m+1} + 
\left( \begin{array}{c} {l+m-2} \\ {m} \end{array} \right) \{ {{l+m-1}\o {m+1}}
- {{n+1}\o 2} \} ({k\o 2}\pa )^{m+1} \eps 
\label{3.26}
\ee
$ l=2,\cdots n-m+1, \;\; m=1, \cdots n-1 $.
The eqs. (\ref{3.20}), (\ref{3.21}), (\ref{3.23}) and (\ref{3.26}) give the
general solution for eqn. (\ref{3.19}) for the case $\eps ^{\pm} = \eta_p =0$,
$p= 3, \cdots n$.  The remaining part of eqns. (\ref{3.15}) represent the
effective infinitesimal transformation for $V^{\pm }$, $T= W_2$ and  $W_p$ we
are looking for, i.e.
\br
\d_{\eps } V^+ &= &\eps _{22} V^+ - {k\o 2} \pa \eps _{21}, \quad \quad
\d_{\eps } V^- =
V^- \sum_{s=1}^{n} \eps _{ss} - {k\o 2} \pa \eps _{1,n+1}, \nonumber \\
\d_{\eps } W_s &=& \eps W_{s+1} -(s-1)W_s ({k\o 2}\pa \eps ) + 
\sum_{p=1}^{s-2}\eps
_{n-s+2,n-s+p+2}W_{s-p} - \eps _{n-s+1,n+1}\nonumber \\
& - &{k\o 2} \pa \eps _{n-s+2, n+1}
\label{3.27}
\er
Substituting the explicit form of the redundant $\eps _{ik}^{\pr}$ in
(\ref{3.27}) and rescaling the conformal parameter $\eps = -{2\o k} \tilde \eps
$ we obtain the desired conformal transformations
\br 
\d_{} V^{\pm} &= &{{n+1}\o 2}V^{\pm} \pa \tilde \eps + 
\tilde \eps \pa V^{\pm}
\nonumber \\
\d_{} W_s & =& sW_s \pa \tilde \eps + \tilde \eps \pa W_s - (-{k\o 2})^{s}
\left( \begin{array}{c} {n-1} \\ {s-1} \end{array}
\right)
{n(n+1)(s-1) \o {2s(s+1)}} \pa ^{s+1}\tilde \eps \nonumber \\
&- &\sum_{p=1}^{s-2} \left( \begin{array}{c} {n-s+p} \\ {p} \end{array}
\right)
[{{n-s+p+1}\o {p+1}} - {{n+1}\o 2}] W_{s-p}({k\o
2})^{p}\pa ^{p+1}\tilde \eps,  
\label{3.28}
\er 
$ s= 2,3, \cdots n, \quad W_2 = T$.
Note that the nonhomogeneous conformal transformations 
of $W_s$, $s=3,\cdots n$,
(the last two terms in $\d W_s$) reflect the fact that we are working
 in the DS
gauge, where the $W_s$ are not primary fields with respect to $T=W_2$.  One can
find an appropriate ``gauge transformation'' that maps the DS gauge in a gauge
where all $\tilde W_s $ are primary fields (see sect. 4 of ref.
(\cite{SSZ})
for the $A_3$-case ).  For example $\tilde W_3 = W_3 - {n-2 \o 2 }\pa T $ is
primary field.  The construction of primary $\tilde W_s $ for $s> 4$ requires
further investigation. 

 In order to find the transformations generated by the
nonlocal currents $V^{\pm}$ we have to solve eqns. (\ref{3.19}) for the
particular case where $ \eps = \eta_p = 0$, leaving this time, $\eps ^{\pm}$
unconstrained.  Following the same strategy we first consider the equations for
the lower triangular part of $\eps _{ik}, \; i>k$.  Starting with the $n^{th}$
row we find 
\be
\eps _{n-k,s} = 0, \quad s=2,\cdots n-k-1; \quad k=0,1,\cdots n-3
\label{3.29}
\ee
and the following recurrence relations for $\eps _{n-s, 1}$
\be
\eps _{n-s, 1} + \eps ^{+}W_{s+1} + {k\o 2} \pa \eps _{n-s+1, 1} = 0, \quad 
\eps _{n,1} = -{k\o 2} \pa \eps ^{+} 
\label{3.30}
\ee
$s=1,2, \cdots n-2 $.  The solution of (\ref{3.30}) is given by
\be
\eps _{l,1} = \sum_{s=0}^{n-l-1} (-1)^{s} ({k\o 2}\pa )^{s} (\eps ^{+}
W_{n-l-s+1} ) + (-{k\o 2}\pa )^{n-l+1} \eps ^{+}, \quad l= 2, \cdots n
\label{3.31}
\ee
For the diagonal elements $\eps _{i,i}$ we get
\br
\sum_{i=1}^{n} \eps _{i,i} &+ &\eps _{n,n} = 0, \quad \eps_{2,2} =  
\cdots =  \eps _{3,3} = \eps _{n,n}
\nonumber \\
{k\o 2}\pa \eps _{1,1} &+& \eps ^{+}V^{-} - \eps ^{-} V^{+} = 0
\nonumber
\er
and therefore ($\pa = 2 \pa _{\sigma} , \;\; \bar \pa \eps _{1,1} = 0 $)
\br 
\eps _{11}(\sigma ) &= &{1 \o 2k} \int \eps (\sigma - \sigma ^{\pr} )[ \eps
^-(\sigma ^{\pr}) V^+(\sigma ^{\pr}) - \eps ^+(\sigma ^{\pr}) V^-(\sigma
^{\pr})]d\sigma ^{\pr} \nonumber \\ 
\eps _{s,s} &=& -{1\o n} \eps _{1,1}, \quad s=2,\cdots n
\label{3.32}
\er
From the upper triangular part of eqn. (\ref{3.19}) we derive
\br
\epsilon_{1,l} &= &({k\o 2}\pa )^{l-2} \epsilon^{-}, \quad l=2,\cdots
n+1\nonumber \\
\epsilon_{l,l+1} &=& {{n-l+1}\o n} \epsilon^{-} V^{+} + {{l-1}\o n}
\epsilon^{+}V^{-}, \quad l=2,\cdots ,n \nonumber \\
\epsilon_{l,l+2}
&=& {k\o 2}(\pa \epsilon^{-} )V^{+} + {k\o 2} (l-1) \pa (\epsilon^{-}V^{+}) +
{{l(l-1)}\o {2n}} {k\o 2}\pa (\epsilon^{+} V^{-} - \epsilon^{-}V^{+})
\label{3.33}
\er
$l=2, \cdots ,n-1$, and for the generic $\epsilon_{l,l+m} $ 
the recursive relation is
\be
\epsilon_{l,l+m} = \epsilon_{1,m+1}V^{+} + \sum_{s=1}^{l-1} ({k\o 2}\pa
)\epsilon_{s+1,s+m},\quad m=2,\cdots , n+1-l
\label{3.34}
\ee
The solution of eqn. (\ref{3.34}) can be written in the following compact form
\br
\epsilon_{l,l+m} &=& {1\o n} \left( \begin{array}{c} {l+m-3} \\ {m-1} \end{array}
\right)
({k\o 2}\pa )^{m-1} 
[(n- {{l+m-2}\o m} )\epsilon^{-}V^{+} + 
{{l+m-2}\o m }\epsilon^{+}V^{-} ] \nonumber \\
&+& \sum_{s=0}^{m-2} \left( \begin{array}{c} {l+s-2} \\ {s} \end{array}
\right)
 ({k\o 2}\pa )^{s} (V^{+} ({k\o 2}\pa )^{m-s-1}
\epsilon^{-} ), \quad m=1, \cdots ,n+1-l
\label{3.35}
\er
The $\epsilon^{\pm}$-transformation laws for $V^{\pm}$, $T$ and $W_s$ derived
from (\ref{3.16}) are
\br
\d V^{+} &= &\eps_{2,2}V^{+} - \eps_{1,1}V^{+} - \eps^{+} W_n - {k\o 2}\pa \eps
_{2,1} \nonumber \\
\d V^{-} &=& (\eps_{1,1} + \sum_{i=1}^{n} \eps_{i,i} )V^{-} + \sum_{s=2}^{n}
\eps_{1,s}W_{n-s+2} - {k\o 2}\pa \eps_{1,n+1} \nonumber \\
\d W_{p} &=& \eps_{n-p+2,1}V^{-} +
\sum_{s=n-p+3}^{n} \eps_{n-p+2,s} W_{n-s+2} - \eps_{n-p+1,n+1} - {k\o 2}\pa
\eps_{n-p+2,n+1}
\label{3.36}
\er
$p=2,\cdots , n$.  Taking into account the explicit form (\ref{3.31}),
(\ref{3.32}), (\ref{3.33}) and (\ref{3.35} ) of the $\eps $'s  that contribute to
(\ref{3.36}) we find the transformations generated by the nonlocal currents
$V^{\pm}$ to be in the form ($\tilde \eps ^{\pm} = - {k\o 2}\eps ^{\pm} $):
\br
\d_{\eps^{\pm}} V^{+} &=& -{(n+1)\o {nk^2}}\int \eps (\sigma -
\sigma^{\pr})[ \tilde \eps ^{+}(\sigma^{\pr})V^{-}(\sigma^{\pr}) -
\tilde \eps ^{-}(\sigma^{\pr}) V^{+}(\sigma^{\pr})]V^{+}(\sigma )d \sigma^{\pr}
\nonumber \\
&+ &\sum_{s=0}^{n-2} ({k\o 2})^{s-1} (-\pa )^{s}(\tilde \eps ^{+} W_{n-s}) - 
({k\o 2})^{n-1}(-\pa )^{n}\tilde \eps ^{+}
\nonumber \\
\d_{\eps^{\pm}} V^{-} &=& {(n+1)\o {nk^2}}\int \eps (\sigma -
\sigma^{\pr})[ \tilde \eps ^{+}(\sigma^{\pr})V^{-}(\sigma^{\pr}) -
\tilde \eps ^{-}(\sigma^{\pr})V^{+}(\sigma^{\pr})] V^{-}(\sigma )d \sigma^{\pr}
\nonumber \\
&- &\sum_{s=0}^{n-2} ({k\o 2})^{s-1} W_{n-s}(\pa )^{s}\tilde \eps ^{-}  + 
({k\o 2})^{n-1}(\pa )^{n}\tilde \eps ^{-}
\nonumber \\
\d_{\eps^{\pm}} T &=& {{n+1}\o 2} V^{-} \pa \tilde \eps ^{+} + {{n-1}\o 2}
\tilde \eps^{+}\pa V^{-}+ {{n+1}\o 2} V^{+} \pa \tilde \eps^{-} + {{n-1}\o 2}
\tilde \eps^{-} \pa V^{+}
 \nonumber \\
\d_{\eps^{+}}W_{p} &=&V^{-} [\sum_{s=0}^{p-3} (-1)^{s}({k\o 2})^{s-1} \pa
^{s}(\tilde \eps ^{+}W_{p-s-1}) + (-{k\o 2})^{p-2} \pa ^{p-1}\tilde \eps
^{+}]\nonumber \\
&+& {{n-1}\o np} \left( \begin{array}{c} {n-2} \\ {p-1} \end{array}
\right)
({k\o 2})^{p-2}\pa ^{p-1}(\tilde \eps ^{+}V^{-}) +
 {{n-1}\o {n(p-1)}}\left( \begin{array}{c} {n-2} \\ {p-2} \end{array}
\right)
 ({k\o 2})^{p-2}\pa ^{p-1}(\tilde \eps ^{+}V^{-})
 \nonumber \\
 &-& \sum_{s=n-p+3}^{n} W_{n-s+2} \left( \begin{array}{c} {s-3} \\ {s+p-n-3} 
 \end{array} \right)
 {{s-2}\o {n(s+p-2-n)}}({k\o
 2})^{s+p-n-4}\pa ^{s+p-n-3}(\tilde \eps ^{+}V^{-})
  \nonumber \\
 \d_{\eps^{-}}W_{p} &=& \sum_{s=0}^{p-2}\left( \begin{array}{c} 
 {n-p+s-1} \\ {s} \end{array} \right)
 {2\o k}({k\o 2}\pa )^{s} 
[V^{+}({k\o 2}\pa )^{p-s-1}\tilde \eps^{-}] \nonumber \\
&+& {2 \o {nk}}[(n-{{n-1}\o p}) 
\left( \begin{array}{c} {n-2} \\ {p-1}   \end{array}
\right)
 + (n-{{n-1}\o {p-1}}) 
 \left( \begin{array}{c} {n-2} \\ {p-2} \end{array}
\right)]({k\o 2}\pa )^{p-1}(\tilde \eps^{-}V^{+}) \nonumber \\
 &+& {2\o k}\sum_{s=0}^{p-3}\left( \begin{array}{c} {n-p+s} \\ {s} \end{array}
\right)
 ({k\o 2}\pa )^{s+1}
[V^{+}({k\o 2}\pa )^{p-s-2}(\tilde \eps^{-}V^{+})] \nonumber \\
&-&{2\o k}\sum_{s=n-p+3}^{n}W_{n-s+2} 
\{ \sum_{l=0}^{s+p-n-4}\left( \begin{array}{c} {n-p+l} \\ {l} \end{array}
\right)
({k\o 2}\pa)^{l}[V^{+}({k\o 2}\pa )^{s+p-n-l-3}
 \tilde \eps^{-}]
\nonumber \\
&+& \left( \begin{array}{c} {s-3} \\ {s+p-n-3} \end{array}
\right)
 (1- {{s-2}\o {n(s+p-n-2)}})({k\o 2}\pa )^{s+p-n-3}
(\tilde \eps^{-}V^{+})\}
\label{3.37}
\er

We next consider the transformations generated by $W_3$ taking $\eps ^+ =
\eps = 0$, $ \eta_p = 0 $ for $p=4, \cdots , n$ leaving $\eta_3 \equiv \eta
$ as a free parameter in eqns. (\ref{3.19}).  Solving  eqn.
(\ref{3.19}) for $i \geq k $ we find 
\br
\eps _{n-k,s} &=& 0, \quad k=0,\cdots ,n-4;\quad s=1, \cdots ,
n-k-3\nonumber \\
\eps _{l,l-2} &=& \eta , \quad l= 4, \cdots ,n; \quad \eps _{3,1} = \eta
V^{+} \nonumber \\
\eps _{l,l-1} &=& (l-n-1){k\o 2} \pa \eta , \quad l=3, \cdots n;\quad \eps
_{2,1} = -(n-1)V^{+} {k\o 2} \pa \eta - {k\o 2} \eta \pa V^{+} \nonumber \\
\eps _{1,1} &=& 0, \quad \eps _{l,l} = -{2\o n}\eta T + \{{(n-1)(n-2) \o 3} -
{(l-2)(2n-l-1) \o 2}\}({k\o 2}\pa )^2 \eta , \;\; l=2, \cdots , n-1 \nonumber
\\
\eps _{n,n} &=& {{n-2}\o n} \eta T - {(n-2)(n-1) \o 6 }({k\o 2}\pa )^2 \eta 
\label{3.38}
\er
The solution of eqn. (\ref{3.19}) for $i<k$ is given by 
\br
\eps _{1,l} &=& 0 \quad l=2, \cdots , n-1; \quad \eps _{1,n} = \eta V^{-},
\quad
\eps _{1,n+1} = {k\o 2} \pa (\eta V^{-})\nonumber \\
\eps _{l,l+m} &=& \eta W_{m+2} \d _{n,l+m} + (\eta W_{m+2} + ({k\o 2}\pa )
(\eta W_{m+1})) \d _{n+1,l+m} - {2\o n} 
\left( \begin{array}{c} {l+m-2} \\ {m} \end{array}
\right) 
({k\o 2}\pa )^{m}(\eta T)
\nonumber \\
&+& \{ \left( \begin{array}{c} {l+m} \\ {m+2} \end{array}
\right)  + {(n^2-1)\o 3 }\left( \begin{array}{c} {l+m-2} \\ {m} \end{array}
\right) 
 -n \left( \begin{array}{c} {l+m-1} \\ {m+1} \end{array}
\right) 
 \} ({k\o 2}\pa )^{m+2} \eta \nonumber \\
& &  m=1,
\cdots ,n-1,\;  \;\; l=2, \cdots ,n-m+1
\label{3.39}
\er
where we denote $V^+V^- = W_{n+1} $ in order to include the case $\eps
_{2,n+1}$ in the general formula for $\eps _{l,l+m}$.  The corresponding
{\it $W_3$-transformation} of the currents $V^{\pm}$, $T=W_2$, $W_p$ 
 calculated by substituting (\ref{3.38}) and (\ref{3.39}) in (\ref{3.15}) has the form 
\br
&&\hskip-0.7cm \d _{\eta }V^{+} = -{2\o n} \eta T V^{+} + {(n^2-1) \o 3} V^{+} ({k\o
2}\pa )^{2} \eta + \eta ({k\o 2}\pa )^2 V^{+} + n ({k\o 2}\pa \eta )({k\o
2} \pa V^{+}) 
\nonumber \\
&&\hskip-0.7cm \d _{\eta }V^{-} ={2\o n} \eta T V^{-} + {{(n+1)(n-4)} \o 6} V^{-} ({k\o
2}\pa )^{2} \eta - \eta ({k\o 2}\pa )^2 V^{-} - 2 ({k\o 2}\pa \eta )({k\o
2} \pa V^{-})
 \nonumber \\
&&\hskip-0.7cm \d _{\eta }W_p = -(p+1)({k\o 2}\pa \eta )W_{p+1} -2\eta ({k\o 2}\pa
)W_{p+1} - ({k\o 2}\pa )^2 (\eta W_p) + {(p-1)(p-2) \o 2}W_p ({k\o 2}\pa )^2
\eta 
 \nonumber \\
&&+{2\o n} \left( \begin{array}{c} {n} \\ {p} \end{array}
\right)({k\o 2}\pa )^p 
(\eta T) - \{\left( \begin{array}{c} {n+2} \\ {p+2} \end{array}
\right)
 + {(n^2-1)\o 3}\left( \begin{array}{c} {n} \\ {p} \end{array}
\right)
  - n\left( \begin{array}{c} {n+1} \\ {p+1} \end{array}
\right)
\} ({k\o 2}\pa )^{p+2}\eta 
\nonumber \\
&&+ \sum_{s=1}^{p-2} W_{p-s}\{ -{2\o n}\left( \begin{array}{c} {s+n-p}
 \\ {s} \end{array} \right)
 ({k\o 2}\pa )^{s}(\eta T)+ 
[\left( \begin{array}{c} {s+n-p+2}
 \\ {s+2} \end{array} \right)
 + {{n^2 -1}\o 3}\left( \begin{array}{c} {s+n-p}
 \\ {s} \end{array} \right)\nonumber\\
 && -n\left( \begin{array}{c} {s+n-p+1}
 \\ {s+1} \end{array} \right)
  ]({k\o 2}\pa )^{s+2}\eta \} 
  +\d_{n,p} ({k\o 2}\pa  \eta ) W_{p+1}
\label{3.40}
\er
Following the same procedure we find the transformations generated by $W_n$,
\br
\d _{\eta_n}V^{+} &=& -{1\o n}V^{+}[(-{k\o 2}\pa )^{n-1}\eta_{n} -
\sum_{s=1}^{n-2}(-{k\o 2}\pa )^{s-1}(\eta_{n}W_{n-s} ] + 
 (-{k\o 2}\pa )^{n-1} (\eta_{n} V^{+}) \nonumber \\
 &+& \sum_{l=1}^{n-1}(-{k\o 2}\pa)^{n-l-1} \{ V^{+}[(-{k\o 2}\pa)^{l}
 \eta_{n} - \sum_{s=1}^{l-1} (-{k\o 2}\pa )^{s-1} (\eta_n W_{l-s+1})]\}
 \nonumber \\
\d _{\eta_n}V^{-} &=& {1\o n}V^{-}\{(-{k\o 2}\pa )^{n-1}\eta_{n} \} - 
 (-{k\o 2}\pa )^{n-1} (\eta_{n} V^{-}) \nonumber \\
 &+& \sum_{s=1}^{n-2}W_{n-s}[ ({k\o 2}\pa )^{s-1}\eta_{n}V^{-}]
 - {1\o n}V^{-} \sum_{s=1}^{n-2}(-{k\o 2}\pa )^{s-1}(\eta_{n}W_{n-s})
 \nonumber \\
\d _{\eta_n}W_p &=&\eps _{n-p+2,1} V^{-} + \eps _{n-p+2,2} W_n +
\sum_{s=1}^{p-2} \eps_{n-p+2, s+n-p+2} W_{p-s} - \eps _{n-p+1,n+1} \nonumber \\
& - &{k\o 2}\pa \eps _{n-p+2,n+1}
 \label{3.41}
 \er
 where
\br
\eps _{n-p+2,1} &=&  (-{k\o 2}\pa )^{p-2}(\eta_n V^{+})+ \nonumber \\
&+ &
\sum_{l=1}^{p-2}(-{k\o 2}\pa )^{p-l-2}[V^{+}\( (-{k\o 2}\pa )^{l}\eta_n -
\sum_{s=1}^{l-1}(-{k\o 2}\pa )^{s-1}(\eta_n W_{l-s+1})\) ] 
\nonumber \\
\eps _{n-p+2,2} &=&  (-{k\o 2}\pa )^{p-1}\eta_n - \sum_{s=1}^{p-2}
(-{k\o 2}\pa )^{s-1}(\eta_n W_{p-s+2}) 
\nonumber \\
\eps _{n-p+2,s+n-p+2} &=& \left( \begin{array}{c} {s+n-p-1}
 \\ {s-1} \end{array} \right)
({k\o 2}\pa )^{s-1}(\eta_n W_n) \nonumber \\
&+&
\sum_{l=1}^{s-1}({k\o 2}\pa )^{s-l-1}[V^{+}({k\o 2}\pa
)^{l-1}(\eta_nV^{-})] \left( \begin{array}{c} {s+n-p-l-1}
 \\ {s-l-1} \end{array} \right)
\nonumber \\
&+& (-1)^{s}[\left( \begin{array}{c} {s+n-p-1}
 \\ {s-1} \end{array} \right)
 -{1\o n}\left( \begin{array}{c} {s+n-p}
 \\ {s} \end{array} \right)
 ][ \sum_{l=1}^{n-2}(-{k\o 2}\pa
)^{l+s-1}(\eta_nW_{n-l}) \nonumber \\
&-& (-{k\o 2}\pa )^{n+s-1} \eta_n ] 
\nonumber \\
\eps _{n-p+1,n+1} &+& {k\o 2}\pa \eps _{n-p+2,n+1} = ({k\o 2}\pa )^{p-1}
(\eta_nW_n) \left( \begin{array}{c} {n-1}
 \\ {p-1} \end{array} \right)
 + V^{+} [({k\o 2}\pa )^{p-2}(\eta_nV^{-})] 
 \nonumber \\
 &+ &(-1)^{p} [n \left( \begin{array}{c} {n-1}
 \\ {p-1} \end{array} \right)
  - \left( \begin{array}{c} {n}
 \\ {p} \end{array} \right) ] [(-{k\o 2}\pa )^{p+n-1}\eta_n  
- \sum_{s=1}^{n-2}(-{k\o 2}\pa )^{p+s-1}( \eta_nW_{n-s}) ]
\nonumber \\
&+& \sum_{s=1}^{p-2}({k\o 2}\pa )^{p-s-1}[V^{+}({k\o 2}\pa )^{s-1}(\eta_nV^{-})
]\left( \begin{array}{c} {n-s-1}
 \\ {p-s-1} \end{array} \right). 
\nonumber
\er
The derivation of the remaining $W_p$-transformations for $3<p\leq n-1 $ is
more complicated and is presented in the appendix B.  

In order to find the explicit form of the classical PB's algebra
$V^{(1,1)}_{n+1} $ generated by the $V^{\pm}$, $T$, $W_p$ we have to remember
the standard relation between the infinitesimal transformations and the
currents PB's : 
\be
\d _{\eps^{\pm}} J(w) = \int dz \eps ^{\pm}(z) \{ V^{\mp}(z) , J(w) \}
\ee
where $J$ is any of the currents  $V^{\pm}$, $T$, $W_p$, $p=3, \cdots ,n $. 
Starting from (\ref{3.37}) we easely derive the algebra of the nonlocal
currents 
\br
&&\{ V^{\pm} (\sigma),   V^{\pm} (\sigma^{\pr})\}
 = - {{n+1} \o nk^2 } \epsilon (\sigma - \sigma^{\pr})V^{\pm}
(\sigma)V^{\pm}(\sigma^{\pr})
\nonumber \\
&&\{ V^{+} (\sigma),   V^{-} (\sigma^{\pr}) \}
  =  {{n+1} \o nk^2 } \epsilon (\sigma - \sigma^{\pr})V^{+}
(\sigma)V^{-}(\sigma^{\pr})+ ({k\o 2})^{n-1} \pa_{\sigma^{\pr}}^{n} \d (\sigma -
\sigma^{\pr})   \nonumber \\
 &&\hskip 3cm  -   \sum_{s=0}^{n-2}({k\o 2})^{s-1} W_{n-s}(\sigma ^{\pr})
  \pa ^{s}_{\sigma^{\pr}}
\d (\sigma - \sigma^{\pr})
\label{3.51}
\er
In the simplest case, $n=1$ ($A_1$-model ) the full $V_2^{(1,1)} $ algebra is
spanned by $V^{\pm}$ ( of spin ${{n+1} \o 2} = 1$ ) only.  It turns out that
this nonlocal $V_2$ algebra coincides with the semiclassical limit of the
Fattev-Zamolodchikov PF- algebra \cite{ZF1} studied in ref.
\cite{Bardakci} (see also our eqns. (\ref{opevv}), (\ref{1.5})).  The
algebra $V_3^{(1,1)}$ of the symmetries of $A_2$ NA Toda model is related to
the semi-classical limit of the Polyakov-Bershadsky $W_3^{(2)} $ algebra
\cite{Bershadsky}, but with the local $U(1)$ current gauged away, i.e. one
additional constraint $J_{\lambda_1H }= 0 $ is imposed in the corresponding
reduction of the $A_2$ WZW model.  The $V_3^{(1,1)} $ algebra is a PF--type
extension of the Virasoro algebra
\be
\{T(\sigma ) , T(\sigma ^{\pr}) \} =
 2[\pa _{\sigma^{\pr}}\d(\sigma -\sigma^{\pr}) ]T(\sigma^{\pr}) +
  \d(\sigma -\sigma^{\pr})\pa _{\sigma^{\pr}}T(\sigma^{\pr}) - 
  {k^2 \o 2}\pa ^{3} _{\sigma^{\pr}}\d(\sigma -\sigma^{\pr}) 
\label{3.43}
\ee
with two spins $s=3/2 $ (nonlocal ) currents 
\be
\{T(\sigma ) , V^{\pm}(\sigma ^{\pr}) \} =
 s[\pa _{\sigma^{\pr}}\d(\sigma -\sigma^{\pr}) ]V^{\pm}(\sigma^{\pr}) +
  \d(\sigma -\sigma^{\pr})\pa _{\sigma^{\pr}}V^{\pm}(\sigma^{\pr})  
\label{3.44}
\ee
The PB's of the $V^{\pm}$ in this case are given by (\ref{3.51}) with $n=2$. 
This algebra is quite similar to the semi--classical limit of the $N=2$ 
superconformal
algebra.  

The $V^{(1,1)}_4 $ algebra of symmetries of $A_3 $-NA-Toda theory provides an
interesting example of  new type of mixed parafermionic-$W_3$-algebra.  It
represents {\it a nonlocal and nonlinear  (non--Lie ) } extension of the Virasoro
algebra (\ref{3.43}) with two spins $s=2$ nonlocal currents $V^{\pm}$ and one
local spin $s=3$, $ {\om}_3 = W_3 - \pa _{\sigma } T$.  Together with
(\ref{3.43}) (with central charge $-2k^2 $) and (\ref{3.51}), it contains two
new PB's, 
\br
\{{\om}_3(\sigma ) , V^{\pm}(\sigma ^{\pr}) \} &= &
 \mp {5k\o 3}V^{\pm}(\sigma^{\pr})\pa^{2}_{\sigma^{\pr}}
 \d(\sigma -\sigma^{\pr}) 
  \mp {5k\o 2}[\pa _{\sigma^{\pr}}\d(\sigma -\sigma^{\pr}) ]
 \pa_{\sigma^{\pr}}V^{\pm}(\sigma^{\pr}) \nonumber \\
 & \pm &
 \d(\sigma -\sigma^{\pr}) \( {2\o {3k}} TV^{\pm} -
  k \pa ^{2}_{\sigma^{\pr}}V^{\pm}(\sigma^{\pr} ) \)
  \label{3.45b}
 \er
 and
\br
&&\{{\om }_3(\sigma ) ,{\om }_3 (\sigma ^{\pr}) \} = 
4\( V^{+}V^{-} + {1\o 6}T^{2} \)(\sigma^{\pr})\pa_{\sigma^{\pr}}
 \d(\sigma -\sigma^{\pr})  + 
 2 \d(\sigma -\sigma^{\pr}) \pa _{\sigma^{\pr}} (V^{+}V^{-} + {1\o 6}T^{2} )
 \nonumber \\
 && \hskip2.5cm -  {k^2\o 6} \d(\sigma -\sigma^{\pr}) \pa ^{3}_{\sigma^{\pr}}T  - {3k^2\o
 4} [\pa _{\sigma^{\pr}}\d(\sigma -\sigma^{\pr}) ]\pa ^{2}_{\sigma^{\pr}}T
  - {5k^2\o 4}[\pa ^{2}_{\sigma^{\pr}}\d(\sigma -\sigma^{\pr})]
 \pa _{\sigma^{\pr}}T - \nonumber \\
 &&\hskip2.5cm -{5k^2\o 6} T(\sigma ^{\pr} )
 \pa ^{3}_{\sigma^{\pr}}\d(\sigma -\sigma^{\pr})+ {k^4\o 6} 
  \pa ^{5}_{\sigma^{\pr}}\d(\sigma -\sigma^{\pr})
  \label{3.45}
  \er
The eqns. (\ref{3.44}), (\ref{3.45b} ) and (\ref{3.45}) are derived  from the infinitesimal
transformations (\ref{3.40}) taking into account that we have introduced the
primary field $ {\om}_3 $ instead of $W_3$ (from the D-S gauge ). It is
straightforward to write the PB's $ \{ W_{p_1}, W_{p_2} \} $ and 
$ \{ W_{p_1}, V^{\pm} \} $
for the arbitrary $V^{(1,1)}_{n+1} $ algebra.  As we have mentioned they are
encoded on the corresponding infinitesimal transformations (\ref{B.16}),(\ref{B.17}) and
and (\ref{3.37}).  In the nonprimary basis (DS-gauge ) for $W_p$'s the algebra 
looks rather
complicated.

 The method we have used in the derivation of the $V^{(1,1)}_{n+1} $ works
equally well for arbitrary $G_n$.  The explicit construction of the solutions 
of eqns.(\ref{3.15}) far say, $B_n$, $C_n$ or $D_n$ however requires a 
bit more work.  

In order to demonstrate that the $V_n^{(1,1)} $ algebra is in fact the algebra
of symmetries of the $G_n^{(1,1)} $-NA-Toda model (\ref{2.17}) we need to know
the transformations for the fields $\psi $, $\chi $ and $ \phi_i $ generated by
the currents $V^{\pm}$, $T$ and $W_p$.  We apply once more the Polyakov method
\cite{Pol1}, this time imposing all constraints (\ref{3.17}) in the chiral
$A_n$-gauge transformations of the WZW field $g$: $\d g_{ik} =-g_{il}(z, \bar
z) \eps _{lk} $.  We have already calculated the redundant gauge parameters. 
What is still missing is the reduced form of $g_{ik}$.  Using the explicit
solutions of the constraint equations (\ref{3.19}) we find
\br
g_{11} &= &e^{R}; \quad g_{22} = e^{\phi_1 -{1\o n}R} (1+ \chi
\psi e^{-\phi_1}); \quad  \quad g_{2l} = ({k\o 2}\pa )^{l-2}g_{22}; 
\quad g_{l2} 
=({k\o 2}\bar \pa )^{l-2}g_{22}, \cdots\nonumber \\
g_{1l} &=& ({k\o 2}\pa )^{l-2} ( e^{{{n-1}\o 2n} R } \psi ); \quad 
g_{l1} = ({k\o 2}\bar \pa )^{l-2} ( e^{{{n-1}\o 2n} R } \chi ),
\cdots  , etc 
\label{3.46} 
\er
We are now able  to write the field transformations generated by $V^{\pm}$
\br
\d_{\eps ^{\pm} }g_{11} = -g_{11} \eps _{11} - \sum_{l=2}^{n+1} g_{1l} \eps
_{l1};  & &\quad \d_{\eps ^{\pm} }g_{12} = {1\o k} g_{11} \eps ^{-} - g_{12} \eps
_{22}; \nonumber \\
 \d_{\eps ^{\pm} }g_{22} = {1\o k} g_{21} \eps ^{-} - g_{22} \eps
_{22};& & \quad \d_{\eps ^{\pm} }g_{21} = -g_{21} \eps _{11} - 
\sum_{l=2}^{n+1} g_{2l} \eps_{l1}, 
\label{3.47}
\er
etc.  All $\eps $'s in (\ref{3.47}) are given by eqns. (\ref{3.31}), 
(\ref{3.32})
and (\ref{3.35}) and are indeed linear functions of $\eps ^{\pm }$.  The
corresponding conformal transformations take the following simple form
\br
\d _{\eps } \psi &=& {{1-n}\o 2} \psi \pa \eps + \eps \pa \psi, 
\quad \d_{\eps
} \chi = \eps \pa \chi , \quad  \d_{\eps }R  = \eps \pa R
\nonumber \\
\d_{\bar \eps }\psi  &=& \bar \eps \bar \pa  \psi; \quad \d_{\bar \eps } \chi =
{{1-n}\o 2} \chi \bar \pa \bar \eps + \bar \eps  \bar \pa \chi , \quad 
\d_{\bar \eps }R =\bar \eps \bar \pa R 
 \nonumber \\
\d _{\eps } \phi _l &=& {{l(l-n)\o 2}} \pa \eps + \eps \pa \phi _l; 
\quad \d _{\bar \eps } \phi _l = {{l(l-n)\o 2}} \bar \pa \bar \eps + 
\bar \eps \bar \pa \phi _l
\label{3.48}
\er
$l=1, \cdots n-1 $.The  eqns. (\ref{3.48}) show that $\psi $ and $\chi $ are
primary  conformal fields  of spin $s = \Delta - \bar \Delta $ and dimension $d
= \Delta + \bar \Delta $: $(s_{\psi}, d_{\psi } ) = ( {{1-n}\o 2} , {{1-n}\o
2}) $ and   $(s_{\chi}, d_{\chi } ) = ( {{1-n}\o 2} , {{1-n}\o
2}) $.   For the vertices $e^{\phi_l} $ we have 
$(s_{l}, d_{l} ) = ( 0 , l(l-n) )$.  The non-local field  $ e^{R} $ is
spinless and dimensionless.  

One can further calculate the corresponding $W_p$ transformations of 
$\psi $, $\chi $ and $\phi_l$ taking into account the explicit form of eqns.
(\ref{3.38}), (\ref{3.39}),(\ref{B.14} ) and (\ref{B.16})  of the $\eps _{ik}$'s in terms of
$\eta_p$ and $V^{\pm}$, $T$ , $W_p$.  Consider, for example, $W_3$
transformations ( $\eta_3 = \eta $)
\br
\d_{\eta} e^{R} &=& (n-1) \pa \eta \pa e^{R} -2\eta 
[\tilde T - {{(n-1)}\o 2}\pa ^2 R + {(n-1)\o 2}
(\pa R)^2 - {1\o 2}\pa ^2]e^{R}, 
\nonumber \\
\d_{\eta} \psi &=& -{(n-1)\o 2n}\psi \pa \eta  \pa R + (n-2) \pa \eta
\pa \psi + {1\o n} \eta [(n+1) T - (n-1)\tilde T - {(n^2-1)\o 2 }\pa ^2
R \nonumber \\
&-& {(n^2-1)\o 4n } (\pa R )^2 - (n-1) \pa R \pa - n \pa ^2 ]
\psi - {(n-1)(n-2)\o 3 } \psi \pa ^2 \eta \nonumber \\
\d_{\eta} \chi &=&-{(n-1)\o 2n} \chi [(n-1) \pa \eta \pa R - 2\eta
(\tilde T -{n\o 2 }\pa ^2 R - {1\o 2n}(\pa R)^2] + 
\nonumber \\
&+& (n-1)\pa \eta (e^{\phi_1} + \psi \chi )e^{-{(n+1)\o 2n}R } - \eta
e^{-{(n+1)\o 2n}R } \pa ( e^{\phi_1 - {1\o n} R } + e^{-{1\o
n}R} \psi \chi )
\label{3.49}
\er
etc. , where $\tilde T = T - \sum_{i=1}^{n-1} [ \pa \phi_i \pa \phi_i - {1\o 2}
\pa \phi_i \pa \phi_{i-1} - {1\o 2} \pa \phi_i \pa \phi_{i+1} + \pa ^2 \phi_i ]
$ and $T$ is given by (\ref{3.1}).  

Although the field equations of the $A_n^{(1,1)} $-NA-Toda model (\ref{2.17}),
\br
\pa \bar \pa \phi_i = ({2\o k})^2 e^{\phi_{i+1} + \phi_{i-1} -2 \phi_i } -
{(n-i)\o n} e^{-\phi_i} {{\pa \chi \bar \pa \psi }\o \Delta ^2} 
\nonumber 
\er
\be
\bar \pa \( {{\pa \chi e^{-\phi_1}} \o \Delta } \) = -{(n+1)\o 2n}{{\chi \pa
\chi \bar \pa \psi e^{-2\phi_1}}\o \Delta ^2},  \quad 
 \pa \( {{\bar \pa \psi e^{-\phi_1}} \o \Delta } \) = -{(n+1)\o 2n}{{\psi \pa
\chi \bar \pa \psi e^{-2\phi_1}}\o \Delta ^2} 
\label{3.50}
\ee
are by construction invariant under all the $V_n^{(1,1)} $-transformations,
 the
proof of the invariance of the action (\ref{2.17}) is rather
 complicated. 
There are however, few exceptions.  The reparametrization (conformal)
invariance of (\ref{2.17}) is straightforward, due to the simple form of the
conformal transformations (\ref{3.48}).  We next verify the invariance of
(\ref{2.17}) under nonlocal transformations (\ref{3.47}), generated by $V^{+} $
 (and $ \bar V^-$), 
 \br
 \d_{\eps ^{-} } \psi &=& {1\o k} e^{{(n+1)\o 2n} R} \eps ^{-} +
  {1\o 2k}
 {(n+1)\o 2n} \psi \int \eps (\sigma - \sigma^{\pr} ) \eps ^{-}(\sigma^{\pr}
 )V^{+}(\sigma^{\pr}) d\sigma^{\pr} \nonumber \\
\d_{\eps ^{-} }\chi  &=& -{1\o 2k} {(n+1)\o 2n}\chi 
 \int \eps (\sigma - \sigma^{\pr} ) \eps ^{-}(\sigma^{\pr}
 )V^{+}(\sigma^{\pr}) d\sigma^{\pr} \nonumber \\
 \d_{\eps ^{-} }\phi_i  &=& 0 
 \label{3.42}
 \er
 Using the definition (\ref{3.11}) of the nonlocal field $R$ and the
 fact that $\tilde \eps _{11} = \eps _{11}|_{\eps^{+} = 0}$ :
 \br
 {k\o 2} \pa \tilde \eps _{11} = \eps ^{-}V^{+}, \quad \bar \pa \eps _{11} = 0
 \nonumber 
 \er
 (see  eqn. (\ref{3.32}) ) we find 
 \br
 \d_{\eps^{-}} S^{NA}_{A_n^{(1,1)}} = {1\o 2\pi } {(n+1)\o 2n} \int d^2z \bar
 \pa (V^{+}\eps ^{-} R) =0 
 \nonumber 
 \er
 Similarly for $\bar V^{-}$-transformation we obtain
 \br
 \d_{\bar \eps^{+}} S^{NA}_{A_n^{(1,1)}} = -{1\o 2\pi } {(n+1)\o 2n} 
 \int d^2z \pa (\bar V^{-}\bar \eps ^{+} R) =0
 \nonumber 
 \er
 The remaining nonlocal transformations ( generated by $V^{-}$ 
 and $\bar V^{+}$ )  are
 not as simple as (\ref{3.42}) and the  check  of the invariance of (\ref{2.17}):
 $\d_{\eps^{+}}S^{NA}_{A_n^{(1,1)}} = \d_{\bar \eps^{-}}S^{NA}_{A_n^{(1,1)}} =
 0$
 is more complicated.  The explicit proof of the  $W_p$ invariance of
  $S^{NA}_{A_n^{(1,1)}}$ is still lacking, except in the simplest case $n=3$,
  i.e. the $A_3$-NA-Toda model.

\sect{$SL(2,R)_q$ Symmetries}
The appearence of {\it nonlocal currents} 
in the theory is always an indication of the
existence of some underlying {\it quantum group structure }(see ref.
\cite{Bern-Leclair},
\cite{Bab-Bern} ).  We shall demonstrate that 
the charges of the {\it chiral} nonlocal
currents  $  Q^{+} = \int V^{+}d \sigma$ and $  Q^{+} = 
\int \sigma ^{n-1}V^{-}d \sigma$ have
nonvanishing PB's with the {\it antichiral} nonlocal charges 
 $ \bar  Q^{-} = \int V^{-}d \sigma$ and $ \bar  Q^{+} = 
 \int \sigma ^{n-1}\bar V^{+}d \sigma $
and together with the nonchiral $U(1)$ charge $Q_0 = \int J_0 d \sigma $
(see eqn. (\ref{3.2})) they generate a {\it q-deformed affine $SL(2,R) $ PB's
algebra}.  The PB's of the charges of  {\it chiral local}
 currents ($T$, $W_p$)
 with the
charges of the {\it antichiral} ($\bar T$, $ \bar W_p$ ) do indeed {\it 
vanish}.  The
presence of the $\hat SL(2,R)_q $ {\it Poisson bracket algebra}
 as  Noether symmetry of the
classical $G_n^{(1,1)} $-NA-Toda theory is one of the basic features of 
these
models.

We begin with the PB's algebra of $V^{+}(z) $ and $\bar V^{-}(\bar z)$.  Using
the explicit form of the conjugate momenta $\Pi_{\psi}$ and $\Pi_{\chi}$
(derived from ((\ref{2.17}), $\Pi_{\rho} = {\d {\cal L}\o {\d \pa_0 \rho}}, \rho =
\psi , \chi $), we eliminate the time derivatives from (\ref{3.12}) and
(\ref{3.11})
\br
V^+ = ({k\o 2}) \(
\chi^{\pr} +{1\o 2}k_{12}\chi \phi_1^{\pr}-({2\pi \o k})
\Pi_{\psi}e^{k_{12}\phi_1}\Delta \)
{{e^{k_{12}\phi_1 - {1\o {2{\cal K}_{11}}}({2\pi \o k})R_0}}\o \Delta^{1\o 2}}
\nonumber \\ 
V^- = ({k\o 2}) \(
-\psi^{\pr} -{1\o 2}k_{12}\psi \phi_1^{\pr} - 
({2\pi \o k})\Pi_{\chi}e^{k_{12}\phi_1}\Delta
\) 
{{e^{k_{12}\phi_1 - {1\o{2{\cal K}_{11}}}({2\pi \o k})R_0}}\o \Delta^{1\o 2}}.
 \label{4.1} 
\er
For further convenience we have splited the nonlocal field $R$ (defined by eqn.
(\ref{3.11}) in two parts
\br
R = -{2\pi \o k}R_0 + {\cal K}_{11} ln \Delta , \quad R_0^{\pr}= {1\o 2} 
(\psi \Pi_{\psi} -\chi \Pi_{\chi} )
\nonumber 
\er
By simple manipulations involving the canonical equal time PB's, 
\br
\{\Pi_{\psi}(\sigma), \psi (\sigma^{\pr})\} = -\d (\sigma - \sigma^{\pr}), \; 
\{\Pi_{\chi}(\sigma), \chi (\sigma^{\pr})\} = -\d (\sigma - \sigma^{\pr})
, \;\{\Pi_{\phi_i}(\sigma), \phi_j (\sigma^{\pr})\} = 
-\d_{ij}\d (\sigma - \sigma^{\pr})
\nonumber
\er
(all other PB vanish) and their space derivatives we find
\br
\{V^{+}(\sigma ), \bar V^{-}(\sigma^{\pr} ) \} &=& -{k\pi \o 2}
e^{k_{12}\phi_1(\sigma ) +k_{12} \phi_1(\sigma^{\pr})- {1\o 2{\cal K}_{11}} 
({2\pi \o k})
(R_0(\sigma ) + R_0(\sigma^{\pr} ))}[e^{-k_{12}\phi_1(\sigma^{\pr})}
 ({\Delta (\sigma^{\pr})\o
\Delta (\sigma)})^{{1\o 2}}\pa _{\sigma}\d(\sigma -\sigma^{\pr}) 
\nonumber \\
&+&e^{-k_{12}\phi_1(\sigma )}({\Delta (\sigma)\o
\Delta (\sigma^{\pr} )})^{{1\o 2}}\pa _{\sigma^{\pr}}\d(\sigma -\sigma^{\pr}) - {\pa
_{\sigma} {(e^{-k_{12}\phi_1} \Delta )}\o {\Delta }}\d(\sigma -\sigma^{\pr})]
\label{4.2}
\er
Integrating (\ref{4.2}) we get the PB's for the charges $Q^{+}$ and $\bar Q^{-}$
\be
\{Q^{+}, \bar Q^{-} \} = {k\pi \o 2} \int_{-\infty}^{\infty} d\sigma \pa
_{\sigma} e^{{1\o {\cal K}_{11}} R + k_{12}\phi_1 - ln \Delta }
\label{4.3}
\ee
One can simplify the r.h.s. of (\ref{4.3}) taking into account the relation of
the field in the exponent 
\be 
\varphi = R +{\cal K}_{11}(k_{12}\phi_1 - ln \Delta )
\label{4.4}
\ee
with the $U(1)$ current (\ref{3.2}), namely
\br
J_{\mu} = {k\o{2\pi }}\eps_{\mu \nu }\pa ^{\nu} 
(\varphi + k_{12}{\cal K}_{11}\phi_1 )
\nonumber 
\er
Note that $I_{\mu} = -{k\o{2\pi }}{\cal K}_{11}\eps_{\mu \nu }\pa ^{\nu} k_{12}\phi_1 $ is
automatically conserved topologically current and its charge $\int I_0 d\sigma  $ have
vanishing PB's with either $V^{+}$ and $\bar V^{-}$.  This fact suggests the
following redefinition of the $U(1)$ charge 
\br
&& H_1 = Q_0 - \int I_0 d\sigma = -{k\o 2\pi }\( \varphi (\infty ) -  
\varphi (-\infty ) \)
\nonumber \\
&&\{H_1 , Q^{+} \} = Q^{+}, \quad \{H_1 , \bar Q^{-} \} = -\bar Q^{-}
\label{4.5}
\er
and the nonlocal charges $ Q^{+}$ and $\bar Q^{-}  $ as well
\be
E_1 = \sqrt {2 \o {k\pi}} { q^{{1+\hat {\kappa}} \o 2} \o { (q^2 -1)}^{1 \o 2}} Q^+,
\quad \quad  F_1 = 
\sqrt {2 \o {k\pi}} { q^{{1+\hat {\kappa}} \o 2} \o { (q^2 -1)}^{1 \o 2}} \bar Q^- 
\ee
where $ q_{(G_n)} = e^{- ({2\pi \o k}){1\o {2{\cal K}_{11}}}} $ and  
$\hat \kappa = -{k \o {2\pi}}( \varphi
(\infty) + \varphi (-\infty))$.  
As a consequence of the PB's of $\varphi $ with $V^{+}$ and $\bar  V^{-}$, 
\br
\{\varphi (\sigma ) , V^{+}(\sigma^{\pr} )\} = 
{\pi \o k}V^{+}(\sigma^{\pr } )\eps (\sigma -\sigma^{\pr} );  \quad
\{\varphi (\sigma ) , \bar V^{-}(\sigma^{\pr} )\} = 
-{\pi \o k}\bar V^{-}(\sigma^{\pr } )\eps (\sigma -\sigma^{\pr} )
\nonumber
\er
we realize that $\hat \kappa $ has vanishing PB's with $Q^+$ and $\bar Q^-$.

The result of all this rearangements of the variables is that the algebra
(\ref{4.3}), (\ref{4.5}) takes the standard form of the $q_{(G_n)}$-deformed
$SL(2,R)$ algebra (for an arbitrary $G_n$):
\be
\{E_1, F_1\} = {{q^{H_1} - q^{-H_1}} \o {q - q^{-1}}}, \quad \quad \{H_1, E_1 \}= E_1
,\quad \quad
\{H_1, F_1 \}= - F_1.
\label{4.7}
\ee

We should mention that the $U(1)$ charge $Q_0$ (or $H_1$) appears as a
{\it topological} charge for the lagrangean derived from  (\ref{2.17}) by the
familiar change of variables:
\br
\psi = \sqrt{ {2 {\cal K}_{11} }}e^{-{1 \o 2} k_{12}\phi_1 - \theta} sh( r) ,\quad \quad 
 \chi = \sqrt{ {2{\cal K}_{11}}}e^{-{1 \o 2} k_{12}\phi_1 + \theta} sh( r)
\nonumber
\er

The computation of the PB's algebra of the remaining nonlocal charges $Q^-$
and $\bar Q^+$ ( as well as the mixed PB's $\{Q^{\pm} ,\bar Q^{\pm} \}$) is
rather difficult problem even in the few cases ($n\leq 3$ ) we know their
explicit form.  The complications arise from the fact that the  currents 
$V^{-}_{(n)} $ and $\bar V^{+}_{(n)} $ contain $n^{th}$ order time
derivatives and their elimination by using the field equations (\ref{2.3})
is a cumbersome task even for $n=3$.  The simplest case $n=1$ is an
exception.  The currents $V^{-}_{(1)} $ and $\bar V^{+}_{(1)} $ given by
(\ref{3.13}) and $V^{+}_{(1)} $ and $\bar V^{-}_{(1)} $ have a very 
similar form.  The calculation of the PB's of the corresponding charges
$Q^+, \bar Q^- $ is straightforward and yields
\be
\{E_0, F_0\} = {{q^{H_0} - q^{-H_0}} \o {q - q^{-1}}} ,
\quad \quad q = q_{(A_1)} = e^{{2\pi \o k}}
\label{4.8}
\ee
where
\br 
E_0 = \sqrt {2 \o {k\pi}} { q^{{1-\hat {\kappa}} \o 2} \o {(1 - q^2)}^{1 \o 2}} Q^- ,
\quad \quad F_0 = 
\sqrt {2 \o {k\pi}} { q^{{1-\hat {\kappa}} \o 2} \o {( 1 - q^2) }^{1 \o 2}} \bar Q^+,\quad
\quad H_o = - H_1
\nonumber 
\er
The two remaining PB's vanish, i.e. $\{Q^{\pm}, \bar Q^{\pm} \} = 0 $.  One
can write the PB's (\ref{4.7} and \ref{4.8}) in the following compact form
\br
&&\{H_i, E_j\}  =  \kappa_{ij}E_j \quad \quad \{H_i, F_j\} = - \kappa_{ij}F_j, 
\quad \quad  i, j = 0, 1
\nonumber \\
&&\{E_i, F_j\}  =  \d_{ij}{{{q^{H_i}_n - q^{-H_i}_n} \o {q_n - q^{-1}_n}}  } , 
~~~~\kappa_{ij} =
\( \begin{array}{cc}
  1 & -1 \\
 -1 & 1
\end{array} \)
\label{4.9}
\er
which is known to be the centerless affine $SL(2,R)_q$ algebra in the 
principal
gradation  \cite{Bern-Leclair} (the Serre relations are omited). 
The question in order {\it is whether the algebra }(\ref{4.8}) ( and therefore the
larger $\hat SL(2,R)_q $ (\ref{4.9})) {\it takes place for all $G_n^{(1,1)}$-NA-
Toda models}.  Starting from the explicit form of the currents $V_{(2)}^{-}$
and $\bar V_{(2)}^{+}$ (see eqn. (\ref{3.14})) and eliminating the $2^{nd}$
time derivative we verify that this is indeed the case for $A_2^{(1,1)}
$-NA-Toda theory.  This leads us to the following {\it conjecture}: {\it the
affine $SL(2,R)_q $ PB algebra (\ref{4.9}) appears to be an algebra of
(Noether ) symmetries of the classical  $G_n^{(1,1)} $-NA-Toda models}.

At this stage one might wonder about the relation of the 
$\hat SL(2,R)_{q_{(G_n)}} $ PB's algebra with the Poisson-Lie group $G_n(r)$
(familiar from the $G_n$-WZW and abelian Toda models \cite{Al-Shat,F-L1}).  One
expect it to take place as a symmetry of the Poisson structure of
$G_n^{(1,1)}$-NA-Toda theories as well.  It is well known that the
symplectic form of the $G_n$-WZW model is invariant with respect to $(a)$ Loop
group $\hat G_n$ generated by $G_n$ current algebra and $(b)$  The Poisson--
Lie group $G_n(r)$ of the monodromy matrices $M \in G_n$ satisfying apart
from 
the $G_n$-multiplication laws, the Sklyanin PB's algebra \cite{Skly}

\be
\{  M(\sigma ) \x M(\sigma^{\pr}) \}  = -{2\pi \o k} 
[r, M(\sigma ) \otimes M(\sigma^{\pr}) ]
\label{4.10}
\ee
as well.   The $G_n$- albelian Toda theory realized as $H= 
 {\cal N}_L \otimes {\cal N}_R $-reduced $G_n$-WZW model manifests similar properties.
   Its symplectic structure is invariant under the action of $G_n(r)$ and
   $WG_n$- algebras \cite{Bab}, \cite{F-L1}, \cite{F-L2}.  As we have shown
   in Sect 3.  the Poisson structure of the $G_n^{(1,1)}$-NA-Toda (generated
   by the Hamiltonian derived from (\ref{2.17})) is invariant under the
   nonlocal algebra $VG_n^{(1,1)}$ and with respect to $\hat
   SL(2,R)_{q_{(G_n)}}$ as well.  In order to find the Poisson-Lie 
   group of the
 $G_n^{(1,1)}$-NA-Toda models we have to construct the
  monodromy matrix, to calculate
 the corresponding classical $r$-matrix and to verify that eqn. (\ref{4.10})
 holds.  As
 usual the monodromy matrix is defined as  a solution of the linear
 problem
 \be
 (\pa - {\cal A} )M(\sigma ,t) =  (\bar \pa - \bar {\cal A} )M(\sigma ,t) =
 0 ,\quad \quad {\cal A}_x = {1\o 2}({\cal A}- \bar {\cal A})
 \label{4.11}
 \ee
 with ${\cal A}$ and $\bar {\cal A} $ given by eqn. (\ref{2.2}).  The
 solution of (\ref{4.11}) normalized by the condition $M(-\infty , t_0) = 1
 $ can be written as the $P$-ordered exponential:
 \be
 M(\sigma ) = 
 Pe^{\int_{-\infty }^{\sigma } A_{x} (\sigma ^{\pr})d\sigma ^{\pr}}
 .
 \label{4.12}
 \ee
We next realize $A_x$ in terms of the fields $\psi, \chi , \phi_i $  and
their momenta $\Pi_{\psi }, \Pi_{\chi }, \Pi_{\phi }$ (derived from eq(\ref{2.17})),
\br
A_x &=& {\pi \o k }[{1\o 2{\cal K}_{11} }(\psi \Pi_{\psi } + \chi \Pi_{\chi
})\lambda_1 \cdot H + {1\o 2}\sum_{m,l=1}^{n-1}{\tilde
{\cal K}}_{lm}\Pi_{\phi_l} \a_{m+1}\cdot H + \Pi_{\chi} e^{-{1\o 2}k_{12}\phi_1}
E_{-\a_1} \nonumber \\
&+& 
 \Pi_{\psi}e^{-{1\o 2}k_{12}\phi_1} E_{\a_1} + {k\o 2\pi }\sum_{i=1}^{n-1}
   e^{-{1\o 2}\tilde k_{ij}\phi_j }(E_{\a_{i+1}} +
 E_{-\a_{i+1}}) ]
\nonumber
\er
 By straightforward calculation (similar to the abelian Toda case ), we
 derive the so called Fundamental Poisson Brackets \cite{Fad}:
 \be
 \{  A_x(\sigma ) \x A_x(\sigma^{\pr}) \}  =-{2\pi \o k}  [ r ,         
 A_x(\sigma )\otimes I + I \otimes A_x(\sigma^{\pr}) ]\d (\sigma - 
 \sigma^{\pr})
 \label{4.13}
 \ee
 where $r \in G_n \otimes G_n $ denote one of the solutions 
 \br
 &&r^+ = {1\o 4}(\sum_{i} H_{\a_i}\otimes H_{\a_i}+ 2 \sum_{\a >0 } E_{\a}
 \otimes E_{-\a}) ,\quad
  r^- = -{1\o 4}(\sum_{i} H_{\a_i}\otimes H_{\a_i}+ 2 \sum_{\a >0 } E_{-\a}
 \otimes E_{\a}) 
 \nonumber 
 \er
 of the classical Yang-Baxter equation
 \br
 [r_{12},r_{13} ] + [r_{12},r_{23} ] + [r_{13},r_{23} ]=0
 \nonumber 
 \er
 Finally the Sklyanin relation (\ref{4.10}) is a simple consequence of 
  (\ref{4.12}) and (\ref{4.13}).  The {\it conclusion}
   is that the Poisson-Lie
  group $G(r)_n$ generated by the monodromy matrices (\ref{4.12}) of the
  $G^{(1,1)}_n$-NA-Toda model coincides with the one that appears in the
  $G_n$-WZW and the $G_n$-abelian Toda models.  With the monodromy matrix at
  hand one can further define the group of (nonlocal ) dressing
  transformations  preserving the form of the Lax connection (\ref{2.2})
  and mapping one  solution of the $G_n$-NA-Toda (with charge $Q_0$) into
  another (of charge $Q^{\pr}_0 \neq Q_0 $) (see \cite{Bab-Bern} for the
  abelian Toda case ).  The {\it nonlocal transformations} (\ref{3.51}),
(\ref{3.42}) generated
  by currents $V^+, \bar V^- $ (and   $\bar V^+,  V^- $) share many of the
  properties of the {\it dressing transformations}.  They leave invariant
  $G_n$-NA-Toda equations (\ref{2.3}), preserve  the form (\ref{2.2}) of
  ${\cal A}$ and $\bar {\cal A}$ and transform solutions with $Q_0$ into
  solutions of charge $Q_0  \pm 1$.  This is a hint that the $\hat
  SL(2,R)_q$ {\it PB's algebra} of the nonlocal (Noether ) charges $Q^{\pm}, \bar
  Q^{\pm}, Q_0 $ might appear as a {\it subalgebra} of the {\it dressing}
   Poisson
  algebra.  However we have no proof of such statement.  It requires the
  explicit construction of the dressing PB'S algebra for the
  $G_n^{(1,1)}$-NA-Toda theory.

  In our derivation of the $\hat SL(2,R)_q$ algebra we left 
  {\it unanswered the
  important question whether the equations (\ref{2.3}) admit solutions such
  that } $ \varphi (\infty ,t_0) \neq  \varphi (-\infty ,t_0)$, i.e. $H_1 \neq
  0 $.  In the next section we  construct the general (exact )
  solutions of eqns. (\ref{2.3}) and study their asymptotics.  

\sect{ General Solutions }
 The solutions of eqn. (\ref{2.3}) can be found  by direct application of the
Leznov-Saveliev method \cite{L-S}.  As it is well known their explicit form
contains multiple integrals and is not appropriate for the analysis of the
asymptotics of a specific combination of these solutions such as $\varphi $
defined by eqn. (\ref{4.4}).  For this purpose we need the NA-Toda analog of
the Gervais-Bilal's solution \cite{Ger-Bil} for the abelian Toda equations.  It
turns out that  the NA-Toda fields $\psi, \chi, \phi_i, i=1,
\cdots , n-1 $ can be realized in 
 terms of the corresponding abelian Toda fields, $\varphi_A,
A=1, \cdots , n$ together with the {\it chiral}  currents $V^+(z) $ and $\bar V^-(\bar
z) $ considered as {\it independent variables}.  The exact {\it statement} is as follows: 
Let $\psi, \chi, \phi_i, i=1, \cdots , n-1 $ satisfy eqn (\ref{2.3}), $R$ is
defined by eqn (\ref{3.11}) and $V^+(z) $ and $\bar V^-(\bar z)$ are given by
(\ref{3.12}).  Then the fields
\be
\varphi_i = \phi_{i-1} + \tilde {\cal K}_{1,i-1}{\tilde \a^2_{i-1}\o 2} R -
{\cal K}_{1,i}{\a^2_i\o 2} ln V^{+}\bar V^{-}, \phi_0 =0, \;\; i=1, \cdots n
\label{5.1}
\ee
with $\a_0^2 = 2$ and ${\tilde {\cal K}_{1,0}} = 1$, satisfy 
the $G_n$ abelian Toda equations
\be
\pa \bar \pa \varphi_i = ({2\o k})^2 e^{-k_{ij}\varphi_j}
\label{5.2}
\ee
The proof is based on the following suggestive form of the first eqn.
(\ref{2.3})
\br
\pa \bar \pa \phi_i = ({2\o k})^2 e^{-\tilde k_{il}\phi_l}-({2\o k})^2
\tilde {\cal K}_{1,l}
 ({\tilde \a^2_l\o 2})                     
e^{-k_{12}\phi_1 - {1\o {\cal K}_{11}}R + 
ln V^{+}\bar V^{-}},
\nonumber
\er
of the eqn (\ref{3.12}) for the nonlocal field R :
\br
\pa \bar \pa R =({2\o k})^2 e^{-k_{12}\phi_1 -{{1\o {{\cal K}_{11}}}}R +
 ln V^{+}\bar V^{-}}
\nonumber
\er
and on the  following algebraic identities for the algebras $A_n, B_n, C_n $
and $D_n$,
\br
&&\hskip-1cm {1\o {{\cal K}_{11}}} = k_{11} + \tilde {\cal K}_{11}k_{12},  
\quad \quad({\a^2_j\o 2}) k_{ij}{\cal K }_{1j}= \d_{1i},\nonumber\\ 
 &&k_{ij} ({\tilde \a^2_{j-1}\o 2})\tilde {\cal K}_{1,j-1} = 0 ,\quad \quad i=2,
\cdots n
\nonumber
\er 
Finally we
observe that as a consequence of (\ref{3.11}) and (\ref{3.12}) the fields $\psi
$ and $\chi $ can be realized in terms of $R$, $V^+(z) $ and $\bar V^-( \bar z)$
only:
\be
\psi V^+ = ({2\o k})e^{{{1\o {2{\cal K}_{11}}}R}} \pa R, \quad 
\chi \bar V^- = ({2\o k})e^{{{1\o {2{\cal K}_{11}}}R}} \bar \pa R
\label{5.4}
\ee
The conclusion is that eqns. (\ref{5.1}) and (\ref{5.2}) allows us to write the
solutions of (\ref{2.3})
in terms of the solutions of eqns. (\ref{5.2}).

We next take the general solutions of the $G_n$-abelian Toda eqns. (\ref{5.2})
in the form proposed by Gervais and Bilal \cite{Ger-Bil}
\be
e^{\varphi_1} =  ({k\o 2})^{-n} F_i \bar F_i,\cdots  
\quad e^{-k_{l-1,l}\varphi_l} = 
 ({k\o 2})^{l(l-n-1)}{1\o {l!}} F_{i_1,i_2,\cdots i_l}(z)
 \bar F_{i_1,\cdots i_l}(\bar z) 
\label{5.5}
\ee
$l=2,\cdots n; i_s = 1, \cdots ,n+1$,
where $F_{i_1,i_2,\cdots i_l}$ are rank $l$ antisymmetric tensors  ( for
example, $F_{i_1, i_2} = F_{i_1}F^{\pr}_{i_2} - F^{\pr}_{i_2}F_{i_1}$). 
 The $(n+1)$-chiral
functions $F_i $ $(\bar F_i)$ are not independent.  The condition they have to
satisfy comes from the last equation ($i=n$) of the system (\ref{5.2}).  For
the $A_n$ case it has the following form:
\br
&&F_{i_1,\cdots i_{n-1}}\bar F_{i_1,\cdots i_{n-1}}  = 
(F^{\pr}_{i_1,\cdots i_n}\bar F^{\pr}_{i_1,\cdots i_n})
(F_{j_1,\cdots j_n}\bar F_{j_1,\cdots j_n}) 
 -  (F^{\pr}_{i_1,\cdots i_n}\bar F_{i_1,\cdots i_n})
(F_{j_1,\cdots j_n}\bar F^{\pr}_{j_1,\cdots j_n})
\nonumber
\er
which turns out to be equivalent to the Wronskian condition 
\be
\eps_{i_1, \cdots  ,i_{n+1}}F_{i_1}F^{(\pr)}_{i_2}\cdots F^{(n)}_{i_{n+1}} = 
\eps_{i_1, \cdots  ,i_{n+1}}
\bar F_{i_1}\bar F^{(\pr)}_{i_2}\cdots \bar F^{(n)}_{i_{n+1}} = 1
\label{5.6}
\ee
The correspondent requirement for $B_n$ is 
\br
2(F_{i_1,\cdots i_{n-1}}\bar F_{i_1,\cdots i_{n-1}})
(F_{j_1,\cdots j_{n}}\bar F_{j_1,\cdots j_{n}})&=& 
(F^{\pr}_{i_1,\cdots i_{n}}\bar F^{\pr}_{i_1,\cdots i_{n}})
(F_{j_1,\cdots j_{n}}\bar F_{j_1,\cdots j_{n}})\nonumber\\
& - & 
(F^{\pr}_{i_1,\cdots i_{n}}\bar F_{i_1,\cdots i_{n}})
(F_{j_1,\cdots j_{n}}\bar F^{\pr}_{j_1,\cdots j_{n}})
\label{5.7}
\er
This way one can write the solutions of the $G_n^{(1,1)}$-NA Toda theory $\phi_i, \psi, \chi $ in terms of the $n+1$-dependent functions $F_i$, ($\bar
F_i$) and the chiral currents $V^+, \bar V^-$.  It is convenient to introduce
a new set of $n+1$ {\it independent} functions $f_i(z)$ and $\bar f_i(\bar z)$ 
\be
f_i = (V^+)^{{\cal K}_{11}}F_i, \quad  \bar f_i = (\bar
V^-)^{{\cal K}_{11}}\bar F_{i}, \quad  
f_{ij} = (V^+)^{2{\cal K}_{11}}F_{ij}, \quad  \bar f_{ij} = (\bar
V^-)^{2{\cal K}_{11}}\bar F_{ij}, \,\, {\rm etc.}
\label{5.8}
\ee
This change of variables is based on the following {\it observation}:  let  us shift
the Toda fields $\varphi_l$ as follows
\br
\varphi_l^{\pr} = \varphi_l + l{\cal K}_{11}ln V^+\bar V^-
\nonumber
\er
where $l=1,2,\cdots ,n$ for the $A_n$ case and $l=1, \cdots n-1$ for $B_n$. 
The last field $\varphi^{\pr}_n$ in $B_n$ is given by
\br
{\varphi_n}^{\pr} = \varphi_n +{n\o 2}{\cal K}_{11} ln   V^+\bar V^-
\nonumber
\er
(${\cal K}_{11} =1$ for $B_n$).  Then the first $n-1$ of the new fields
$\varphi_j^{\pr}$, $(j=1, \cdots n-1)$ satisfy the same abelian Toda eqns.
(\ref{5.2}), while the last one becomes
\br
\pa \bar \pa \varphi^{\pr}_n &=& (V^+ \bar V^- )^n e^{-2\varphi_n^{\pr} +
\varphi_{n-1}^{\pr}}, {\rm for }\;\; A_n \nonumber \\
\pa \bar \pa \varphi^{\pr}_n &=& (V^+ \bar V^- ) e^{-2\varphi_n^{\pr} +
\varphi_{n-1}^{\pr}}, {\rm for }\;\; B_n
\nonumber 
\er
The general solutions are given again by eqn (\ref{5.5}) ( but now with $f_i,
\bar f_i$, etc ).  The fact that the last equation of  the system (\ref{5.2})has
been modified leads to the evident changes in eqns. (\ref{5.6}) and (\ref{5.7})
\be
(V^+)^n = \eps_{i_1, \cdots  ,i_{n+1}}f_{i_1} \cdots f^{(n)}_{i_{n+1}}, \quad\quad 
(\bar V^-)^n = 
\eps_{i_1, \cdots  ,i_{n+1}}\bar f_{i_1} \cdots \bar f^{(n)}_{i_{n+1}}
\label{5.9} 
\ee
for $A_n$ and 
\be
2(V^+\bar V^-) (f_{i_1, \cdots i_{n}}\bar f_{i_1, \cdots i_{n}}) = 
(\eps_{i_1, \cdots  ,i_{n+1}}f_{i_1} \cdots f^{(n)}_{i_{n+1}})
(\eps_{j_1, \cdots  ,j_{n+1}}\bar f_{j_1} \cdots \bar f^{(n)}_{j_{n+1}})
\label{5.10}
\ee
for $B_n$.
 
Substituting (\ref{5.5}) in (\ref{5.1}) and (\ref{5.4}) we obtain the general
solutions of eqns. (\ref{2.3}) in the following form{\footnote{We are
sistematicaly omiting the explicit solutions for $\phi_1, \; i=2,\cdots n-1$
which contains rank $i-1$ antisymmetric tensors since we do not need them in
the calculations of the asymptotics of $\varphi $.}}
\br
&&e^{R} =({k\o 2})^{-n} f_i \bar f_i, \quad \quad e^{-k_{12}\phi_1} = ({k\o 2})^{1-n}{1\o 2}
 f_{ij} \bar f_{ij}{(f_i \bar f_i)}^{{1\o {{\cal K}_{11}}}-2}( V^{+}\bar
 V^{-})^{-1}, 
 \nonumber \\
&&\hskip -1cm \psi = ({k\o 2})^{{1-n}\o 2} {(f_i \bar f_i)}^{{1\o {2{\cal K}_{11}}}-1}
 {(f_i^{\pr} \bar f_i)}(V^{+})^{-1}, \quad  
  \chi =({k\o 2})^{{1-n}\o 2} {(f_i \bar f_i)}^{{1\o {2{\cal K}_{11}}}-1}
 {(f_i \bar f_i^{\pr})}(\bar V^{-})^{-1} , \, {\rm etc.}
 \label{5.11}
 \er
 We are now able  to write the explicit solution for the field $\varphi $
 given in eqn. (\ref{4.4})
 \be
 \varphi = -{\cal K}_{11} \ln \{ ({k\o 2}) {{(f_i^{\pr}\bar f_i^{\pr})
 (f_j \bar f_j) + ({{1\o {2{\cal K}_{11}}}-1})(f_i^{\pr}\bar f_i)(f_j \bar
 f_j^{\pr})}\o {(f_i \bar f_i)^2V^{+}\bar V^{-}}} \} \equiv  -{\cal K}_{11} ln G
 \label{5.12}
 \ee
 whose asymptotics are under investigation.  The scaling properties of the
 function $G$ suggest to look for a class of solutions such that $V^{+}(z)\bar
 V^{-}(\bar z) = A(t)$ i.e. its fixed time limit for $\sigma \longrightarrow
 \pm \infty $ are constants, $A_+ = A_- = A(t=0)$.  This leads to the following
 ansatz:
 \be
 f_i = \a_i e^{a_i (t+\sigma )}, \quad \bar f_i = \bar \a_i e^{a_i (t-\sigma )}
 \label{5.13}
 \ee
 Taking into account the explicit form (\ref{5.9}),(\ref{5.13}) of $V^+\bar V^-$ we conclude
 that the requirement  $V^{+}(z)\bar
 V^{-}(\bar z) = A(t)$ is satisfied only if
 \be
 \sum_{i=1}^{n+1} (a_i-\bar a_i) =0 \quad \quad for \;\; A_n
 \label{5.14}
 \ee
 and
 \be
  a_i= \bar a_i  \quad ,i=1,2, \cdots ,n+1 \quad \quad for \;\; B_n
 \label{5.15}
 \ee 
 For the $A_n$ case it is convenient to parametrize the solutions of
 (\ref{5.14}) as follows
 \br
 a_1 -\bar a_1 = b_1+b_2+ \cdots +b_n, \quad \quad a_l-\bar a_l = -b_{l-1},
 \;\; l=2,3,\cdots n+1
 \nonumber
 \er
 and for simplicity we specify $b_n>b_{n-1}> \cdots >b_1>0$.  Then taking the
 limits $\sigma \longrightarrow \pm \infty $ at $\tau =0$ in eqn. (\ref{5.12})
 we find
 \br
 \varphi (\infty ,0) = -{\cal K}_{11} ln({1\o{2{\cal K}_{11}}}a_1\bar a_1 A({k\o
 2})), \quad 
 \varphi (-\infty ,0) = -{\cal K}_{11} ln({1\o{2{\cal K}_{11}}}a_{n+1}\bar
 a_{n+1} A({k\o 2}))
 \nonumber
 \er
 Since $a_1\bar a_1 \neq a_{n+1}\bar a_{n+1}$, therefore $\varphi (\infty ,0)
 \neq \varphi (-\infty ,0)$, (i.e. $H_1 \neq 0$) for the class of solutions
 (\ref{5.13}),(\ref{5.14}) of $A_n^{(1,1)}$-model we have chosen.  This makes
 complete our statement that $\hat SL(2,R)_q$ PB algebra appears as a symmetry
 of the $A_n^{(1,1)}$-NA Toda models.
 
 The condition (\ref{5.15}) for $B_n$-case gives $\varphi (\infty ,0)
 = \varphi (-\infty ,0)$ and therefore $H_1 = 0$ for the class of solutions (\ref{5.13}).  Whether there
 exist $B_n^{(1,1)}$-solutions such that $H_1 \neq 0$ is still an open
 question.

 \sect{$W_{n+1}$-structures in NA-Toda models}

$\frac{}{}$

The way we have constructed the solutions of the $G_{n}^{(1,1)}$-NA-Toda
models, address the {\it question about the origin of the relation 
between the
abelian and NA-Toda theories}. As we have mentioned in the introduction, the
explanation of this phenomena can be found by realizing both as gauged
$G_{n}/H_{i}$-WZW models, ($N^{+}\setminus G_n/ N^{-})$ and
$H^+\setminus G_n/ H^{-}$ and looking for $G_{n}$-gauge transformation
$h=h(V^{+})\bigotimes \bar{h}(\bar{V}^{-})\in G_{n}\bigotimes G_{n}$ 
mapping  $H^{NA}$ in $H^{A}$. This
 transformation can
be considered as a map between the constraints, gauge fixing conditions and
the remaining currents of NA-Toda (\ref{naw}) into the corresponding ones of
the abelian Toda (\ref{aw}). Therefore, it should satisfy the eqns.
(\ref{transf}):
\be
-J_{ij}^{(A)}H_{jk}+H_{ij}J_{jk}^{(NA)}=\frac{k}{2}\partial H_{ik},\quad 
\bar{J}_{ij}^{(NA)}\bar{H}_{jk}-\bar{H}_{ij}\bar{J}_{jk}^{(A)}
=\frac{k}{2}\bar{\partial}\bar{H}_{ik},
\label{6.1}
\ee
where 
 ${H}_{ik}=({h}^{-1})_{ik}$, $\bar{H}_{ik}=(\bar{h}^{-1})_{ik}$
 and, for  simplicity, we are
considering the $A_{n}$-case only (in the convenient Weyl basis --see eqns.
(\ref{3.17}), (\ref{3.18})).

In order to derive the solutions of eqn. (\ref{6.1}), we apply once more the
method we have used in Sect. 3. Due to the specific forms of $J_{ij}^{A}$ and
$J_{ij}^{NA}$, the eqn.(\ref{6.1}) for $i>k$ $(i,k=1,2,...,n+1)$ imply that
\begin{eqnarray}
H_{ik}=0, \quad i>k
\label{6.2}
\end{eqnarray}
 i.e., $H=h^{-1}$ is an upper triangular matrix. For the diagonal
elements $H_{ii}$ we find 
\begin{eqnarray}
H_{ss}=H_{11}(V^{+})^{-1},\quad s=2,3,...,n+1
\label{6.3}
\end{eqnarray}
 Imposing the condition $\det H=1$:
\begin{eqnarray}
\prod_{i=1}^{n+1}H_{ii}=1
\label{6.4}
\end{eqnarray}
(we have used (\ref{6.2}) in deriving (\ref{6.4})), we find that 
\be
H_{ss}=e^{ -\frac{1}{n+1}\Phi } ,\quad \quad 
H_{11}=e^{ \frac{n}{n+1}\Phi } ,\quad \quad \Phi=\ln{V^{+}}
\label{6.5}
\ee
$s=2,3,...,n+1$, is the solution of eqns.
(\ref{6.3}) and (\ref{6.4}). We next analize the equations for the elements
of the first row $H_{1k}$:
\br
H_{1k}=\frac{k}{2}\partial H_{1,k-1},\quad \quad k=3,4,...,n+1\quad \quad
H_{12}V^{+}=\frac{k}{2}\partial H_{11}
\nonumber
\er
The solutions of these recursive relations is given by
\begin{eqnarray}
H_{1s}=-n\left( \frac{k}{2}\partial \right)^{s-1}
e^{ -\frac{1}{n+1}\Phi } ,\quad \quad s=2,3,...,n+1
\label{6.6}
\end{eqnarray}

The elements $H_{l,l+1}$ satisfy more complicated recursive relations:
\br
H_{l,l+1}=H_{l-1,l}+\frac{k}{2}\partial H_{ll},\quad l=2,3,...,n,\quad
\quad H_{12}=\frac{k}{2}\partial H_{11}=
-n\left( \frac{k}{2}\partial \right) H_{22},
\nonumber
\er
which can be solved by taking
\br
H_{l,l+1}=\sum_{s=1}^{l}H_{ss}=(l-1-n)\frac{k}{2}
\partial \left( e^{ -\frac{1}{n+1}\Phi } \right), \quad l=2,3,\cdots ,n .
\label{6.7}
\er
Similarly, for generic $H_{l,l+m}$, we find
\begin{eqnarray}
H_{l,l+m}=\frac{k}{2}\sum_{s=1}^{l}\partial H_{s,s+m-1},\quad 
l=1,2,...,n-m+1, \quad m=1,2,...,n-1
\label{6.g}
\end{eqnarray}
 Taking into account (\ref{6.7})
and (\ref{3.25}), we obtain the solution of (\ref{6.g}), in the form:
\begin{eqnarray}
H_{l,l+m} = \left[ \left( \begin{array}{c} l+m-1 \\ m \end{array} \right)
- (n+1) \left( \begin{array}{c} l+m-2 \\ m-1 \end{array} \right) \right]
\left( \frac{k}{2}\partial \right)^{m}
e^{ -\frac{1}{n+1}\Phi } .
\label{6.8}
\end{eqnarray}
The remaining part of the eqns. (\ref{6.1}):
\begin{eqnarray}
W_{n-l+2}^{(A)}H_{n+1,n+1}+H_{l-1,n+1}-\sum_{s=l}^{n}H_{ls}W_{n-s+2}^{(NA)}
=-\frac{k}{2}\partial H_{l,n+1}
\label{6.9}
\end{eqnarray}
provides the explicit realization of the conserved currents of the {\it
abelian}
Toda theory $W_{n-l+2}^{(A)}$, in terms of the ones of the {\it NA-Toda }
$W_{n-s+2}^{(NA)}$ and $V^{\pm}$. By means of (\ref{6.5}) and (\ref{6.8}), we
can write (\ref{6.9}) in the following compact form:
\begin{eqnarray}
&&W_{n+1}^{A}=V^{+}V^{-}+ne^{ \frac{1}{n+1}\Phi } \left(
\frac{k}{2}\partial \right)^{n+1}e^{ -\frac{1}{n+1}\Phi }
-n\sum_{s=2}^{n}W_{n-s+2}^{NA} e^{ \frac{1}{n+1}\Phi } 
\left(\frac{k}{2}\partial \right)^{s-1}e^{ -\frac{1}{n+1}\Phi} ,
\0\\
&&W_{n-l+2}^{A}=W_{n-l+2}^{NA}-(\Gamma_{l-1,n-l+2}+\Gamma_{l,n-l+1})e^
{ \frac{1}{n+1}\Phi } \left( \frac{k}{2}\partial
\right)^{n-l+2}e^{ -\frac{1}{n+1}\Phi }
\nonumber
\\
&&\hskip 1.5cm +\sum_{s=l+1}^{n}\Gamma_{l,s-l}W_{n-s+2}^{NA} e^{
\frac{1}{n+1}\Phi } \left( \frac{k}{2}\partial \right)^{s-l}
e^{ -\frac{1}{n+1}\Phi }  ,
\label{6.10}
\end{eqnarray}
for $l=2,3,...,n$, where the coefficients $\Gamma_{l,m}$ are given by
\begin{eqnarray}
\Gamma_{l,m}=\left( \begin{array}{c} l+m-1 \\ m \end{array} \right) -(n+1)
\left( \begin{array}{c} l+m-2 \\ m-1 \end{array} \right) .
\nonumber
\end{eqnarray}

Let us give two explicit examples of the relations (\ref{6.10}):
\begin{eqnarray}
T^{(A)}&=&T^{(NA)}-\left( \frac{k}{2}\right)^{2}\frac{n}{2}\left(
\partial^{2}\Phi -\frac{1}{n+1}( \partial \Phi )^{2}\right) ,
\nonumber
\\
W_{3}^{(A)}&=&W_{3}^{(NA)}+\frac{2}{n+1}T^{(NA)}\left( \frac{k}{2}\partial
\right) \Phi +(3n-4)e^{ \frac{1}{n+1}\Phi } \left(
\frac{k}{2}\partial \right)^{3}e^{ -\frac{1}{n+1}\Phi } .
\nonumber
\end{eqnarray}
(for $n=2, W_3^{NA}=V^+V^-$).  
Comparing the form of the eqns. (\ref{6.1}) for the chiral $H_{jk}(z)$ with
the one for the antichiral $\bar{H}_{ik}(\bar{z})$, we conclude that
\begin{eqnarray}
\bar{H}_{ik}=H_{ki}(\partial \rightarrow \bar{\partial},\Phi
=\ln{V^{+}}\rightarrow \bar{\Phi}=ln{\bar{V}^{-}}).
\nonumber
\end{eqnarray}
The corresponding relations between the antichiral currents $\bar{W}_{p}^{A}$
and $\bar{W}_{p}^{NA}$, $\bar{V}^{\pm}$ have the same form as (\ref{6.10}),
with $\partial \rightarrow \bar{\partial}$ and $\Phi \rightarrow \bar{\Phi}$.

Our initial motivation of studying the solutions of eqns. (\ref{6.1}) was to
find an explanation of the change of variables (\ref{5.1}), (\ref{5.2}), that
transforms part of the NA-Toda into the abelian Toda equations. Denoting by
$g_{ik}^{NA}$ the $A_{n}$-WZW field $g_{ik}\in A_{n}$, constrained by eqns.
(\ref{3.17}), and by $g_{ik}^{A}$, the reduced form of $g_{ik}$ by the abelian
Toda constraints (\ref{aw}), we realize that
\begin{eqnarray}
g^{NA}=\bar{H}g^{A}H.
\label{6.11}
\end{eqnarray}
The explicit form of $g_{ik}^{NA}$ is given by eqns. (\ref{3.46}), while
$g_{ik}^{A}$ are known to be (see Sect. 2 of ref. \cite{Orai})
\br
&&g_{11}^{A}=e^{ \varphi_{1}^{A}},\;\;
g_{22}^{A}=e^{ \varphi_{2}^{A}-\varphi_{1}^{A}} -e^{
\varphi_{1}^{A} } ( \frac{k}{2}\partial  \varphi_{1}^{A})(
\frac{k}{2}\bar{\partial} \varphi_{1}^{A}),
\nonumber\\
&&g_{1l}^{A}=\left( \frac{k}{2}\partial \right)^{l-1}e^{
\varphi_{1}^{A}},\quad \quad g_{l1}^{A}
=\left( \frac{k}{2}\bar{\partial}\right)^{l-1}e^{
\varphi_{1}^{A}},\quad \ldots \quad {\rm etc}.
\er

More generally $\phi_{i}^{A}=\ln{D_{i}}$, where $D_{i}$ are certain
subdeterminants fo the matrix $g_{ik}^{A}$ \cite{Orai}. With the explicit form
of $H$ and $\bar{H}$ at hand, we verify that (\ref{6.11}) indeed reproduces
eqns. (\ref{5.1}).

The most {\it important consequence} of the $H$ and $\bar{H}$-transformations
(\ref{6.1}) is the explicit realization (\ref{6.10}) of the {\it abelian Toda
conserved currents} $W_{p}^{A}$ ($p=2,3,...,n+1$) in terms of the {\it NA-Toda
currents} $V^{\pm}$, $W_{p}^{NA}$ ($p=2,3,...,n$). As it is well known, the
$W_{p}^{(A)}$'s generate the $W_{n+1}$-algebra \cite{Ger-Bil}.
 On the other hand,
we have shown in Sect. 3 that $V^{\pm}$ and $W_{p}^{NA}$ are the generators
of the {\it non-local (non-Lie) algebra} $V_{n+1}^{(1,1)}$, which is the algebra of
the symmetries of the $A_{n}$-NA Toda theory. The eqns. (\ref{6.10})
suggest that the {\it $W_{n+1}$-algebra \cite{Ger-Bil} 
lies in the universal enveloping of the
$V_{n+1}^{(1,1)}$-algebra}, i.e., the $W_{n+1}$-generators are specific
combinations of certain products of the $V_{n+1}^{(1,1)}$-generators. Using
the $V_{n+1}^{(1,1)}$-PB's only, we verify that for $n=1,2,3$ the
$W_{n+1}$-generators, constructed by the $V_{n+1}$-generators, according to
the rule (\ref{6.10}), indeed satisfy the standard $W_{n+1}$-PB's relations.
The shortest way to prove this for arbitrary $n$ is to derive the
$W_{n+1}$-infinitesimal transformations $\delta_{\eta_{p}}W_{n-l+2}^{(A)}$
from the $V_{n+1}$-transformations $\delta_{\epsilon^{\pm}}W_{p}^{(NA)}$,
$\delta_{\epsilon^{\pm}}V^{\pm}$, $\delta_{\eta_{p}}W_{n-l+2}^{(NA)}$,
$\delta_{\epsilon^{\pm}}H_{ik}$ etc, solving explicitly the eqn. (\ref{6.9}),
written this time for the infinitesimal transformations. It is not difficult
to verify that
\begin{eqnarray}
\delta_{\epsilon^{-}}W_{n-l+2}^{(A)}=0,\quad i.e. \quad 
\left\{ V^{+},W_{n-l+2}^{(A)}\right\} =0,
\nonumber
\end{eqnarray}
and that
$\delta_{\epsilon^{+}}W_{n-l+2}^{(A)}=\delta_{\eta^{(A)}}W_{n-l+2}^{(A)}$.
However, the complete proof is still missing.

One might wonder whether these $W_{n+1}$-transformations that appear in the
NA-Toda theories are in fact symmetries, i. e., whether the action (\ref{2.17})
is invariant under the $W_{n+1}$-transformations of the NA-Toda fields
$\psi$, $\chi$, $\Phi_{l}$. It is indeed the case, but again our proof is
restricted to the particular cases $n=1,2,3$.

\sect{$G_{n}^{(j,1)}$-NA-Toda Models }

The $G_{n}^{(j,1)}$-NA-Toda models are straightforward generalization of the 
$G_{n}^{(1,1)}$-NA-Toda ones  (\ref{2.3}), (\ref{2.17}).  They are defined as
the Hamiltonian reduction of the $G_n$-WZW model by the constraints 
\br
J_{-\alpha_{i}}&=&\bar{J}_{\alpha_{i}} = 1, \quad i=1, \cdots n\quad i\neq j
\nonumber
\\
J_{-[\alpha]}&=&\bar{J}_{[\alpha]} = 0, \quad \a \;\; {\rm non \;\; simple \;\; root }
\nonumber
\\
J_{\lambda_{j}\cdot H}&=&\bar{J}_{\lambda_{j}\cdot H} = 0 
\label{7.1}
\er
i.e. the current $J_{-\a_j} $ $(\bar J_{\a_j} )$, ($j$ is arbitrary fixed ) is
now left unconstrained.  Similarly to the $j=1$ case the $G_{n}^{(j,1)}$
models can be realized as gauged  $G_n /H^{(j)}$-WZW models.
  The 
subgroups $ H^{(j)}_{\pm} $ with $H^{(j)}_{+} =
N_{+}^{(j)}\otimes {H_{0}^{(j)}}^{0}$ and $H^{(j)}_{-} =
N_{-}^{(j)}\otimes {H_{0}^{(j)}}^{0}$  are introduced by means of the grading
operator  $
Q_j = \sum_{k \neq j}^{n} \lambda_k \cdot H$.  
The nilpotent subgroup $N^{(j)}_{\pm} $ are generated by $Q_j$-positive
(negative) step operators (i.e. all except $E_{\pm \a_j}$ since 
$[Q_j, E_{\pm \a_j} ] = 0 $).  The $U(1)$-subgroup ${H_{0}^{(j)}}^{0}$  
is generated
by ${{2\lambda_j \cdot H}\o {\a_j^2}}$.  This $Q_j$-gradation of $G_n$ 
reflects
the algebraic structure of the constraints (\ref{7.1}) and  suggests the
following ``{\it nonabelian Gauss decomposition}'' for each $g \in G_n$ 
( valid for
the connected part of $G_n$):
\br
g = g_- g_0^f g_+, \quad g_{\pm} \in H_{\pm} ^{(j)}, \quad g_0^f \in
{H_0^f}^{(j)}
\nonumber 
\er
\begin{eqnarray}
g_{-}&=&\exp \left\{
\sum_{[\alpha] \neq \a_j}\chi_{[\alpha]}E_{-[\alpha]}\right\}
\exp \left\{ \frac{1}{{\cal K}_{jj}}{{\lambda_{j}\cdot H\phi_{j}}\o {\a_j^2}}
\right\}
\nonumber
\\
g_{+}&=&\exp \left\{ \frac{1}{{\cal K}_{jj}}{{\lambda_{j}\cdot H}\o {\a_j^2}}
\phi_{j}\right\}
\exp \left\{ \sum_{[\alpha] \neq \a_1}\psi_{[\alpha]}E_{[\alpha]}\right\}
\nonumber
\\
g_0^f&=&\exp \left\{ \chi E_{-\alpha_{j}}\right\}
\exp \left\{ \sum_{i \neq j}^{n} \phi_{i}{2{\a_i \cdot H}\o {\a_i^2}}
\right\}
\exp \left\{ \psi E_{\alpha_{j}}\right\}
\label{7.2}
\end{eqnarray}
The action of the $G_n/H^{(j)}$-WZW model that describes the 
$G_{n}^{(j,1)}$-NA-Toda theory is given again by eqn{(2.10}) where now,
\br
A^{j}&=&h_{-}^{-1}\partial h_{-},\quad 
\bar A ^{j}=\bar {\partial }h_{+}h_{+}^{-1},\quad
 h_{\pm}(z,\bar{z})\in H_{\pm}^{j}
\nonumber \\
A^{j} & = & A_0 + A_- ,\quad \bar A^{j}= \bar A_0 + \bar A_-,
\quad A_{-} \in N^{j}_{-} , \quad \bar A_{+} \in N^{j}_{+}\nonumber \\
A^{j}_0 & = & {2\o {{\cal K}_{jj} \a_{j}^2}}a_0\lambda_j \cdot H,
 \quad \bar A^{j}_0= 
{2\o {{\cal K}_{jj} \a_j^2}}\bar a_0\lambda_j \cdot H , \quad \eps_{\pm}
 = \sum_{i\neq j}({ {{\a_i} ^2}\o 2}) E_{\pm \a_i}
\nonumber
\er

Following the recipe developed in Sect 2, we
integrate out the auxiliary gauge fields $A^{j}$
and  $\bar A^{j}$ in order to obtain the
corresponding action for the $G_n^{(j,1)}$-NA
Toda models
\br
S^{(j)}_n &=& -{k\o {2\pi}} \int dz d\bar z  ({1\o
2} \eta^{(1)}_{ab} \pa \rho^{(1)}_a \bar \pa
\rho^{(1)}_{b} +{1\o 2}
 \eta^{(2)}_{a^{\pr} b^{\pr} }
\pa \rho^{(2)}_{a^{\pr}} \bar \pa 
\rho^{(2)}_{b^{\pr}}\nonumber \\
&  - &({2\o k})^2  \sum_{a=1}^{j-1}
{2\o \a_a^2} e^{-{k^{(1)}_{ab} \rho^{(1)}_b}}  
- ({2\o k})^2\sum _{a^{\pr}=1}^{n-j} {2\o
\a_{a^{\pr}}^2} e^{-{k^{(2)}_{a^{\pr} b^{\pr }}
\rho^{(2)}_{b^{\pr}}}} +
{2 \o {\a_j^2}}e^{k_{j,j-1}\rho^{(1)}_{j-1} + 
k_{j,j+1}\rho^{(2)}_{1}} {{\pa \chi \bar \pa \psi
}\o {\Delta_j}})
\label{7.3}
\er
where $\rho^{(1)}_a = \phi_a$, ($a=1, \cdots ,j-1$), 
$\rho^{(2)}_a = \phi_{a^{\pr}+j}$, ($a^{\pr}=1, \cdots
,n-j$) and 
\br
\Delta_j = 1 + {1\o {2{\cal K}_{jj}}}
\psi \chi e^{k_{j,j-1}\rho^{(1)}_{j-1} + 
k_{j,j+1}\rho^{(2)}_{1}}
\nonumber
\er  
We have assumed for
simplicity that deleting the $j^{th}$ vertex of
the $G_n$ Dynkin diagram the resulting
$G^{\pr}_{n-1}$ algebra is a direct product of two
subalgebras $G_1$ and $G_2$ of rank $j-1$ and
$n-j$ respectively, i.e. $ G^{\pr}_{n-1} = G_1
\otimes G_2$.  The exception arises when a
specific vertex of $D_n, E_6, E_7$ or $E_8$ is
deleted.  In such cases,$ G^{\pr}_{n-1} = G_1
\otimes G_2 \otimes G_3$ and the generalization
of (\ref{7.3}) is evident.

 As in the $j=1$ case (see Sect. 3), the
symmetries of the action (\ref{7.3}) are
generated by the $n+1$-chiral ``remaining
currents'' $W_{s({\a})}$, $V^{\pm}$ (and 
$\bar W_{s({\a})}$, $\bar V^{\pm}$) and the global
(nonchiral) $U(1)$ current 
\be
J_{\mu}^{j} = -{k\o {4\pi }}{ e^{{k_{ji}\phi_i}}\o
\Delta_j } \( \psi \pa_ {\mu }\chi - \chi \pa_
{\mu }\psi - \psi \chi \pa _{\mu}k_{ji}\phi_i \)
\nonumber 
\ee
Their conformal spin (dimension ) are given by
the following $j$-analog of eqn. (\ref{3.9}):
\be
s(\a ) = 1 + X_j {{2\lambda_j \cdot \a }\o
{\a_j^2}} + \sum_{i \neq j} {{2\lambda_i \cdot \a
}\o {\a_i^2}}, \quad X_j = -{1\o
{\cal K}_{jj}}\sum_{i\neq j} {\cal K}_{ji}
\label{7.4}
\ee
Choosing a specific Drinfeld-Sokolov (DS) type gauge we find the
remaining currents for the $A^{(j,1)}_n$ case to be 
\br
W_2 &=& J_{\a_n} \quad W_3 = J_{\a_n +
\a_{n-1}}, \quad  \cdots , W_{n-j+1} = J_{\a_n
+\cdots + \a_{j+1}}\nonumber \\
\tilde W_2 &=& J_{\a_1} \quad \tilde W_3 =
J_{\a_1+ \a_{2}}, \quad  \cdots , \tilde
W_{j} = J_{\a_1 +\cdots + \a_{j-1}} \nonumber \\
V_j^{+} &=& J_{-\a_j}, \quad V_j^{-} = J_{\a_1
+\cdots + \a_n }
\label{7.5}
\er
where the index $s(\a )$ in $W_s$ and in 
$\tilde W_s$ denote their spin ($X_j = -{{n-1}\o
2}$ for $A_n$).  The nonlocal currents
$V_j^{\pm}$  are both of spin $s^{\pm} = {{n+1}\o
2}$.  For $B_n^{(j,1)}$-models we have $X_j={(1-j)\o 2}$ and
\br 
 W_2 &=& J_{\a_n} \;\;  W_4 = J_{2\a_n +
\a_{n-1}}, \;\;  W_6 = J_{2\a_n +
2\a_{n-1} + \a_{n-2}}, \;\; \cdots ,  W_{2(n-i+1)} = J_{2\a_n +2\a_{n-1}
+\cdots + 2\a_{i+1} + \a_i} \nonumber \\
\tilde W_2 &=& J_{\a_1} \quad  \tilde W_3 =
J_{\a_1+ \a_{2}}, \quad  \cdots , \tilde
W_{j} = J_{\a_1 +\cdots + \a_{j-1}} \nonumber \\
V_j^{+} &=& J_{-\a_j}, \quad V_j^{-} = J_{\a_1
+\cdots + \a_n} 
\label{7.6}
\er 
The spin of $V_j^{\pm}$ is now $s^{\pm}_j =
n-{1\o 2}(j-1)$.

The structure of the constraints (\ref{7.1})
allows to choose one of the nonlocal currents to
be $V_j^{+} = J_{-\a_j}$, ($\bar V_j^{+} =
\bar J_{\a_j}$) which have the explicit form  (c.f.
eqn. (\ref{3.12}) for $j=1$)
\be
V_j^{+}(z) = {k\o 2} 
{\pa \chi \o \Delta_j }e^{k_{ji}\phi_i + {1\o
{2{\cal K}_{jj}}}R_j}, \quad 
\bar V_j^{-}(z) = {k\o 2} {\bar \pa \psi \o \Delta_j}e^{k_{ji}\phi_i + {1\o
{2{\cal K}_{jj}}}R_j}
\label{7.7}
\ee

Applying the method we have used in Sect 3. in
the derivation of the $V_n^{(1,1)}$-algebra one
can find the algebra of the symmetries $V_n^{(j,1)} 
$( $\bar V_n^{(j,1)}) $ of the
$A_n^{(1,1)}$-NA-Toda model.  The corresponding
recursive (differential ) relations and their
solutions are quite similar to those of $j=1$
obtained in Sect 3.  We present here the explicit
form of the {\it simplest nontrivial example} of such
$j\neq 1$ type of algebra namely,
$V_4^{(2,1)}$-algebra. According to eqn.
(\ref{7.5}) it consist of four spin two currents
$V^{\pm} $, $T= \tilde W_2 + W_2$ and $V^{0} = \tilde W_2 - W_2$ satisfying, 
\br
&&\hskip -0.5cm \{ T(\sigma ), T(\sigma^{\pr} ) \}  = 2T(\sigma^{\pr} )\pa _{\sigma^{\pr}} \d
(\sigma - \sigma^{\pr} ) + \d (\sigma - \sigma^{\pr} )\pa _{\sigma^{\pr}}
 T(\sigma^{\pr} ) - 4 \pa ^{3}_{\sigma^{\pr}}\d (\sigma - \sigma^{\pr} )
 \nonumber \\
 &&\hskip -0.5cm \{ T(\sigma ), V^{\a }(\sigma^{\pr} ) \}  =
 2V^{\a }(\sigma^{\pr} )\pa _{\sigma^{\pr}} \d
(\sigma - \sigma^{\pr} ) + \d (\sigma - \sigma^{\pr} )\pa _{\sigma^{\pr}}
 V^{\a }(\sigma^{\pr} ), \quad \a = 0,\pm
 \nonumber \\
 &&\hskip -0.5cm \{ V^{\pm }(\sigma ), V^{\pm }(\sigma^{\pr} ) \}  = {1\o 8} 
 \eps ( \sigma - \sigma^{\pr} )V^{\pm }(\sigma ) V^{\pm }(\sigma^{\pr} ),
\quad
\{ V^{0}(\sigma ), V^{\pm }(\sigma^{\pr} ) \}  ={1\o 8} 
 \eps ( \sigma - \sigma^{\pr} )V^{0 }(\sigma ) V^{\pm }(\sigma^{\pr} )
\nonumber \\
&&\hskip -0.5cm \{ V^{0}(\sigma ), V^{0}(\sigma^{\pr} ) \}  = -{1\o 4} 
\eps ( \sigma - \sigma^{\pr} )[V^{+ }(\sigma ) V^{- }(\sigma^{\pr} ) +
V^{-}(\sigma ) V^{+}(\sigma^{\pr} )] \nonumber \\
&&\hskip3cm + 2 T(\sigma ^{\pr} )\pa _{\sigma^{\pr}} \d (\sigma - \sigma^{\pr} ) +  
\d (\sigma - \sigma^{\pr} ) \pa _{\sigma^{\pr}}T(\sigma^{\pr} ) - 
4 \pa ^{3}_{\sigma^{\pr}}\d (\sigma - \sigma^{\pr} )
 \nonumber \\
&&\hskip -0.5cm \{ V^{- }(\sigma ), V^{+ }(\sigma^{\pr} ) \}  = -{1\o 8} 
 \eps ( \sigma - \sigma^{\pr} )[V^{0 }(\sigma ) V^{0 }(\sigma^{\pr} ) +
V^{-}(\sigma ) V^{+}(\sigma^{\pr} )] \nonumber \\
&&\hskip3cm  +  T(\sigma ^{\pr} )\pa _{\sigma^{\pr}} \d (\sigma - \sigma^{\pr} )
+ {1\o 2}\d (\sigma - \sigma^{\pr} ) \pa _{\sigma^{\pr}}T(\sigma^{\pr} )
- 2\pa ^{3}_{\sigma^{\pr}}\d (\sigma - \sigma^{\pr} )
\label{7.8}
\er
(k is fixed to 2 in eqn. (\ref{3.15})).  It turns out that $V_4^{(2,1)}$ has
 the same 
 structure  as the $V_{2,2}$- algebra of ref. \cite{Bil} (see eqn.
(2.37) of ref. \cite{Bil}).  In our case the $V_4^{(2,1)}$ algebra
(\ref{7.8}) appears as the algebra of symmetries of the $A_3^{(2,1)}$-NA-Toda
model
\br
{\cal L}_{(3)}^{(2,1)} &=& \pa A \bar \pa A + \pa B \bar \pa B - e^{2A} -
e^{2B} 
+ {1\o 2}e^{A+B}{{ \(\bar \pa \psi \pa \chi + \bar \pa \chi \pa \psi \)}\o 
{1 +
{1\o 2}e^{A+B} \psi \chi }}
\nonumber \\
& + & {1\o 4} {{\( \bar \pa (A+B) (\chi \pa \psi - 
\psi \pa
\chi ) - \pa (A+B) (\chi \bar \pa \psi - \psi \bar \pa \chi ) \)}\o {1 +
{1\o 2}e^{A+B} \psi \chi }}
\label{7.9} 
\er
The $ {\cal L}_{(3)}^{(2,1)} $ differs from the one derived from (\ref{7.3}) 
by
an appropriate total derivative term similar to the one introduced in the 
$j=1$
action (\ref{2.17}).

As one might expect the charges $Q^+_{(j)}$, $\bar Q^-_{(j)}$, of the
nontrivial nonlocal currents $V^+_{(j)}$, $\bar V^-_{(j)}$ have nonvanishing
PB's and together with the $U(1)$ charge, $Q_0 = \int J_0^{(j)}$ close  
$SL(2,R)_{q_{(j)}}$ PB's algebra (${q_{(j)}}= e^{-{2\pi \o k}({1\o
{2{\cal K}_{jj}}})} $).  The calculation is identical to the case $j=1$ case
considered in Sect 4.  The final result is 
\be
\{ Q_{(j)}^{+}, \bar Q_{(j)}^{-} \} = {k\pi \o 2}\int _{-\infty}^{\infty}
d\sigma \pa_ {\sigma } e^{{1\o {\cal K}_{jj}}\varphi  ^{(j)}}, \quad 
\varphi  ^{(j)} = R^{(j)} + {\cal K}_{jj} (k_{ji} \phi_i - ln \Delta _j )
\label{7.10}
\ee
The derivation of the PB's of the remaining nonlocal charges $\{ Q_{(j)}^{-} , 
\bar Q_{(j)}^{+} \}$ is an open problem.  It is important to note that as
 in the
$j=1$ case we have used the ${\cal L}_n^{(j)}$ modified by a specific total
derivative term in the calculation of the conjugate momenta $\Pi_{\psi , \chi ,
\rho_i }$.

We next consider the problem of mapping the $G_n^{(j,1)}$-models into the 
$G_n$
abelian Toda models.  Our starting point is again the observation that there
exists a transformation of variables
\br
\varphi_i &=& \rho_i^{(1)} - {2\o {\a_j^2}}(k_{ja}{\eta_{ia}^{(1)}}^{-1} )R_j
-{\cal K}_{ij} lnV^{+}_j \bar V_j^{-}, \quad i= 1, \cdots ,j-1
\nonumber \\ 
\varphi_j &=& R_j -{\cal K}_{jj}lnV^{+}_j \bar V_j^{-} ,
\nonumber \\
 \varphi_{j+l} &=& \rho_l^{(2)} - {2\o {\a_j^2}}(k_{ja}{\eta_{la}^{(2)}}^{-1} )R_j
-{\cal K}_{l+j,j} lnV^{+}_j \bar V_j^{-}, \quad l= 1, \cdots n -j
 \label{7.11}
 \er
 which transforms part of the equations of motion of (\ref{7.3})
  into the abelian
 $G_n$ Toda equations (\ref{5.2}) for the new fields $\varphi_l$, ($ l=1,
 \cdots n$), where the identity 
 \br
 {1\o {{\cal K}_{jj}}} = \sum _{i=1}^{j-1}{2k_{ja}\o {\a_j^2}} k_{j,i}{\eta_{i,a}^{(1)}}^{-1}
  +\sum_{b=1}^{n-j}{2k_{j,j+a}\o {\a_j^2}} k_{j,j+b}{\eta_{b,a}^{(2)}}^{-1} - k_{jj}
  \nonumber
  \er
  was verified for $A_n, B_n, C_n $ and $D_n$.  The nonlocal field $R_j$
  satisfy now
 \br
 \bar \pa \pa R_j = {{\bar \pa \psi \pa \chi }\o {\Delta_j^2}} e^{k_{jb}\phi_b }
 \nonumber 
 \er
   For example in the $A_3^{(2,1)}$ case (\ref{7.9}) we have 
 \br
 \varphi_1 =  {1\o 2}R_{(2)} + A - {1\o 2} ln V^{+}\bar V^{-}, \quad \;
 \varphi_2 =  R_{(2)} - ln V^{+}\bar V^{-} \quad \;
 \varphi_3 =  {1\o 2}R_{(2)} + B- {1\o 2} ln V^{+}\bar V^{-} 
 \nonumber 
 \er
 
 Following the same line of argument presented in Sect.6, we seek a $G_n$
 gauge transformation mapping the constraints (\ref{7.1}) and the remaining
 currents (\ref{7.5}) into the constraints and remaining currents
 (\ref{aw}) leading to the abelian Toda theory:
 \be
 \bar H \bar J^{(A)} = \bar J^{(NA)}_{(j)} \bar H - {k\o 2} \bar \pa \bar H
 \label{7.12}
 \ee
 For the $A_n$  case, $ \bar J^{(A)}$ and $ \bar J^{(NA)}_{(j)}$  have the
 following matrix form:
 \br
  (\bar J^{(A)})_{il} &= & \d_{i,l-1} + \bar W_{n-l+2}\d_{i,n+1}, \quad \bar W_1
  =0 \nonumber \\
  (\bar J^{(NA)}_{(j)})_{il} &=& (\bar V^{-})^{\d_{ij}\d_{l,j+1}}\d_{i,l-1}
   + \d_{l,1}
  \sum_{s=2}^{j} \tilde {\bar W_s }\d_{is} + \d_{i,n+1} \d_{l,1}\bar
  V^{+}\nonumber \\
  & +&  
  \sum_{p=j+1}^{n} \bar W_{n-p+2}\d_{p,l} \d_{i,n+1}
  \label{7.13}
  \er
  Substituting (\ref{7.13}) in (\ref{7.12}) and requiring $det \; \bar H = 1$
  and $\bar H_{i,l} =0, \;\; i<l$ we obtain:
  \br
  \bar H_{ii} &= &\left\{\begin{array}{cc} (\bar V^{-} )^{{n-j+1}\o {n+1}}, & i=1, \cdots j \\
  \hskip-0.3cm (\bar V^{-} )^{{-j}\o {n+1}}, & i=j+ 1, \cdots n+1 \end{array} \right.
  \nonumber 
  \er
  We next consider the equations for $\bar H_{i,1}$:
  \br
  (\bar V^{-} )^{\d_{ij}} \bar H_{i+1, 1} = - \sum_{p=2}^{j} \tilde {\bar W_p}
  \bar H_{11} \d_{ip} + {k\o 2} \pa \bar H_{i1} 
  \nonumber 
  \er
  Their  solutions are given by
  \br
  (\bar V^{-} )^{\d_{ij}} \bar H_{i+1, 1} &=& -\sum_{l=0}^{j-2}
   ({k\o 2}\pa )^{l}
  ({\tilde {\bar W}}_{j-l} \bar H_{11}) + ({k\o 2}\pa )^{i}\bar H_{11}, 
  \quad
  i=1,\cdots j \nonumber \\
  \bar H_{j+r, 1} &=&({k\o 2}\pa )^{r-1} \bar H_{j+1,1}, 
  \quad\quad r=2, \cdots ,n-j+1
  \label{7.14}
  \er
 The general solution of (\ref{7.12}) can be written in terms of $H_{i1}$ as
 follows
\br
 &&(\bar V^{-} )^{\d_{sj}} \bar H_{s+1,k} = \sum_{p=0}^{s-k+1} {(p+k-2)! \o
 {(k-2)! p!}} ({k\o 2}\pa )^{p}\bar H_{s-p+k+2,1}, \quad s= k-1, \cdots j
 \nonumber \\
 &&\hskip-0.8cm \bar H_{j+r,k} = \sum_{p=1}^{k} {(r-1)! \o {(k-p)!(r+p-k-1)!}}({k\o 2}\pa
 )^{r-k-1+p} \bar H_{j+1,p}, r=2,\cdots, n-j+1, \quad k=1,\cdots ,n+1
 \nonumber
 \er
 Note that $\bar H_{il} = \bar H_{il}(\bar V^{-}, \tilde {\bar W_p } )$
 contrary to the $j=1$ case are now functionals of the currents
  $\tilde {\bar W_p }$ as well.  This reflects  a specific (mixed type ) of
  gauge (\ref{7.5}) we have chosen.  In the DS gauge where all $W_{n-s+2}=
  J_{\a_s + \a_{s+1} \cdots + \a_n }$, $ V^{-} = J_{\a_1 + \cdots + \a_n }$
  lie
  on the last column ( all $\bar W_s$, $\bar V^{+}$ on the last row)  the
  corresponding $H_{il} \;\; (\bar H_{il})$ indeed depend on $V^{+}\;\; (\bar V^{-})$
  only.
  The spins of $W_{n-s+2} $ in the DS gauge is $n-s+2$ for $s \geq j+1 $ and
  ${{n+3} \o 2} - s $ for $s \leq j$.
  
  The main advantage of this current dependent $H$-transformation  is that it
  provides an explicit realization of the $A_n$- NA-Toda currents
  ($W_{n-s+2}^{(A)}$)  in terms of the $A_n^{(j,1)}$- NA-Toda currents $W_p$,
  $\tilde W_p$, $V^{\pm}$ (see eqn. (\ref{6.10}) for $j=1$ case ) and
  vice-versa.  For our $A_3^{(2,1)} $ example (\ref{7.9}) we have $h_{il}=
  (H^{-1})_{il} $ , $H_{il} = \bar H_{li}(\pa \rightarrow \bar \pa, \bar
  V^{-}, \tilde {\bar W_2}\rightarrow V^{+}, \tilde { W_2})$ ($k$ is taken
  to be 2 in (\ref{7.12})):
  \br
  h_{11} &=&  h_{22} = h_{33}^{-1} = h_{44}^{-1} = e^{-{1\o 2}ln V^{+}} \nonumber \\
  h_{12} &=& {1\o 2} h_{23} = \pa h_{11}, \quad  h_{34} = -\pa h_{33}, \quad
  h_{13} = \pa ^2 h_{11}+ \tilde W_2 h_{11} \nonumber \\
  h_{14} &=& \pa ^3 h_{11} + h_{11} \pa \tilde W_2 + 3 W_2 \pa h_{11}, \quad
  h_{24} = 3 \pa ^2 h_{11} + \tilde W_2 h_{11} \nonumber 
  \er
  The corresponding abelian Toda currents $T^{A}$, $W_3^{A}$, $W_4^{A}$ are
  expressed in terms of $V^{\pm}$, $W_2$, $\tilde W_2$ as follows:
  \br
  &&\hskip -0.8cm T^{A} = T^{NA} - 2 \pa ^2\ln V^{+} + {1\o 2} (\pa \ln V^{+})^2, 
  \quad \quad\quad T^{NA} =W_2 + \tilde W_2 \nonumber \\
  &&\hskip -0.8cm W_3 =(W_2 - \tilde W_2)\ln V^{+} + 2 \pa \tilde W_2 - 2 \pa ^3 \ln V^{+} +
  (\pa ^2 \ln V^{+}) \pa (\ln V^{+}) \nonumber \\
  && \hskip -0.8cm W_4 = V^{+} V^{-} - W_2 \tilde W_2 + {1\o 2} (W_2 - \tilde W_2) 
   \pa ^2 (\ln V^{+})  + {1\o 4 } (W_2 + \tilde W_2)(\ln V^{+} )^2 + 
   \pa ^2 \tilde W_2 - {1\o 2} \pa ^4\ln V^{+}-\nonumber \\
   && \hskip0.2cm - (\pa \tilde W_2)(\pa \ln V^{+})  -
   {1\o 8} (\pa  \ln V^{+})^4 
   - {1\o 4} (\pa ^2  \ln V^{+})^2 
   + {1\o 2} 
   (\pa ^2  \ln V^{+})(\pa   \ln V^{+})^2    
\label{7.16}
   \er
  One can verify by direct calculation that if $V^{\pm}$, $W_2$, $\tilde W_2$
  satisfy {\it the $V_4^{(2,1)}$ algebra} (\ref{7.8}),
   then $T^{A}$, $W_3^{A}$ and
  $W_4^{A}$ given by eqns. (\ref{7.16}) indeed close the (classical ) $W_4$
  algebra
  \cite{F-L} .  Therefore the $A_3^{(2,1)}$-NA-Toda model
  has together with the  $V_4^{(2,1)}$ algebra, 
  also the {\it $W_4^{A}$ as its
  algebra of symmetries}.
  
  We now address the question about the relation between
  $A_n^{(j_1,1)}$ and the $A_n^{(j_2,1)}$-NA-Toda models ( $j_1 \neq j_2 $). 
  In terms of transformations $H(j_1)$ and $H(j_2)$ mapping them into
   the $A_n$
  abelian  Toda theory we compose the new transformation $H(j_1, j_2) = H(j_1)
  H^{-1}(j_2) $.  By construction $H(j_1, j_2)$ transforms the constraints and
  remaining currents of the $A_n^{(j_1,1)}$into the corresponding ones of the
  $A_n^{(j_2,1)}$  model:  
  \be
  J^{NA}_{j_2} = H^{-1}(j_1, j_2) J^{NA}_{j_1}H(j_1, j_2) + 
  {k\o 2} H^{-1}(j_1,
  j_2)\pa H(j_1, j_2)
  \label{7.17}
  \ee
  As a byproduct $H(j_1, j_2) $ realizes a map of $V_{n+1}^{(j_1,1)}$-algebra into 
  $V_{n+1}^{(j_2,1)}$ and vice-versa.  The simplest example 
  is given by $n=3, \; j_1 = 2, \;
  j_1 = 1$ i.e.  the transformation of the  $A_3^{(2,1)}$ into the
  $A_3^{(1,1)}$-NA-Toda model $(H(2,1) \equiv H)$
  \br
  H_{11}&=& H_{33} = H_{44} = (V^{+}) H_{22} = e^{{1\o 4} ln V^{+}}, \quad
  H_{12} = -{1\o 3} \pa H_{22} = -{1\o 2}  H_{23} \nonumber \\
  H_{13} &=& \pa H_{12} + H_{22} \tilde W_2, \quad H_{24} = 
  -\pa H_{12} +  H_{22} \tilde W_2 \nonumber\\
  H_{14}&=& \pa ^3 H_{12} - {5\o 4}  (\tilde W_2H_{22})\pa ln V^{+}
  + (\pa \tilde W_2 ) H_{22}
  \nonumber 
  \er
  (we have chosen $V^{+}_{(1)} \equiv V^{+}_{(2)} $).  The current
  transformations take the form:
  \br
  T^{(1)} &=& T^{(2)} -{1\o 2} \pa^2  lnV^{+} + {1\o 8} ( \pa ln V^{+} )^2,
  \quad T^{(2)} = W_2 + \tilde W_2 \nonumber \\
  W_{3}^{(1)} &=& 2 \pa \tilde W_2 +{1\o 2} (\pa ln V^{+} ) 
  (W_2 - 3 \tilde W_2)
  - {1\o 4} (\pa ^2 ln V^{+} ) \pa ln V^{+} + {1\o 16} (\pa ln V^{+} ) ^3
  \label{7.18}
  \er
and $V_{(1)}^{-}$ has a rather complicated form in terms of $V^{-}_{(2)},
V^{+}, W_2$ and $ \tilde W_2$.  

The models $A_n^{(j_1,1)}$ and $A_n^{(j_2,1)}$ have identical field contents
but their lagrangeans represent different interactions between the neutral
fields $\rho^{(1)}_{a}$ and  $\rho^{(2)}_{a^{\pr}}$, (compare for example
 $A_3^{(1,1)}$ and $A_3^{(2,1)}$ models).  The transformation $ H(j_1, j_2)$ 
 changes the ${(j_1)}$-constraints into the  ${(j_2)}$ ones and according 
 to the
 hamiltonian reduction procedure it maps the field equations of  $A_n^{(j_1,1)}$
 to those of $A_n^{(j_2,1)}$.  If we denote by $g_{il}(j_1)$ and 
$g_{il}(j_2)$ the constrained WZW matrix field $g_{il} \in A_n$ ( 
i.e. $g_{il}(j_{\a}) $  depending on the fields $\psi_{j_{\a}}, \chi_{j_{\a}},
$ and $ \rho_a^{(i)}(j_{\a}) $ only ) then $H(j_1, j_2) $ induces the following
field transformations:
\br
&&\phi_1 = -{2\o 3} A + {2\o 3} R_{(2)} - {2\o 3} ln V^{+}\bar V^{-}, 
 \quad 
\phi_2 = B -{1\o 3}A + {1\o 3}R_{(2)} - {1\o 3} ln V^{+}\bar V_{-} \nonumber \\
&&\hskip3cm R_{(1)} =  A + {1\o 2} R_{(2)} + {1\o 4} ln V^{+}\bar V^{-}\nonumber \\
\er
($R_{(1)}$ and $R_{(2)}$ are nonlocal in terms of $\psi, \chi$ and $ A,B$
respectively).  The
remaining $\psi, \chi $ transformations are quite implicit.  Although the 
$A_n^{(j_1,1)}$ and 
  $A_n^{(j_2,1)}$-NA-Toda models represent different interactions and have
  different algebras of symmetry (but equal number of generators )  the
  arguments presented above indicate that they are classically equivalent
  models.  The proof of such statement requires however further investigations.

\sect{Weyl Group Families of $A_n^{(j_1,1)}$-Models}

This section is devoted to the {\it problem}
 of the relation between the NA-Toda
models that have {\it identical algebras of symmetries}. 
 Our starting point is the
{\it following fact}: 
 The $V_3^{(1,1)} $- algebra (\ref{3.51}),(\ref{3.43}) and (\ref{3.44}), ($n=2,
s={3\o 2}$) appears as the symmetry algebra of the $ A_2^{(1,1)}$-NA-Toda
model (\ref{2.17}), ($n=2$) as well as of the reduced Bershadsky-Polyakov 
(BP)
$A_2^{(2)} $-model \cite{Pol1}, \cite{Bershadsky}.  The latter is defined by
the set of constraints imposed on the $A_2$-WZW currents,
\be
J_{-\a_2} =  \bar J_{\a_2} = 0, \quad \quad 
J_{-\a_1-\a_2} = \bar J_{\a_1 + \a_2} =1, 
\label{8.1}
\ee
\be
J_{(\lambda_1 - \lambda_2 )\cdot H } = 
\bar J_{(\lambda_1 - \lambda_2 )\cdot H } =0
\label{8.2}
\ee
It differs from the standard BP model \cite{Bershadsky} by the additional
constraint (\ref{8.2}).   This new constraints is responsible for the
reduction of the $W_3^{(2)}$-algebra (symmetry of the BP model defined by
(\ref{8.1})) to the nonlocal algebra $V_3^{(2)} \equiv V_3^{(1,1)}$. 
Following the methods of Sect. 2 we first derive the lagrangean of the reduced
$A_2^{(2)}$-BP-model:
\be
{\cal L}_3^{(2)} = -{k\o 2\pi }\{ \pa \varphi \bar \pa \varphi  +
 e^{\varphi } {{\bar \pa
\psi_0 \pa \chi_0 }\o { 1 + {3\o 4} e^{\varphi } \psi_0 \chi_0 }} - 
e^{-2\varphi } (
1+ \psi_0 \chi_0 e^{\varphi }) \}
\label{8.3}
\ee
The algebra of symmetries $V_3^{(2)}$ of (\ref{8.3}) obtained by direct
application of the recipe described in Sect. 3 turns out to be identical
 to $V_3^{(1,1)}$.  The lagrangean of the $A_2^{(1,1)}$-NA-Toda model possess
 however quite a different form, 
 \be
 {\cal L}_3^{(1,1)} = -{k\o 2\pi }\{ \pa \phi \bar \pa \phi  + e^{-\phi }
 {{\bar \pa
\psi \pa \chi }\o { 1 + {3\o 4} e^{-\phi } \psi \chi }} - e^{-2\phi } \}
\label{8.4}
\ee
We shall prove that the models (\ref{8.3}) and (\ref{8.4}) are (classically )
equivalent since their Lagrangeans are related by the following change of
field  variables, 
\br
&&\psi = \chi_0 e^{\varphi }(1+ e^{\varphi } \psi_0 \chi_0 )^{-{1\o 4}}, \quad \quad 
\chi = \psi_0 e^{\varphi }(1+ e^{\varphi } \psi_0 \chi_0 )^{-{1\o 4}}
\nonumber \\
&&\hskip 2cm \phi = \varphi  - {1\o 2} \ln (1+ e^{\varphi } \psi_0 \chi_0 )
\label{8.5}
\er
i.e.  ${\cal L}_3^{(2)} ={\cal L}_3^{(1,1)} +$ total derivative.  This can be
verified by direct calculation.  Our derivation of transformation (\ref{8.5})
is based on the following {\it observation}:  The constraints (\ref{8.1}) and
(\ref{8.2}) are the {\it image} of 
\be
J^{\pr}_{-\a_2} = \bar J^{\pr}_{\a_2} = 1, \quad 
J^{\pr}_{-\a_1-\a_2} = \bar J^{\pr}_{\a_1+\a_2} = 0, \quad
J^{\pr}_{-\lambda_1 \cdot H } = \bar J^{\pr}_{-\lambda_1 \cdot H } = 0
\label{8.6}
\ee
 together with the gauge fixing condition
 $ J^{\pr}_{\a_1} = \bar J^{\pr}_{-\a_1} = 0$ (defining the model (\ref{8.4}))
 under the action of a particular  $A_2$-{\it Weyl reflection} 
 \br
 \omega _{\a_1} (\a ) = \a_1 - (\a \cdot \a_1 ) \a, \quad \omega^2_{\a_1} =
 1 
 \nonumber 
 \er
 In fact $\omega _{\a_1} $ maps all the algebraic (Hamiltonian reduction )
 data of the model (\ref{8.3}), constraints, gauge fixing condition and
 remaining currents  into those of model (\ref{8.4}): $ J^{\pr} =
 \omega_{\a_1} (J)$.  The change of variables (\ref{8.5}) is a consequence of
 the relation between the reduced $A_2$-WZW matrix fields $g_{(3)}^{(2)}$
  and $g_{(3)}^{(1,1)}$:
  \be
  g_{(3)}^{(2)} = \omega_{\a_1}(g_{(3)}^{(1,1)} )
  \label{8.7}
  \ee
   The explicit form of $g_{(3)}^{(1,1)}$ in terms of fields $\psi, \chi, $
 and $\varphi $ is given by eqn. (\ref{3.46}).  Solving the constraints
 (\ref{8.1}) and (\ref{8.2}) we find the matrix elements of $g_{(3)}^{(2)}$:
\br 
\( g^{(2)} \) _{11} &=& e^{\varphi - {1\o 2}R_0}, \quad 
\( g^{(2)} \) _{13} = \pa e^{\varphi - {1\o 2}R_0}, \quad 
 \( g^{(2)} \) _{31} = \bar \pa e^{\varphi - {1\o 2}R_0}
 \nonumber \\
 \( g^{(2)} \) _{22} &=& e^{R_0} ( 1+ e^{\varphi }\psi_0 \chi_0 ), \quad 
   \( g^{(2)} \) _{12} = e^{{1\o 4}R_0 + \varphi }\psi_0, \quad 
 \( g^{(2)} \) _{21} = e^{{1\o 4}R_0 + \varphi }\chi_0, 
 \nonumber \\
 \( g^{(2)} \) _{23} &=& \pa  \( g^{(2)} \) _{21}, \quad 
   \( g^{(2)} \) _{32} = \bar \pa  \( g^{(2)} \) _{12}, \cdots etc
\nonumber 
\er
 
 We next write eqns. (\ref{8.7}) in a matrix form
 \be
 \( g^{(2)} \) _{ik} =  \( \omega_{\a_1}\) _{il} \( g^{(1,1)} \) _{lm}
 \( \omega_{\a_1}\) _{mk}, \quad i, k = 1,2,3
 \label{8.8}
 \ee
 where $ \( \omega_{\a_1}\)= 
\left(
\begin{array}{lll}
0 &1 & 0 \nonumber \\
1 & 0 & 0 \nonumber \\
0 & 0 & 1 \nonumber \\
 \end{array} \right)$.

As a solution of (\ref{8.8}), (i.e. $\psi, \chi, \phi $ in terms of $\psi_0,
\chi_0, \varphi $) we find    the change of variables (\ref{8.5}) as well as
the relation between the nonlocal fields $R$ and $R_0$:
\br
R = R_0 + \ln (1+e^{\varphi } \psi_0 \chi_0 ).
\nonumber
\er

The generalization of such results for a generic $A_n^{(1,1)}$-model
(\ref{2.17}) is straightforward.  Reflecting the  $A_n^{(1,1)}$ constraints
(\ref{2.4}) by $\omega_{\a_1} $ we find the set of constraints that define the
new model $\omega_{\a_1} $ ($A_n^{(1,1)}$):
\br
&&\hskip -1.5cm J_{-\a_i} =  \bar J_{\a_i} = 0, \quad i=3,4,\cdots ,n;\quad \quad
J_{-[\a ]} =  \bar J_{[\a ]} = 0, \quad [\a ]\,\, {\rm all\,\, other\,\,  roots} \nonumber \\
&&J_{-\a_1-\a_2} = \bar J_{\a_1 + \a_2} =1,
\quad \quad
J_{(\lambda_1 - \lambda_2 )\cdot H } = 
\bar J_{(\lambda_1 - \lambda_2 )\cdot H } =0
\label{8.9}
\er
 It is important to note the reflection $\omega_{\a_1} $ keeps unchanged
 the gradation operator, $\hskip1.5cm$ $ Q_{(1)} = \sum_{i=2}^{n} \lambda_i \cdot H $ and
 the nilpotent subgroups $N_{(1)}^{\pm} \subset H_{(1)}^{\pm} $, 
 $\omega_{\a_1}(Q_{(1)}) = Q_{(1)}, \quad 
\omega_{\a_1}(N_{(1)}^{\pm} ) = N_{(1)}^{\pm}$
but it does change ${H_{0}^{(1)}}^0 $ and $\eps^{(1)}_{\pm}$
\br
\omega_{\a_1}({H^{(1)}_{0}}^0 ) =exp \( {1\o {2{\cal K}_{11}}}[(\lambda_2 - \lambda_1
)\cdot H]\varphi_1 \)  , \quad \omega_{\a_1}(\eps_{(1)}^{\pm} ) =E_{\pm (\a_1+\a_2)} +
\sum_{i=3}^{n} E_{\pm \a_i} 
\nonumber 
\er
With all this data we derive the Lagrangean for the $\omega_{\a_1}
(A_n^{(1,1)})$ model:
\be
{\cal L}_{n, \omega_{\a_1}}^{(1,1)} = -{k\o 2\pi }\{ {1\o 2} \tilde k_{ij}\pa
\varphi_i  \bar \pa \varphi_j + e^{\varphi_1 } {{\bar \pa \psi_0 \pa \chi_0 }\o
{1+ {{n+1}\o 2n}\psi_0\chi_0 e^{\varphi_1 }}} - \sum_{i=2}^{n-1}
 e^{-\tilde k_{ij}\varphi_j }  - (1+ e^{\varphi }\psi_0 \chi_0 ) e^{-\tilde
 k_{1i} \varphi_i } \}
 \label{8.10}
 \ee
 The change of variables that follows  from the $A_n$-analog of eqn.
 (\ref{8.8}) is now given by
 \br
 &&\hskip1.5cm \phi_i = \varphi_i - \tilde {\cal K}_{1i} \ln (1+ e^{\varphi_1} \psi_0 \chi_0 ),
 \quad \tilde {\cal K}_{1i} = {{n-i}\o n}
 \nonumber \\
 &&\psi = \chi_0 e^{\varphi_1}(1+ e^{\varphi_1} \psi_0 \chi_0 )^{-{1\o 2}
  {\tilde {\cal K}_{11}}}, \quad
 \chi = \psi_0 e^{\varphi_1}(1+ e^{\varphi_1} \psi_0 \chi_0 )^{-{1\o 2} 
 {\tilde {\cal K}_{11}}}
 \label{8.11}
 \er
Substituting eqn. (\ref{8.11}) in (\ref{2.17}) we
find (\ref{8.10}) modulo  total derivative.  The
following identity 
\br
&& {{e^{\varphi } \o {1+ e^{\varphi }\psi \chi }}}
\( \bar \pa \chi \pa \psi - \bar \pa \psi \pa
\chi + {1\o 2}(\chi \bar \pa \psi - \psi \bar
\pa \chi )\pa \varphi - {1\o 2} (\chi \pa \psi -
\psi \pa \chi )\bar \pa \varphi \) =\nonumber \\
&& = {1\o 2} \pa \( \ln
 ( {1+ e^{\varphi }\psi \chi })\bar \pa \ln {\psi
\o \chi } \) - 
{1\o 2} \bar \pa \( \ln
 ( {1+ e^{\varphi }\psi \chi }) \pa \ln {\psi
\o \chi }\) \nonumber 
\er
is crucial in the proof of the above statement. 
The {\it conclusion }is that the pair of NA-Toda models
$A_n^{(1,1)}$ and $\omega_{\a_1}(A_n^{(1,1)}) $
{\it sharing the same algebra of symmetries} $
V_{n+1}^{(1,1)}$ are classically {\it equivalent}.  The
same is true for the pair of generic $A_n^{(j_1,1)}$ and 
$\omega_{\a_j}(A_n^{(j_1,1)})$-models.   The simplest
example of such doublet is given by the
$A_3^{(2,1)}$ and the
$\omega_{\a_1}(A_3^{(2,1)})$-NA-Toda models. 
Their lagrangeans are given by:
\be
{\cal L}^{(2,1)}_{3} = -{k\o 2\pi } (\pa A\bar
\pa A + \pa B\bar \pa B - e^{2A} - e^{2B}
+ e^{A+B} {{\bar \pa \psi \pa \chi }\o {1+
e^{A+B} \psi \chi }})
\label{8.12}
\ee
and 
\br
{\cal L}^{(2,1)}_{{3, \omega_{\a_1 }}} &=&
 -{k\o 2\pi}\{ \pa {\cal A}\bar
\pa {\cal A} + \pa {\cal B}\bar \pa {\cal B} -
e^{2{\cal A}} - e^{2{\cal B}} - ({1+ e^{-{\cal
A}-{\cal B}} \psi_0 \chi_0 })(e^{2{\cal A}} +
e^{2{\cal B}})
\nonumber \\
& + & e^{-{\cal A}-{\cal B}} {{\bar \pa \psi_0 \pa
\chi_0 }\o {1+ e^{-{\cal A}-{\cal B}} \psi_0
\chi_0 }}\}
\label{8.13}
\er
The change of variables that makes their
equivalence evident has the form:
\br
A &=& {\cal A} + {1\o 2} \ln ({1+ e^{-{\cal
A}-{\cal B}} \psi_0 \chi_0 }), \quad B= {\cal B}
+ {1\o 2} \ln ({1+ e^{-{\cal
A}-{\cal B}} \psi_0 \chi_0 })
\nonumber \\
\psi &=& \chi_0 e^{-{\cal A}-{\cal B}}
({1+ e^{-{\cal A}-{\cal B}} \psi_0 \chi_0
})^{-{1\o 2}}, \quad 
\chi = \psi_0 e^{-{\cal A}-{\cal B}}
({1+ e^{-{\cal A}-{\cal B}} \psi_0 \chi_0
})^{-{1\o 2}}\nonumber 
\er
Our observation that {\it each pair of $A_n^{(1,1)}$-
models, whose constraints are related by
$\omega_{\a_j}$  are equivalent}, adresses the
{\it question} about the {\it family of models} obtained from
$A_n^{(j,1)}$ by transforming its constraints
(\ref{7.1}) (or (\ref{2.4}), for $j=1$) under the
{\it whole Weyl group} $S_{n+1}$ of $A_n$.

We first consider a subset of models $\Pi_i
(A_n^{(1,1)})$ from the $A_n$ family, whose
constraints are the images of (\ref{2.4}) under
the action of the following (composite )
reflections,
\br
\Pi_i = \om_{\a_i} \om_{\a_{i-1}} \cdots 
\om_{\a_1}, \quad i= 1, \cdots , n
\nonumber 
\er
In order to derive their lagrangeans we need the
explicit form of $\Pi_i (H_{\pm }) = \Pi_i
(N_{\pm })\otimes \Pi_i ({H_{0}^{{(1)}}}^0)$, $\Pi_i
(g_0^f) $ and $\Pi_i(\eps ^{(1)}_{\pm})$ ( see Sect.
 2).  It is convenient to first calculate the
corresponding grading operator $\Pi_i (Q_{(1)}) = Q_{(1)} - \sum_{j=2}^{i} (j-1)
\a_j $
 where $Q_{(1)} = \sum_{l=2}^{n} \lambda_l \cdot H
$.  Then the nilpotent subgroups $\Pi_i
(N_{\pm}) $ are spanned by elements of grade $\pm
1$ with respect  to $\Pi_i (Q_{(1)} ) $ and $ \Pi_i
(g_0 ) $ is the subgroup of zero 
$\Pi_i (Q_{(1)} ) $-grade.  The diagonal subgroup
$\Pi_i ({H_{0}^{(1)}}^0) $ has the form 
\br
\Pi_i ({H_{0}^{(1)}}^0) = exp \{ {1\o 2{\cal K}_{11}}
(\lambda_1 - \sum_{l=1}^{i} \a_l )\cdot H \}
\nonumber 
\er
Following the recipe of Sect. 2 we find that the
lagrangeans  of $\Pi_i (A_n^{(1,1)})$ models
coincide with (\ref{8.10}) for $i<n$  and with $A_n^{(n,1)}$ (see eqn
(\ref{7.3}) for $j=n$).  Note that
due to the $Z_2$ symmetry of the $A_n$ Dynkin
diagram 
\be
\Pi^{+} (\a_j ) = \a_{n+1-j}, \quad {\cal K}_{jj} =
{\cal K}_{n+1-j, n+1-j}
\label{8.14}
\ee
the $A_n^{(j,1)}$ and the  $A_n^{(n+1-j,1)}$
models are {\it identical}.  In particular ${\cal
L}_{n}^{(1,1)} = {\cal L}_{n}^{(n,1)}$, ($\phi_i
^{\pr}= \phi_{n-i} $) and therefore all the
$\Pi_i (A_n^{(1,1)}) $-models are equivalent to
the $A_n^{(1,1)}$.

Another subfamily of the $A_n^{(1,1)}$ models is
defined by transforming the constraints
(\ref{2.4}) under the composite Weyl reflections
\be
\Pi_i^{-} = \Pi_{n+1-i} \Pi^{-}, \quad
\Pi^{-}(\a_j ) = - \a_{n+1-j}
\label{8.15}
\ee
($\Pi^{-}$ is an element of the Weyl group,
 contrary  to the $\Pi^{+}$ from (\ref{8.14})). 
Repeating once more the procedure, we have used
in the construction of  $\Pi_i
(A_n^{(1,1)})$ models, we realize that
all the $\Pi_i^{-} (A_n^{(1,1)})$ models are
equivalent to the $A_n^{(n,1)}$ model (modulo
$PC$-transformations: $C \psi = \chi,  \; C^2 =
1; \; P \sigma = - \sigma $).  Taking into account (\ref{8.14}) {\it the
conclusion} is that the {\it family} of $A_n^{(1,1)}$
models obtained by $\Pi_i$ and $\Pi_i^{-}$- Weyl
reflections ($i=1, \cdots n$) of $A_n^{(1,1)} $
constraints (\ref{2.4}) have as lagrangeans
(\ref{8.10}) or (\ref{2.17}).  They  are
related by the change of variables (\ref{8.11}). 
Hence all this family of models can be
represented by the original $A_n^{(1,1)} $ model
only.

Note that this (  $\Pi_i$ , $\Pi_i^{-}$ ) family
of equivalent models coincides with the complete
Weyl group  $A_n^{(1,1)} $ family only for
$n=2$.  Whether the other models generated by
Weyl reflections different from $\Pi_i$ and
$\Pi_i^{-}$ are also equivalent to $A_n^{(1,1)} $
is still an open problem for generic $n \neq 2 $.

\sect{What does the  Quantum $V$-algebras look
like}

The main obstacle for quantizing the NA-Toda models (\ref{2.17}) 
( and their
algebra of symmetries $V^{(1,1)}_{n+1}$ ) is the additional PF-type 
constraint
$J_{\lambda_1H}=0$.  As we have shown in Sect. 3 the latter is
 responsible for
the nonlocal terms in the $V^{(1,1)}_{n+1}$-algebra.  This suggests the
following {\it strategy}:  First consider an {\it intermediate}
 ``local''  NA-Toda model
defined by the set of constraints (\ref{2.4}) but with   the {\it current  
$ J_{\lambda_1H}$  left unconstrained}.  Its Lagrangean as well as the 
PB
symmetry algebra $W_{n+1}^{(1,1)} $ can be easely derived by the methods of
Sect.2 and 3.  The $W_{n+1}^{(1,1)} $ is generated by $n+2$ local currents:
$n+1$ of them $G^{(\pm )}$, $W_{n-k+2}, k=2,3,\cdots ,n$ have the same  
spins as
the  $V^{(1,1)}_{n+1}$-currents and the last one is the chiral 
$U(1)$-current 
$J_{\lambda_1H}\equiv J$ of spin one.  An {\it important difference}
 with respect to
the original NA-Toda model and its nonlocal $V^{(1,1)}_{n+1}$-algebra is  
that
 $W_{n+1}^{(1,1)} $ is a {\it local} quadratic (non-Lie) algebra.  It has a structure
 similar to the $W_{n+1}$ and $W_{n+1}^{(l)}$-algebras \cite{Bershadsky}  
  and its
 quantization can be realized by the standard methods of
  ref. \cite{F-L},
 \cite{Bershadsky}, \cite{Ber-Oo}.
 
The {\it problem} we address in this section is the following: {\it 
 Given the quantum
$W_{n+1}^{(1,1)}$ algebra and its irreducible representations, to derive  
the
quantum $V_{n+1}^{(1,1)} = W_{n+1}^{(1,1)}/ U(1)$- algebra and its
representations by implementing the constraint $ J_{\lambda_1 H}=0$}.  The
method we are going to use is an appropriate generalization of the 
derivation
of the PF-algebra from the affine $SU(2)$ by imposing the constraint 
$J_3 =0$
\cite{ZF1}.  The crucial ingredient of this approach is the
free field representation of the $W_{n+1}^{(1,1)}$-currents.  In the
 framework
of the quantum Hamiltonian reduction \cite{Ber-Oo}, \cite{Bershadsky} 
 this is
rather difficult problem even for $n=3$.  To start with it requires the
explicit bosonization of the $SL(4,R)$ currents.  One expects the 
nonabelian
analog  of the quantum Miura transformation \cite{F-L}
 to be
the effective tool for the solution of this problem.  The $n=2$ case is an
exception.  According to the arguments of Sect. 8, the 
$W_3^{(1,1)}$ algebra
coincides with the $W_3^{(2)}$-one.  The free field representation of the
$W_3^{(2)}$  currents $G^{\pm }, T$ and $J$ is well known
\cite{Bershadsky}:
\br
J(z)& =& \sqrt {{{2k+3}\o 3}} \pa \tilde \Phi, \;\; 
\sqrt {{{2k+3}\o 3}}  \tilde \Phi = {\a_{+} \o 3}(\vec{\a_2} - \vec{\a_1})
\vec{\varphi } + \phi_0, \;\; \a_{+} = \sqrt{2k+6} \nonumber \\
G^{-}(z) &=& [(k+3) \pa \phi_0 + \a_{+} \vec{\a_2}\pa \vec{\varphi} - (k+2)\pa
\chi ]e^{\phi_0 - \chi} \nonumber \\
G^{+}(z) &=&[-\pa \chi((k+3)\pa \phi_0 - \a_{+}\vec{\a_1 }\pa \vec{\varphi })
+ (k+2)(\pa^2 \chi + (\pa \chi )^2]e^{-\phi_0 + \chi } \nonumber \\
T_{W}(z) &=& {2 \o 3}[(\vec{\a_1}\pa \vec{\varphi })^2 + 
(\vec{\a_2}\pa \vec{\varphi })^2 +
(\vec{\a_1}\pa \vec{\varphi })(\vec{\a_2}\pa \vec{\varphi })] \nonumber \\
&+& {{k+1}\o {\a_+}} (\vec{\a_1} + \vec{\a_2}) \pa ^2 \vec{\varphi} + {1\o
2}[(\pa \chi )^2 + \pa ^2 \chi - (\pa \phi_0)^2],
\label{9.1}
\er
where $\varphi_i, i=1,2, \;\; \phi_0 $ and $\chi $ are free bosonic fields, 
\be
<\varphi_i(z_1) \varphi_j(z_2)> = {1\o 2}\d_{ij}ln z_{12},\;\; 
<\phi_0(z_1) \phi_0(z_2)> = -ln z_{12},\;\;  <\chi (z_1) \chi (z_2)> = ln z_{12}
\label{9.2}
\ee
and $\vec{\a_1 }, \vec{\a_2 }$ are the simple roots of $A_2$.  The difference
of our eqn. (\ref{9.1}) from the original Bershadsky's ones (see eqns.
(\ref{3.7}) of ref. \cite{Bershadsky}) is due to the fact that we have bosonized
the pair of {\it bosonic}  ghosts ( $\Phi,  \Phi^{ \dagger} $ ) of spin ($1/2, 1/2 $):
\br
\Phi = e^{\phi_0 - \chi }, \;\;\;\;  \Phi^{ \dagger} = e^{-\phi_0 + \chi }\pa \chi
\nonumber 
\er
With eqns. (\ref{9.1}) at hand one can easely derive the explicit form for the
$W_3^{(1,1)}$-algebra, 
\br
J(z_1)J(z_2) &=& {{2k+3}\o {3z_{12}^2}}+ O(z_{12}), \;\;\;  
 G^{\pm }(z_1)G^{\pm }(z_2) =  O(z_{12}) \nonumber \\
 J(z_1)G^{\pm }(z_2) &= &\pm {1 \o {z_{12}}}G^{\pm }(z_2) 
  + O(z_{12}),\nonumber \\
G^{+}(z_1)G^{-}(z_2) &=&  {{(k+1)(2k+3)}\o { z_{12}^3 }} + 3{(k+1) \o
{z_{12}^2}}J(z_2) \nonumber\\
&+&{1 \o {z_{12}}}[3J^2(z_2) - (k+3)T(z_2) +
 3{(k+1) \o {2}} \pa J(z_2) ] +  O(z_{12})\nonumber \\ 
 \label{9.2a}
\er 
etc.    The central charge of the $W_3^{(1,1)}$ algebra is 
\be
c_W = {8k \o {k+3}} -6k -1
\label{9.3} 
\ee
According to the definition of the $V_3^{(1,1)}$ algebra 
$V_3^{(1,1)} = \{ W_3^{(1,1)} : J =0 \}$
its  generators $V^{\pm}(z), T_V(z)$ have to commute with $J(z)$,
\br
J(z_1) V^{\pm}(z_2) =O(z_{12}) = J(z_1) T_V(z_2)
 \nonumber\\
\er
i.e. they are related to the $\tilde \Phi $ independent parts ($J =
\sqrt{{{2k+3}\o 3}} \pa \tilde \Phi $) of the $W_3^{(1,1)}$-currents.  This
suggests to seek for a specific change of the field variables $\phi_0,
\varphi_i \longrightarrow \tilde \Phi , \rho_i $ that makes explicit the $J$ (
or $\tilde \Phi $) dependence of $G^{\pm}$ and $T_W$:
\be
G^{\pm} = V^{\pm}e^{\pm a\tilde \Phi }, \;\;\; T_W = T_V + T_{\tilde \Phi }
\label{9.4}
\ee
An important condition required for the new fields $\rho_i$ to  satisfy is the
orthogonality to $\tilde \Phi $:
\be
\tilde \Phi (z_1) \rho_i(z_2) = O(z_{12}) 
\label{9.5}
\ee
One solution for such requirement is given by 
\br
\tilde\rho_1 = -(k+3) \phi_0 + \a_+ \vec{\a_1 }\vec{\varphi }, \quad 
\tilde\rho_2 = (k+3) \phi_0 + \a_+ \vec{\a_2 }\vec{\varphi }
\nonumber
\er
 If we further impose orthonormality among the $\rho_i$'s, 
 \be
 \rho_i(z_1) \rho_j(z_2) = -\d_{ij}ln z_{12} +O(z_{12}) ,\quad\quad i,j=1,2 
 \label{9.6}
 \ee
 the unique change of variables satisfying both (\ref{9.5}) and (\ref{9.6}) has
 the form
 \br
 \phi_0 &=& -{1 \o {2k+3}}\sqrt{{{k+3}\o {k-3}}}[(1+\b_1)g_1\rho_1 -
 (1+\b_2)g_2\rho_2] - \sqrt{ {3 \o {2k+3}}} \tilde \Phi \nonumber \\
 \vec{\a_1} \vec{\varphi } &=& 
 -{1\o {(2k+3)\sqrt{ 2k-6}}}[ (k+3-k\b_1)g_1\rho_1
 -(k+3-k\b_2)g_2\rho_2] - \sqrt{{{3(k+3}\o {2(2k+3)}}}\tilde \Phi \nonumber \\
 \vec{\a_2} \vec{\varphi } &=& 
 -{1\o {(2k+3)\sqrt{ 2k-6}}}[ (k-(k+3)\b_1)g_1\rho_1
 -(k-(k+3)\b_2)g_2\rho_2] + \sqrt{{{3(k+3}\o {2(2k+3)}}}\tilde \Phi 
 \nonumber \\
 \label{9.7}
 \er
 where 
 \br
 \b_j = {1\o 2}(k+1-(-1)^j \sqrt{(k+1)(k-3)}), \;\; g_j^2 = {1\o
 2}[(k-3)(k+1) + (-1)^j (k^2 -3)\sqrt{{{k-3}\o {k+1}}}],
 \nonumber
 \er
 $ j=1,2$.  Substituting (\ref{9.7}) in (\ref{9.1}) we realize that the reduced
 $W_3^{(1,1)}$ currents can be written in the form (\ref{9.4}) with $a=
 \sqrt{{3\o {2k+3}}}$ and $T_{\tilde \Phi} = {1\o 2}(\pa \tilde \Phi )^2$. 
 This provides us with the free field representation of the $V_3^{(1,1)}$
 currents we were looking for, 
 \br
 &&V^-(z) = [(k+2) \pa \eta - \eta \sqrt{{{k+3}\o {k-3}}}(g_1\pa \rho_1 -
 g_2\pa \rho_2)]e^{-a_1\rho_1 + a_2\rho_2}, \nonumber\\
 &&V^+(z) = [(k+2) \pa ^{2} \xi + \pa \xi \sqrt{{{k+3}\o {k-3}}}
 (\b_1g_1\pa \rho_1 -
 \b_2g_2\pa \rho_2)]e^{a_1\rho_1 - a_2\rho_2}, \nonumber\\
 &&T_V = -{1\o 2}[(\pa \rho_1)^2 + 
 (\pa \rho_2)^2] + \sum_{i=1}^{2}\gamma_i \pa ^2 \rho_i +
 T_{\xi \eta },\quad  T_{\xi \eta } = 
 \eta \pa \xi = {1\o 2} (\pa \chi )^2 + {1\o 2} \pa
 ^2 \chi \nonumber\\
 \label{9.8}
 \er
 where  
 \br
  \gamma_i = {{(-1)^{i+1}\o {4}}}{(k+1)\o {\sqrt{k^2-g}}} (k-1
 +(-1)^{i+1}\sqrt{(k+1)(k-3)})g_i;\quad a_i = \sqrt{{{k+3}\o
 {k-3}}}{{(1+\b_i)g_i}\o {2k+3}}
 \nonumber
 \er
  and we have denoted
 $ \xi = e^{\chi}, \quad  \eta = e^{-\chi } $. One can easely 
 recognize $(\xi, \eta )$ as a pair of fermionic ghosts of spins
 $(0,1)$.  
 
 We are now prepared to derive the explicit form of the $V_3^{(1,1)}$ OPE
 algebra ($k\neq -3, -{3\o 2}, -1$):
 \br
 V^{\mp}(z_1)V^{\mp}(z_2)&=& (z_{12})^{-{3\o {2k+3}}}V^{\mp}_{2}(z_2) +
  O(z_{12}) \nonumber \\
  V^{+}(z_1)V^{-}(z_2)&=&(z_{12})^{{3\o {2k+3}}}[{{(2k+3)(k+1)}\o {z_{12}^3}}-
  {{k+3}\o {z_{12}}}T_V(z_2) +{{k+3}\o 2}(2W_3 -\pa T_V)] +  O(z_{12})
  \nonumber \\
 T_V(z_1)V^{\mp}(z_2)&=& {\Delta^{\mp}_1 \o {z_{12}^2}}V^{\mp}(z_2) + {1\o
 {z_{12}}}\pa _{z_2}V^{\mp}(z_2) + O(z_{12})
 \label{9.9}
 \er
 where $\Delta^{\mp}_{1} = {3\o 2}(1- {1\o {2k+3}})$ are the {\it 
 renormalized spins}
 (dimensions ) of the {\it quantum currents} $V^{\pm}_1 $ ($\equiv V^{\pm}$) and 
 $V^{\pm}_2 (z)$ are new currents of spins 
$\Delta^{\mp}_{2} = {3\o 2}(1- {2\o {2k+3}})$.  For example $V^{-}_2 (z)$ has
the form
\br
V^{-}_2 (z) &=& \{[(k+2)\pa \chi +\sqrt{{k+3}\o{k-3}} (g_1\pa \rho_1 -g_2\pa
\rho_2 )]^2 -\pa [(k+2)\pa \chi \nonumber\\
&+& \sqrt{{k+3}\o{k-3}}(g_1\pa \rho_1 -g_2\pa
\rho_2 )]\} e^{-2\chi -2a_1 \rho_1 +2a_2\rho_2 }
\nonumber 
\er
The stress tensor $T_V(z)$ of spin $2$ satisfies the standard Virasoro algebra
OPE's with specific central charge
\br
c_V = c_W -1 = -6{{(k+1)^2}\o {k+3}}
\nonumber 
\er
The OPE's of $V_2^+ ,V_2^-$ and $V_1^{\pm}$  ($\equiv V^{\pm}$) introduce more new
currents $V_l^{\pm} $ of spins ($L=2k+3$)
\br
\Delta_l = {3\o 2}l(1-{l\o L}), \quad l=1,2, \cdots 
\nonumber 
\er
and with $U(1)$ charges $Q_0 =l$ ( see Sect. 4 for the definition of $Q_0$):
\be
V_l^{\pm}(z_1) V_{l^{\pr}}^{\pm}(z_2) = C_{l, l^{\pr}} z_{12}^{-{{3ll^{\pr}}\o
L}}V^{\pm}_{l+l^{\pr}}(z_2) + O(z_{12}) 
\label{9.10} 
\ee
The OPE's $V^+_l V^-_{l} $ give rise to new $Q_0$-neutral  currents $W_{l+1}$, 
($l=2, \cdots ,L-4$) of spins $\Delta_l = l+1$
\be
V^+_l(z_1) V^-_{l}(z_2) = z_{12}^{{3l^2}\o L} ({1\o {z_{12}^{3l}}} +
{{2\Delta_l}\o {c_V z_{12}^{3l-2}}}T_V(z_2) + \cdots + {d_l \o
{z_{12}}}W_{3l-1} + O(1) )
\label{9.11}
\ee
For example the $W_3$ current that appears in the finite part of eqn.
(\ref{9.9}) and in the singular  part of (\ref{9.11}) (for $l\geq 2 $) has the
form 
\be
W_3 = B_{\a} \pa ^3 \rho_{\a} + B_{\a \b} \pa ^2\rho_{\a} \pa \rho_{\b} + C_{\a
\b} (\pa \rho_{\a})^2 \pa \rho_{\b},\quad\quad\a, \b = 1,2,3
\label{9.12}
\ee
 where we have denoted $\rho_3 = \chi $ and
\br
B_{33} &=& 6B_3 = {3\o 2}C_{33} = -{{2k+5}\o 2}, \;\; B_{i3} = C_{3i} = 
(-1)^{i} (k+2)a_i, \; i=1,2 \nonumber \\
B_{11} &=& -B_{22} = {1\o 2}\sqrt {{{k+1}\o {k-3}}}, \;\; B_i = (-1)^{i} {1\o
6} (\gamma_i + {{k+1}\o {\sqrt {k^2-9}}}g_i)\nonumber \\
C_{ij}&=& (-1)^{j+1}{{3^{|i-j|}\o 6}}a_j[k+4-2|i-j| + (-1)^i \sqrt {{k+1}\o
{k-3}}] \nonumber \\
B_{ij}&=& {1\o 2}{{g_i g_j} \o {(k-3)(2k+3)}} [k(k+1) +(-1)^i (k+2) \sqrt
{(k+1)(k-3)} ], \; i\neq j
\nonumber 
\er
For generic $k $ ($= {{L-3}\o 2}$) this $W_3$-current and $T_V$ ($=W_2$) do not
close the standard $W_3$ algebra.  In the OPE of $W_3(z_1)W_3(z_2)$ the new
$W_4$-current contributes together with the $\Lambda = :T^2: - {3\o {10}}\pa ^2
T$, etc.  It is not difficult to verify that all the $W_{l+1}$ ($l=1, \cdots $)
form the standard $W_{\infty }$-algebra which appears as a subalgebra of the
{\it parafermionic extension
 of the $W_{\infty}$-algebra  $W_{\infty}^{(pf)}$ of
central charge $c_V$ spanned by $V^{\pm}_l$ and $W_{l+1}$,
 $l=1, 2, \cdots $}. As we
shall show latter this is the case when $L$ is positive, rational and
{\it noninteger}.  The natural question to ask is {\it whether exist
 values of $L$ for
which the algebra of the currents $V_l^{\pm}$, $W_{l+1}$ closes 
for finite $l$}. 
The most interesting case would be if this happens for $l=1$, i.e. when
$V_1^{\pm}$ and $T_V$ alone form a closed algebra.  To answer this
 question one
has to analyze the $L$ dependence of the singularities in the OPE's 
(\ref{9.9}),
(\ref{9.10}) and (\ref{9.11}) ( the order of the poles  and cuts ) and to
calculate the structure constants $C_{l, l^{\pr}}$ and $d_l$. 
 The exponents
$\pm {3\o L} $ in (\ref{9.9}) that gives the $L$-dependent 
singularities of the
OPE's suggest to consider separately the following intervals 
of values of $L$:

{\it Case $(1)$}:  $L<3$.  No singularities in the $V^{\pm}V^{\pm}$ OPE's.  Only poles of
order $\leq 3$ and ($-{3\o {|L|}}$)-cut in the $V^+V^- $ OPE.   Hence $V^{\pm}$
($\Delta ^{\pm} = {3\o 2}(1+{1\o {|L|}})$) and $T_V$ alone close an algebra.

{\it Case $(2)$}:  $-3<L<0$.   Poles of order higher then 3 and ($-{3\o {|L|}}$)-cut
in the $V^+V^-$ OPE.  No singular terms in  $V^{\pm}V^{\pm}$.
For each $|L|$ in the interval $s-1\leq {3\o {|L|}} <s$  ( $s=2,3, \cdots $ ) the
$V^+V^-$ contains poles of order $\leq 2+s$ and involve $s-2$ new currents
$W_p$ $(p=3,4, \cdots s+1)$.  Whether $V^{\pm}$, $T$ and $W_p$ $ (p=3, \cdots
s)$  close an algebra is an open question.  In the simplest case $s=2$ the
straightforward calculation based on the bosonized form (\ref{9.8}) and
(\ref{9.12}) of $V^{\pm}$, $T$ and $W_3$ show us that the $W_3(1)W_3(2)$ OPE
introduce new $W_4$ current.  This only proves that $V^{\pm}$, $T$ and $W_3$
 do not close an algebra in this case.  It might happen that for some specific value of $L$
 (and $s=2$) the algebra spanned by finite number of currents 
$V^{\pm}$, $T_V$ and $W_{l+1}, l=2,\cdots ,M)$  closes.  However it is more
natural to expect that $T_V$ and $W_{l+1}, l=2,\cdots ,M)$ form
$V^{\pm}$-{\it extension of the $W_{\infty}$-algebra}.

{\it Case $(3)$}:  $0<L\leq 3$.   Pole of order $[{3\o L}]$ and a cut in the
$V^{\pm}V^{\pm}$ OPE.  Poles of order $<3$ and $({3\o L})$-cut in $V^+V^-$. 
Therefore no $W_{l+1}$ currents can appear.

 $(3a)$ For $L$ {\it rational
noninteger} one has to consider all
 the $V^{\pm}_l$'s $(l=1,2, \cdots ,)$ and $T_V$
in order to close an algebra.  The latter 
is an example of {\it purely parafermionic}
$W_{\infty}$ algebra. 
 Note that $\Delta^{\pm}_l <0$ for  $l>L$, i.e. for $L<1$
all the $V^{\pm}_L$ have negative dimensions; for $1<L<2$-all,
 but $V_1^{\pm}$ 
and for $2<L<3$-all but $V^{\pm}_1$ and $V^{\pm}_2$. 

$(3b)$  The case $L$ integer $\leq 3$ provides two simple examples 
of quantum
$V^{(1,1)}_3(L)$-algebras generated by finite number of currents: for
 $L=2$ these
are $V^{\pm}$ of $\Delta^{\pm}= {3\o 4}$ and $T_V$ ($c_V= -{3\o 5}$)
 and for
$L=3$, $T_V$ and  $V^{\pm}_1 $ ,$V^{\pm}_2 $ 
 of $ \Delta^{\pm}_1=\Delta^{\pm}_2 = 1$ and
$c_V(3) =-2$.  We shall give the explicit form of these 
two algebras later in
this section.

{\it Case $(4)$}:$L>3$.   No poles in $V^{\pm}V^{\pm}$.  
 Poles of order $\leq 3$ and
$({3\o L})$-cut in $V^+V^-$.  Therefore the Laurent modes of $V^{\pm}$ and
$T_V$ have to close an algebra.  As in the standard PF case \cite{ZF1} the 
first
of the OPE's (\ref{9.9}) define the modes of $V^{\pm}_2$ as an infinite sum of
bilinears of the $V^{\pm}$ modes.  The OPE's $V^+_{[{L\o 2}]}V^-_{[{L\o 2}]}$
represent poles of order $L-2$ (higher than 3 for $L>5$) and thus introduce new
currents $W_{p+1}, p=2, \cdots , L-4$. 

 $(4a)$ For $L$ {\it integer} ($L>3$),
$\Delta^{\pm}_l = \Delta^{\pm}_{L-l}$ (i.e. $\Delta^{\pm}_L =0$) the  
$V^{\pm}_l V^{\pm}_{l^{\pr}}$ subalgebra involve $L-1$ currents only, i.e.
$l=1, \cdots ,L-1$.  Whether the algebra of $2(L-1) +L-4$ currents  $V^{\pm}_l,
W_{p+1}$,($l=1, \cdots ,L-1$, $p=1,2, \cdots L-4$)  
  closes for $L>5$ or one has to consider an infinite set of
$W_{p+1}$, ($p=1,2 \cdots $), $V^{\pm}_l, l=1,2, \cdots ,L-1$ 
is an open question.  For $L=4,5$ the $V_l^{\pm}$ 
and $T_V$ do form closed algebras. 

$(4b)$ For $L$ rational {\it noninteger}
 the OPE's $V^{\pm}_l V^{\pm}_{l^{\pr}}$
introduce infinite number of PF-currents of $U(1)$-charges $l=1,2, \cdots $.
Together with the neutral currents $W_{l+1}$ they span a kind of {\it PF}
$W_{\infty}$-algebra.  We have to note that the {\it difference}
 between the algebras
$(4a)$ and $(4b)$ can be {\it summarized} in the fact that for $L$-integer the
corresponding OPE's represent the multiplication rules of the discrete group
$Z_L\otimes Z_2$ (with the identification $V^+_l = V^-_{L-l}$), $l=1, \cdots
,L-1$ being the $Z_L$-charges.  In the noninteger $L$ case $(4b)$ the
symmetry encoded in the OPE's is $U(1)\otimes Z_2$.

 Our analysis of the singularities of the OPE's (\ref{9.10}) and (\ref{9.11})
  shows
 that the quantum $V_3^{(1,1)}(L)$-algebra shares many of the properties of the (both
 unitary and nonunitary) PF-algebras \cite{ZF1}. This fact allows
  us to apply the methods developed in ref. \cite{ZF1} 
  in the derivation of the explicit (Laurent
 modes) form of the $V_3^{(1,1)}(L)$.  We restrict ourselves to consider cases (1),
 $(3b)$ and $(4a)$ only.  The allowed boundary conditions \footnote {We are not
 considering here the twisted (or C-disorder) boundary conditions: $V^{\pm}(ze^{2\pi
 i})\phi_s(0) = e^{2\pi i b_s} V^{\mp}(z) \phi_{s \pm 2}(0)$} 
 (for $L$-integer) of the
 currents $V^{\pm}$:
 \br
 V^{\pm}(ze^{2\pi i})
  \phi_s^{\eta} (0) = e^{2\pi i (bs + \eta) } V^{\pm}(z) \phi_s^{\eta}(0)
 \nonumber 
 \er
 where $b= {3\o {2L}}$, $\eta =0,{1\o 2}$, $s=1, \cdots L-1$ lead to 
 the following mode
 expansion
\be
V^{\pm} (z)  \phi_s^{\eta} (0) = \sum_{m=-\infty}^{\infty} z^{\pm {{3s}\o
{2L}} +m-1 \mp \eta} V^{\pm}_{-m \pm \eta -{1\o 2}+{{3(1\mp s)}\o {2L}}}
\phi_s^{\eta} (0)
\label{9.13} 
\ee
The $ \phi_s^{\eta} (0)$ denote certain Ramond ($\eta = {1\o 2}$, $s$-odd)
 and the
Neveu-Schwarz ($\eta=0$, $s$-even) fields.  Following the arguments of
ref. \cite{ZF1} (Sect. 4) we derive the $V_3^{(1,1)}(L)$-algebra (for $|L|>3$)
from the OPE's (\ref{9.9}):
\br
& &{2\o {L+3}}\sum_{p=0}^{\infty }C_{(-{3\o L})}^p [V^{+}_{-{{3(s+1)}\o
{2L}}+m-p-\eta +{1\o 2}}V^{-}_{{{3(s+1)}\o {2L}}+n+p+\eta -{1\o 2}}
+V^{-}_{-{{3(1-s)}\o {2L}}+n-p+\eta -{1\o 2}}V^{+}_{{{3(1-s)}\o
{2L}}+m+p-\eta +{1\o 2}}] \nonumber\\
&=& -L_{m+n} + {1\o 2}{{(L-1)L}\o {(L+3)}}({{3s}\o {2L}}+n+\eta )
({{3s}\o {2L}}+n+\eta -1)\d_{m+n,0}
\label{9.14}
\er
where $C^p_{(M)} = {{\Gamma (p-M)}\o {p!\Gamma (-M)}}$, 
 $m,n =0, \pm 1, \pm 2,
\cdots $ and: 
 \be
\sum_{p=0}^{\infty }C_{({3\o L})}^p [V^{\pm}_{{{3(3\mp s)}\o
{2L}}-p+m+\eta -{1\o 2}}V^{\pm}_{{{3(1\mp s)}\o {2L}}+p+n+\eta -{1\o 2}}-
V^{\pm}_{{{3(3\mp s)}\o
{2L}}-p+n+\eta -{1\o 2}}V^{\pm}_{{{3(1\mp s)}\o {2L}}+p+m+\eta -{1\o 2}}]=0
\label{9.15}
\ee
In fact eqns. (\ref{9.14}) and (\ref{9.15}) together with the Virasoro algebra of the
$L_n$'s, ($L_n = \oint z^{n+1}T(z)dz$)
\br
[L_n,L_m] = (n-m)L_{m+n} + {c_V \o{12}}n(n^2-1)\d_{m+n,0}
\nonumber
\er
represent the entire $V_3^{(1,1)}$-algebra for $L<-3$ only.  In the case
$(4a.)$ (i.e. for $L>3$) they give 
the subalgebra spanned by $V^{\pm}_1$ and $T$ of the larger
$V^{(1,1)}_3$-algebra which also includes $V^{\pm}_l$, $l=2, 3, \cdots ,L-1$
and certain $W_p$'s. The algebra of the $V_l^{\pm}$'s charges ($l=1, \cdots
,L-1$):
\be
V^{\pm(l)}_{m-{l\o 2} +\eta_l + {{3l(l\mp s)}\o {2L}}} \phi_s^{\eta_l}(0) =
\oint dz z^{\mp {{3ls}\o {2L}}+m+l+\eta_l-1}V_l^{\pm}(z)\phi_s^{\eta_l}(0)
\label{9.16}
\ee
 can be easely derived from the OPE's (\ref{9.10}) and (\ref{9.11}).  Note
 that as a consequence of (\ref{9.10}) and (\ref{9.13}) the
 $V^{\pm}_{(2l)}$-currents have only Neveu-Schwarz boundary conditions, i.e.
 $\eta_{2l} = 0$, $\eta_{2l+1}=0,{1\o 2}$.  All algebraic relations that
 follows from (\ref{9.10}) are either in the form (\ref{9.15}) or give
 $V^{\pm (l)}_{-m}$ as an infinite sum of bilinears of 
 $V^{\pm (l_1)}_{-m_1}V^{\pm (l_2)}_{-m_2}$, $l=l_1+l_2$ as for example:
 \be
 V^{\pm (2)}_{m+n+{{3(2\mp s)}\o L}} = \sum_{p=0}^{\infty}C^{p}_{({{3-L}\o
 L})}[V^{\pm (1)}_{m-p+{{3(3\mp s)}\o {2L}}}
 V^{\pm (1)}_{n+p+{{3(1\mp s)}\o {2L}}} 
 +V^{\pm (1)}_{n-p-1+{{3(3\mp s)}\o {2L}}}
 V^{\pm (1)}_{m+p+1+{{3(1\mp s)}\o {2L}}}]
 \label{9.17}
 \ee
 
To give an idea how the entire $V^{(1,1)}_3(L)$-algebra looks like we
 consider in more detail the simplest case $L=4$. 
  The $V^{(1,1)}_3(4)$-algebra
 is generated by $V_1^{\pm}, V^{\pm}_2$ and $T_V$ of spins $\Delta^{\pm}_1
 ={9\o 8}, \Delta^{\pm}_2 = {3\o 2 }$ and $\Delta_T =2$.  We have to complete
 the algebraic relations (\ref{9.14}), (\ref{9.15}) of the $V^{\pm}_{(1)}, T$
 subalgebra with those involving $V^{\pm}_{(2)}$.  Taking $s=2$ sector for
 simplicity we find
 \be
 V^{+(2)}_m V^{-(2)}_n + V^{-(2)}_n V^{+(2)}_m = {1\o 2}(n^2 -{1\o
 4})\d_{m+n,0} - {7\o 9}L_{m+n}
 \label{9.18}
 \ee
 \be
 V^{\pm (2)}_{m+n} = \sum_{p=0}^{\infty}C_{(-{1\o 4})}^{p}[V^{\pm
 (1)}_{m-p+{3\o 8}-\eta_{\pm}}V^{\pm (1)}_{n+p-{3\o 8}+\eta_{\pm}}+
 V^{\pm
 (1)}_{n-p-{5\o 8}+\eta_{\pm}}V^{\pm (1)}_{m+p+{5\o 8}-\eta_{\pm}}]
 \label{9.19}
 \ee
 where $\eta_+ = 0$ and $\eta_- = {1\o 2}$.  The remaining relations involving
 $V^{\pm (2)}_{m_1}V^{\pm (2)}_{m_2}$, $V^{\pm (1)}_{m_1}V^{\pm (2)}_{m_2}$
 and $V^{\pm (2)}_{m_1}V^{\mp (1)}_{m_2}$ have a form similar to (\ref{9.15})
 and (\ref{9.17}).  Note that due to the identification $V^+_l = V^-_{L-l}$ we
 have 
 \br
 V^+_{(2)}(1)V^-_{(1)}(2 )=  {{c_{2,3}}\o {z_{12}^{{3\o 2}}}}V^+_{(1)}(z_2)
 + O(z_{12}), \quad 
 V^{\pm }_{(2)}(1)V^{\pm }_{(1)}(2 )=  {{c_{2,1}}\o {z_{12}^{{3\o 2}}}}
 V^{\mp }_{(1)}(z_2)
 + O(z_{12})
 \nonumber
 \er
 The eqns. (\ref{9.18}) and (\ref{9.19}) are an indication to consider the
 $V_3^{(1,1)}(4)$-algebra as `` a square root'' of the Virasoro superalgebra.
 
 Our last example is the $L=2$  $V^{(1,1)}_{3}$-algebra.  It is spanned by
 $V^{\pm}_1$ of $\Delta^{\pm} = {3\o 4}$ and $T_V$ and its central
  charge is
 $c_V= -{3\o 5}$.  This is one of the cases  $(3b)$ when the OPE 
$V^{\pm}_1 V^{\pm}_1 $ has a pole and therefore the `` commutation relations''
(\ref{9.15}) are {\it not valid}.  Instead we obtain:
\be
\sum_{p=0}^{\infty} C_{({1\o 2})}^{p}[V^-_{-p+m+(\eta -{1\o 2})- {1\o 4}}
V^-_{p+n+(\eta -{1\o 2})- {3\o 4}} + V^-_{-p+n+(\eta -{1\o 2})- {1\o 4}}
V^-_{p+m+(\eta -{1\o 2})- {3\o 4}}] = \d_{m+n+2\eta , 0}
\label{9.20}
\ee
and similar one for the $V^+V^+$'s .  The eqn. (\ref{9.14}) in this case take
the form ($s=2$):
\be
{2\o 5}\sum_{p=0}^{\infty} C_{(-{3\o 2})}^{p}[V^+_{-{3\o 4}+m-p-\eta }
V^-_{{3\o 4}+n+p+\eta } + V^-_{-{3\o 4}+n-p+\eta }V^+_{{3\o 4}+m+p-\eta
}]=-L_{m+n} + {1\o 5}(n+\eta +{1\o 2})(n+\eta + {3\o 2})\d_{m+n,0}
\label{9.21}
\ee

The complicated structure of the $V^{(1,1)}_3(L)$-``commutation relations''
makes the problem of the construction of irreducible highest weight 
representations rather difficult.  One could however further extend the
relation between $W_3^{(2)}$ and $V^{(1,1)}_3$-algebras on their
representations.  The way we have derived the explicit form of the
$V_3^{(1,1)}$-generators (\ref{9.8}) out of the $W_3^{(2)}$-ones (\ref{9.1}):
\be
G^{\pm}= V^{\pm}e^{\pm \sqrt {{3\o L}}\tilde \Phi },
 \quad T_W = T_V + {1\o
2}(\pa \tilde \Phi )^2, \quad J=\sqrt {{L\o 3}}\pa \tilde \Phi 
\label{9.22}
\ee
suggests the following form of the $W_3^{(2)}$ chiral vertex operators 
$\phi^{r_1 s_1}_{r_2 s_2} (z) \equiv \phi^{W}_{(r_i,s_i)}(z) $  in terms of
the $V_3^{(1,1)}$-ones, $\phi^{V}_{(r_i,s_i)}(z)$ and $\tilde \Phi (z)$:
\be
\phi^{W}_{(r_i,s_i)}(z) = 
\phi^{V}_{(r_i,s_i)}(z)e^{\b_{r,s}\tilde \Phi}
\label{9.23}
\ee
Taking into account the basic OPE's that define the $\phi^{W}$'s:
\br
T^{W}(z_1)\phi ^{W}_{(r,s)}(z_2) & = &
{{\Delta^{W}_{r,s}}\o {z_{12}^2}}\phi ^{W}_{(r,s)}(z_2)  +
 {1\o {z_{12}}}\pa \phi ^{W}_{(r,s)}(z_2) + O(z_{12})
 \nonumber\\
J(z_1)\phi ^{W}_{(r,s)}(z_2) & = &{{q_{r,s}}\o {z_{12}}}
 \phi^{W}_{(r,s)}(z_2) + O(z_{12}), \quad \quad 
 J(z_1)\phi^{V}_{(r,s)}(z_2)= O(z_{12})
 \nonumber 
\er
and from eqn (\ref{9.22}) we conclude that:
\be
\b_{r,s} = q_{r,s}\sqrt {{3\o L}}, \quad \quad \Delta^{V}_{r,s} =   
\Delta^{W}_{r,s}- {{3\o {2L}}}q_{r,s}^2
\label{9.24}
\ee

The dimensions and the charges of the fields $\phi ^{W}_{(r,s)}$ that
represent the so called ``complete degenerate'' highest weight
representations of $W_3^{(2)}$ are given by \cite{Bershadsky} (for {\it
rational } levels $L+3 = {4p\o q}$):
\br
&&\hskip -1cm \Delta_{r,s}^{W} = {{1\o {24(L+3)}}}[((L+3)r_{12}-4s_{12})^2 +
3((L+3)r_1+4s_1)((L+3)r_2-4s_2)-3(L-1)^2] -{1\o 8}\eta^{W},\nonumber\\
&&q_{r,s} = {1\o {12}}[(L+3)r_{12}-4s_{12}] \pm {1\o 2}\eta^{W}, \quad  r_{12} = r_1-r_2, \quad s_{12}= s_1-s_2, 
\label{9.25}
\er
where $\eta^{W} =0$, $r_i$-odd integers for NS-sectors, $\eta^{W} ={1\o 2}$,
$r_i$-even integers for R-sectors and $r_i, s_i (i=1,2)$ take their values in
the interval $1\leq r_i \leq 2p-1,$ $1\leq s_i \leq 2q-1$.  According to eqn.
(\ref{9.24}) the corresponding representations of $V_3^{(1,1)}(L)$-algebra
have the following dimensions:
\br
&&\Delta_{r,s}^{V} = {{1\o {32(L+3)}}}[(L-3)((L+3)r_{12}-4s_{12})^2
+4((L+3)r_1-4s_1)((L+3)r_2-4s_2)- \nonumber\\
&&\hskip0.7cm-4L(L-1)^2]-{{\eta^{W}}\o {8L}}[L+3\eta^{W} \pm ((L+3)r_{12}-4s_{12})].
\label{9.26}
\er
We have to  note that $\eta^{W} = {1\o 2}- \eta^{V}$, ($\eta^{V} = \eta $). 
The analog of the $W_3^{(l)}$--chiral fields $\phi_{s_0}$:
\br
&&G^{\pm}_{-1/2}\phi_{s_0}|0\rangle = 0 \quad \quad \Delta^{W} = {{6\o
{L+3}}}q_{\pm} (q_{\pm} \pm {{L-1}\o 4}),
\nonumber\\
&&q_+ = q_-={1\o 6}(L+5-2s_0),\quad r_1=3r_2=3, s_1=s_0, s_2=1
\nonumber
\er
 are the order-disorder parameter fields
$\sigma^{\eta}_{s}$ in the $V_3^{(1,1)}$-models:
\br
V^{\pm}_{{{3(1\pm s)}\o {2L}}- 1/2 + \eta}\sigma^{\eta}_{s}|0\rangle = 0
\nonumber
\er
One can find their dimensions directly from (\ref{9.14}):
\br
\Delta^{(1/2)}_s = {1\o 8}{{L(L-1)(3s+L)(3s-L)}\o {(L+3)}},\quad \quad s-odd
\nonumber
\er
They are a particular case of the  (\ref{9.26}) when 
\br
2s_0=3s+2L+5, \quad \quad s_0 =1,3,\cdots L-1
\nonumber\\
r_1 = 3r_2 =3, \quad \quad s_2=1, \quad s_1 = s_0.
\nonumber
\er
The complete description of the representations of the
$V_3^{(1,1)}(L)$-algebra  (even for the case $L$-positive integer $L>3$)
however requires much more work.  The construction of representations for the
other {\it Cases } $(1), (2)$ and $(3)$ is an interesting open problem.

The {\it purpose} of this rather detail discussion of the {\it quantum}
$V^{(1,1)}_3(L)$-algebras was to point out the {\it differences} with the
quantization of the $W_3$ and $W_3^{(2)}$ algebras and the {\it similarities}
 with
the PF-algebra.  The origin of all these complications is the renormalization
of the spins of the $V^{\pm}$-currents
\br
\Delta^{{\pm}^{quantum}}_{1} = \Delta_1^{{\pm}^{class}} - {3\o {2L}}
\nonumber
\er
 which makes the singularities of the OPE-algebra (\ref{9.9}) $L$-dependent.  
 For certain values of $L$ this requires to introduce new currents $V^{\pm}_l$
 and $W^{\pm}_p$ in order to close the OPE-algebra.   An important new
 phenomena is the breaking of the (classical ) $U(1)$-symmetry to the discrete
 $Z_L$-symmetries for $L$ positive integer.  The typical PF feature is the
 replacing of the commutators or anticommutators with an infinite sum of
 bilinears of generators as in eqns. (\ref{9.14}), (\ref{9.15}), (\ref{9.16})
 , (\ref{9.20}).   One might wonder whether the $V^{(1,1)}_{n+1}$-algebras
 exhibit similar features.  Our preliminary result shows that the
 renormalization of the spins of the nonlocal currents $V^{(1,1)}_{n+1}$ is a
 common property of all $V^{(1,1)}_{n+1}$'s: $\Delta^{\pm (q)}_{(n)} = {{n+1}\o
 2}(1- {1\o {2k+n+1}}) $. As  usual the spins of the local currents $W_{l+1}$
 remain equal to the classical ones.  All this indicates that the quantum
 $V^{(1,1)}_{n=1}$-algebras obey many of the properties of the quantum
 $V^{(1,1)}_3$-algebra.

${\bf Acknowledgments }$

One of us (GS) thanks the Department of Theoretical Physics,
UERJ-Rio de Janeiro  for the hospitality and  financial support. GS also thanks  IFT-UNESP,Laboratoire de Physique Mathematique,Universite de Montpellier II, DCP-CBPF and Fapesp 
 for the partial financial support at the initial and the final stages  of this work.
(JFG) thanks ICTP-Trieste for hospitality and support where part
 of this work
was done. This work was partially supported by CNPq.

\appendix
\section{Appendix A }
\renewcommand{\theequation}{\thesection.\arabic{equation}}
\setcounter{equation}{0}
\setcounter{subsection}{0}
\setcounter{footnote}{0}

Here we define a generic Lie algebra $\lie $ in the Chevalley basis by the
commutation relations
\be
[h_i, h_j ] =0, \quad [h_i, E_{\pm \a_j} ] =\pm k_{ji}E_{\pm \a_j}, \quad 
 [E_{\a_i}, E_{ -\a_j} ] = \d_{ij}h_j
\label{A.1}
\ee
$i,j=1, \cdots ,rank \; \lie $, where $h_i = {{2\a_i \cdot H }\o {\a_i^2 }} $ and
$H_i$ define  $\lie $ in the Cartan-Weyl basis, i.e.
\be
[H_i, H_j ] =0, \quad [H_i, E_{\pm \a} ] =\pm (\a )^{i}E_{\pm \a }, \quad 
 [E_{\a_i}, E_{- \a_j} ] = {{2\a \cdot H }\o {\a^2 }}
\label{A.2}
\ee
and $ (\a )^{i} $ denote the $i^{th}$ component of the root $\a $ and $k_{ij} =
{{2\a_i \cdot \a_j } \o {\a_j^2}} $ is the Cartan matrix.

The rank $\lie $ fundamental weights are defined by 
\be
{{2\a_i \cdot \lambda_j }\o {\a_j^2}} = \d_{ij}, \quad i,j = 1, ..., rank \;\;\lie 
\label{A.3}
\ee
and may be written in terms of simple roots as 
\be
\lambda_i = {\cal K}_{ij} \a_j
\label{A.4}
\ee
An invariant scalar product on the Lie algebra is defined by rescaling the
trace of two generators in some finite dimensional representation such that
\be
{\rm Tr}(H_i H_j ) = \d_{ij}, \quad {\rm Tr}(H_i E_{\a}) = 0, \quad Tr(E_{\a} E_{\b} ) ={2\o \a^2} \d_{\a +\b ,0} 
\label{A.5}
\ee
 
 A general Lie algebra $\lie $ may be decomposed into a graded structure
 generated by a grading operator $Q$, such that
 \be
 [Q,\lie _{\pm i} ] = \pm i \lie _{{\pm i}}, \quad  
[\lie _{ i},\lie _{ j} ] \in \lie _{i+ j}
\label{A.6}
\ee
and $\lie = \oplus_i \lie _{i}$.

Throughout this paper several different gradings are used.  We shall restrict
ourselves to integer gradings.    The most familiar
is defined by 
\be
Q= \sum_{k=1}^{rank \lie}{2\lambda_k \cdot H \o {\a_k^2}}
\label{A.7}
\ee
In this case $\lie _{\pm i}$ contain positive/negative step operators composed
of $i$ simple roots.  It then follows that 
\be
\lie _{<} = \oplus _{i<0} \lie _{i}, \quad 
\lie _{>} = \oplus _{i>0} \lie _{i}
\label{A.8}
\ee
are nilpotent subalgebras generated by positive/negative step operators, while
the zero grade $\lie _0$  is an abelian subalgebra and is spanned  by the
Cartan subalgebra of $\lie $, i.e. $ \lie _0 = U(1)^{rank \lie }$.
A general group element of $\lie $ can be decomposed according to these
nilpotent subalgebras together with the exponentiation of the abelian
subalgebra $\lie _0$ using the Gauss decomposition formula,
\be
g = g_{-}g_{0}g_{+}
\label{A.9}
\ee
where $g_{-}$ and $g_{+}$ are obtained by
 exponentiation of $\lie _< $ and $\lie _>$
respectively.
Other gradings extensively used in this paper are defined by:
\be
Q_j = \sum_{ k \neq j}^{rank \lie } {2\lambda_k \cdot H \o {\a_k^2}}
\label{A.10}
\ee
The absence of the $i^{th}$ fundamental weight in (\ref{A.10}) generates a non
abelian structure in the zero grade subalgebra $\lie _0$ which is now generated
by the Cartan subalgebra together with $E_{\pm \a_i }$, i.e. $\lie _0 =
SL(2)\otimes U(1)^{rank \lie -1}$.  The nilpotent subgroups $g_{-}^{(j)}$ 
and $g_{+}^{(j)}$ are
generated by exponentiation of the negative and positive  grades respectively
(according to $Q_j$). 
The non abelian Gauss decomposition formula now reads
\be
g=g_{-}^{(j)}g_{0}^{(j)}g_{+}^{(j)}
\label{A.11}
\ee
where $g_{0}^{(j)}$ is the $ SL(2)\otimes U(1)^{rank \lie -1}$subgroup generated by
exponentiation of $\lie _0$.

Following the same line of thought, more and more complicated non abelian
structure can be introduced in $\lie _0$ by defining grading operators  with
the form
\be
Q_{i_1,i_2,\cdots i_l } = \sum_{ k\neq i_1, \cdots i_l}^{rank \lie } {2\lambda_k \cdot H \o {\a_k^2}}
\label{A.12}
\ee
where $\lie _0$ has now the form
\be
\lie _0 = g_1\otimes g_2 \otimes \cdots \otimes g_m \otimes (U(1))^{rank \lie -
\sum_{a=1}^{m} rank g_a}
\label{A.13}
\ee
and the nonabelian Gauss decomposition is formally given by eqn. (\ref{A.11}).

These are useful concepts in order to derive a simple and compact form for the
equations of motion.  Let the WZW currents be decomposed according to a
gradation $Q_i$.  From (\ref{A.11}) we find, 
\br
J &=& g^{-1}\pa g = {g_{+}^{(j)}}^{-1} 
\( {g_{0}^{(j)}}^{-1} {g_{-}^{(j)}}^{-1} \pa g_{-}^{(j)} g_{0}^{(j)}
 + {g_{0}^{(j)}}^{-1} \pa g_{0}^{(j)} + \pa g_{+}^{(j)} {g_{+}^{(j)}}^{-1}
\) g_{+}^{(j)} \nonumber \\
\bar J &=& -\bar \pa \pa g g^{-1} = -g_{-}^{(j)}\(
 {g_{-}^{(j)}}^{-1} \bar \pa g_{-}^{(j)} +
  \bar \pa g_{0}^{(j)} {g_{0}^{(j)}}^{-1}
+  g_{0}^{(j)}\bar \pa g_{+}^{(j)} {g_{+}^{(j)}}^{-1} {g_{0}^{(j)}}^{-1} \)
 {g_{-}^{(j)}}^{-1}
\label{A.14}
\er
 We now define the reduced model by giving the constant element $\eps_{\pm}$ 
 responsible for constraining the currents 
  in a general manner to  
 \br
 J_{constr } &=& \eps_- + j \nonumber \\
 \bar J_{constr } &=& \eps_+ + \bar j
 \label{A.15}
 \er
 where $j$ (and $\bar j$) contain only positive and zero (negative and zero )
 grades, $\eps_{\pm}$ are constant generators of grade $\pm 1$ respectively. 
 From the graded structure (\ref{A.14}) and (\ref{A.15}) and the fact that the
 grades are integers,  we find that 
 \br
 {{g_{0}^{(j)}}^{-1} {g_{-}^{(j)}}^{-1} 
 \pa g_{-}^{(j)} g_{0}^{(j)}}\vert_{constr } &=& 
 \eps_-  \nonumber \\
 {g_{0}^{(j)} \bar \pa g_{+}^{(j)} {g_{+}^{(j)}}^{-1} 
 {g_{0}^{(j)}}^{-1}}\vert_{constr } 
 &=& \eps_+
 \label{A.16}
 \er
 and the equations of motion $\bar \pa J = \pa \bar J =0 $ become 
 \br
 \bar \pa K + [K, \bar \pa g_{+}^{(j)}{g_{+}^{(j)}}^{-1} ] &=& 0 \nonumber \\
 \pa \bar K - [\bar K , {g_{-}^{(j)}}^{-1}\pa g_{-}^{(j)} ] &=& 0
 \label{A.17}
 \er
 where 
 \br
 K &=& {g_{0}^{(j)}}^{-1}{g_{-}^{(j)}}^{-1} \pa g_{-}^{(j)} g_{0}^{(j)}
  + {g_{0}^{(j)}}^{-1} \pa g_{0}^{(j)}  + 
  \pa g_{+}^{(j)} {g_{+}^{(j)}}^{-1} \nonumber \\
 \bar K &=& {g_{-}^{(j)}}^{-1} \bar \pa g_{-}^{(j)} + 
 \bar \pa g_{0}^{(j)} {g_{0}^{(j)}}^{-1}  + 
 g_{0}^{(j)} \bar \pa g_{-}^{(j)} {g_{-}^{(j)}}^{-1}  {g_{0}^{(j)}}^{-1}  
 \label{A.18}
 \er
 
 Imposing (\ref{A.16}) into (\ref{A.17}) we find the nontrivial equations to
 correspond to the zero grade components of eqns. (\ref{A.17}), i.e.
 \br
 \bar \pa ({g_{0}^{(j)}}^{-1} \pa g_{0}^{(j)} ) &+& 
 [\eps_-, {g_{0}^{(j)}}^{-1} \eps_+ g_{0}^{(j)} ] = 0 \nonumber \\
 \pa (\bar \pa g_{0}^{(j)} {g_{0}^{(j)}}^{-1} ) &-&
  [\eps_+ , g_{0}^{(j)}\eps_- {g_{0}^{(j)}}^{-1} ] = 0 
 \label{A.19}
 \er
 
 We now point out that there is a subalgebra of $\lie _0$, namely $\lie _0^0 $,
 commuting with $\eps_{\pm}$ such that
 \be 
 \bar \pa {\rm Tr} ({g_{0}^{(j)}}^{-1} \pa g_{0}^{(j)} {\cal H}) = 
 \pa {\rm Tr} (\bar \pa g_{0}^{(j)} {g_{0}^{(j)}}^{-1} {\cal H}) = 0 
 \label{A.20}
 \ee
 where ${\cal H } \in \lie _0^0 $.  We therefore impose an additional subsidiary
 constraint within the subalgebra $\lie _0^0$, 
 \br
 &&J_{{\cal H}} = {\rm Tr}({g_{0}^{(j)}}^{-1} \pa g_{0}^{(j)} {\cal H}) = 0,\quad \quad  \bar J_{{\cal H}} = {\rm Tr}(\bar  \pa g_{0}^{(j)} {g_{0}^{(j)}}^{-1} {\cal H}) = 0 
\label{A.21}
\er
For the simple case of gradation (\ref{A.10}) ${\cal H}$ correspond to 
${{2\lambda_j\cdot H}\o {\a_j^2}}$ .  In terms of
fields defined by
\be
g_{0}^{(j)} = 
e^{\chi E_{-\a_j}} e^{{{2\lambda_j\cdot H}\o {\a_j^2}}R_j +\sum_{l\neq
j}H_l \phi_l } e^{\psi E_{\a_j}}
\nonumber
\ee
we find
\br
&&\hskip-1.1cm {g_{0}^{(j)}}^{-1} \pa g_{0}^{(j)} = 
\(  \pa \chi e^{R_j + \sum_{l\neq j}k_{jl}\phi_l}\)  E_{-\a_j}
+ \sum_{l\neq j} H_l\( \pa \phi_l + {{\a_l^2 {\cal K}_{jl}  }\o
{\a_j^2 {\cal K}_{jj}}}\psi \pa \chi e^{R_j + \sum_{l\neq j}k_{jl}\phi_l} \) \nonumber \\
&& + {{2\lambda_j\cdot H}\o {\a_j^2}} \( \pa R_j -{ 1\o {{\cal K}_{jj}}}\psi \pa
\chi e^{R_j + \sum_{l\neq j}k_{jl}\phi_l} \)
+ \( \pa \psi + \psi \pa R + \psi k_{jl}\pa \phi_l - \psi^2 \pa \chi 
e^{R_j + \sum_{l\neq j}k_{jl}\phi_l} \) E_{\a_j}
\nonumber
\er
and 
\br
&&\hskip-1.1cm \bar \pa g_{0}^{(j)} {g_{0}^{(j)}}^{-1} =
 \(  \bar \pa \psi e^{R_j + \sum_{l\neq j}k_{jl}\phi_l}\)
  E_{\a_j}
+ \sum_{l\neq j} H_l\(\bar \pa \phi_l + {{\a_l^2 {\cal K}_{jl}  }\o
{\a_j^2 {\cal K}_{jj}}}\chi \bar \pa \psi e^{R_j + \sum_{l\neq j}k_{jl}\phi_l} \) \nonumber \\
&&+  {{2\lambda_j\cdot H}\o {\a_j^2}} \( \bar \pa R_j 
-{ 1\o {{\cal K}_{jj}}}\chi \bar \pa
\psi e^{R_j + \sum_{l\neq j}k_{jl}\phi_l} \)
+ \( \bar \pa \chi + \chi \bar \pa R + \chi k_{jl}\bar \pa \phi_l - \chi^2 \bar \pa \psi 
e^{R_j + \sum_{l\neq j}k_{jl}\phi_l} \) E_{-\a_j}
\nonumber
\er
Moreover, 
\br
&&J_{{{2\lambda_j \cdot H}\o {\a_j^2}}} =  \pa R_j  \Delta_j - {1\o
{{\cal K}_{jj}}} \tilde \psi \pa \tilde \chi 
e^{ \sum_{l\neq j}k_{jl}\phi_l},\quad \quad 
\bar J_{{{2\lambda_j \cdot H}\o {\a_j^2}}} = \bar \pa R_j  \Delta_j - {1\o
{{\cal K}_{jj}}} \tilde \chi \bar \pa \tilde \psi 
e^{ \sum_{l\neq j}k_{jl}\phi_l} \nonumber
\er
where $ \Delta_j = 1 + {1\o {2{\cal K}_{jj}}} \tilde \psi \tilde \chi 
e^{ \sum_{l\neq j}k_{jl}\phi_l}$.
The subsidiary condition (\ref{A.21}) yields,
\br
&&\pa R_j = {1\o {{\cal K}_{jj} \Delta_j }}\tilde \psi \pa \tilde \chi 
e^{ \sum_{l\neq j}k_{jl}\phi_l}, \quad \quad \quad 
\bar \pa R_j = {1\o {{\cal K}_{jj} \Delta_j }}\tilde \chi \pa \tilde \psi 
e^{ \sum_{l\neq j}k_{jl}\phi_l} \nonumber
\er
where $\tilde \psi = \psi e^{{1\o 2}R_j}$ and $\tilde \chi = 
\chi e^{{1\o 2}R_j}$.  When inserted into (\ref{A.19}), leads to the equations
of motion (\ref{2.3}) of the NA-Toda model described by the action (\ref{7.3}) and (\ref{2.3}) for the case$j=1$.

Henceforth, given an integer gradation $Q$ and the constant elements
$\eps_{\pm} $,  in the Lie algebra, we are able to decompose the currents
according to $Q$, implement consistently the constraints (\ref{A.15}) to
obtain (\ref{A.16}) and therefore the equations of motion (and hence the
action, (see \cite{LAF})).  This construction highlights the fact that the
first set of constraints in (\ref{2.4}) are fully determined by $Q$ and
$\eps_{\pm}$.  The subsidiary condition (\ref{A.21}) has to be implemented by
fiat since it constraints the subalgebra $\lie_0^0$.

Conversely, since 
the structure of the constraints is such that allows for dynamical degrees of
freedom only those contained in the zero grade subgroup $\lie_0 $ ( or more
precisely $ \lie_0 / \lie_0^0 $), the 
constraints (\ref{2.4}) suggest a construction of a gauged WZW action
invariant under the constraint subgroup $ H_+ \otimes H_-$ , as in sect. 2.
   The gauge
invariance suggests decomposing the group element as 
\be
g = \tilde g_{-}^{(j)}  g_{0}^{f} \tilde g_{+}^{(j)}= 
 g_{-}^{(j)}  g_{0}^{(j)}  g_{+}^{(j)}
\label{A.22}
\ee
where $\tilde g_{-}^{(j)}$, ($\tilde g_{+}^{(j)}$ )
 are generated by exponentiation of
negative (positive ) grade generators together with those in $\lie _0^0$, 
\be
\tilde{ g_+^{(j)}} = g_+^{(j)} e^{{\cal H}(R_j)}, 
\quad \tilde {g_-^{(j)}} = e^{{\cal H}(R_j)} g_{-}^{(j)}
\nonumber
\ee
where $R_j$ are nonlocal, nonphysical  fields eliminated by the constraints
(\ref{A.21}).The  $g_0^f $ is now generated by 
exponentiating ${{\lie  }\o 
{{\lie }_0^0}}$.  The decomposition (\ref{A.22}) is important
 to point out the symmetry
structure of the NA-Toda models as described in sect. 2.

\section{Appendix B }
\renewcommand{\theequation}{\thesection.\arabic{equation}}
\setcounter{equation}{0}
\setcounter{subsection}{0}
\setcounter{footnote}{0}

We shall derive the general
solution of eqns. (\ref{3.19}) for generic
$W_s$-transformations ($ 3 \leq s \leq n-1 $, fixed
), i.e. $\eps^{\pm} = \eps = 0$, $ \eta_p = 0 $ for
all $p \neq s $, ($\eta_s \equiv \eta $).  Let us
first consider the equations for $\eps_{ik}$'s
such that $i > k$. We have,
\br
\eps_{l+m,l} &=& 0, \quad l=1, \cdots s-1, \quad 
l+m = n-s+3, \cdots ,n+1 \nonumber \\
\eps _{l+m, l} &=& 0, \quad l=s+1, \cdots n,
\quad  l+m = s+2, \cdots ,n+1
\label{B1}
\er
and all the remaining $\eps _{ik}$'s ($ i>k$ )
satisfy the following recursive relations
\be
\eps_{n-p,s} + \eta W_{p+1} + {k\o 2} \pa
\eps_{n-p+1,s} = 0 \quad p=0,\cdots ,n-s-1, \;\; s\neq 1
\label{B.2}
\ee
\be
\eps_{n-p,1} =
\sum_{l=1}^{p-s+3} (-{k\o 2}\pa )^{l-1}
(V^{+}\eps_{n-p+l,2} )
\label{B.3}
\ee
\be
\eps_{n-p,s-r} = \sum_{l=1}^{p-s+2}
(-{k\o 2}\pa )^{p-l-r+2} \eps_{n-l-r+3,s-r+1}, 
\label{B.4}
\ee
$ n-p= 2, \cdots n-s+2;\quad r=0, \cdots ,s-2$. 
The solution of (\ref{B.2}) is given by
\be
\eps_{n-p,s} = (-{k\o 2}\pa )^{p+1} \eta -
\sum_{l=1}^{p}(-{k\o 2}\pa )^{l-1} (\eta
W_{p+2-l}), \quad p=0, \cdots , n-s-1
\label{B.5}
\ee

The eqn. (\ref{B.4}) can be simplified to:
\be
\eps_{n-p,s-r} = \sum_{l=1}^{p-r+2} \left( \begin{array}{c} {l+r-2} \\ {r-1}
 \end{array}
\right) (-{k\o
2}\pa )^{l-1} \eps_{n-p+l+r-1,s}
\nonumber 
\ee
and taking into account (\ref{B.5}) we find the
general solution of (\ref{B.4}):
\be
\eps_{n-p,s-r} = \left( \begin{array}{c} {p+1} \\ {r} \end{array}
\right)
 (-{k\o 2}\pa )^{p-r+1}\eta
- \sum_{l=1}^{p-r} \left( \begin{array}{c} {l+r-1} \\ {r} \end{array}
\right)
 (-{k\o 2}\pa )^{l-1}(\eta
W_{p-r-l+2}), 
\label{B.6}
\ee
$r=0, \cdots , s-2, \;\; n-p = 2, \cdots , n-s+2$.
As a consequence of (\ref{B.3}) and (\ref{B.6})
we obtain
\br
\eps_{n-p, 1} &=& \sum_{l=1}^{p-s+3} (-{k\o 2}\pa
)^{l-1} \{ V^{+} [\left( \begin{array}{c} {p-l+1} \\ {s-2} \end{array}
\right)
 (-{k\o 2}\pa
)^{p-l-s+3}\eta \nonumber \\
&-& \sum_{m=1}^{p-l-s+2}
 \left( \begin{array}{c} {m+s-3} \\ {s-2} \end{array}
\right)
(-{k\o 2}\pa )^{m-1} \eta W_{p-l-m-s+4} ]
\}
\label{B.7}
\er
The equations for the diagonal elements $ \eps_{ll}$
read
\br 
&&\hskip-1cm 2\eps_{n,n} + \eps_{22} + \cdots + \eps_{n-1,n-1}
=0, \quad \eps_{nn} = \eps_{n-1,n-1} = \cdots =
\eps_{s+1, s+1}, \quad \eps_{11}=0 
\nonumber \\
&&\eps_{s+1,s+1} = \eps_{ss} + \eta W_{n-s+1} +{k\o
2}\pa  \eps_{s+1,s},\quad
\eps_{pp} = \eps_{22} + \sum_{r=3}^{p} ({k\o
2}\pa ) \eps_{r,r-1}, 
\label{B.8}
\er
$p=3, \cdots , s$.  The solution of (\ref{B.8}) can be written in
the following compact form
\br
\eps_{ll} &=& [\left( \begin{array}{c} {n-l+1} \\ {n-s+1} \end{array}
\right)
 - {1\o n} \left( \begin{array}{c} {n} \\ {n-s+2} \end{array}
\right)
 ](-{k\o 2}\pa
)^{n-s+1}\eta 
\nonumber \\
&+& \sum_{p=0}^{n-s-1} [{1\o n}
\left( \begin{array}{c} {p+s-1} \\ {p+1} \end{array}
\right) - \left( \begin{array}{c} {p+s-l} \\ {p} \end{array}
\right)
 ]
(-{k\o 2}\pa )^{p}(\eta W_{n-s-p+1})
\label{B.9}
\er 
$l=2, \cdots , s,s+1, \cdots ,n$. We next consider the recursion relations
 for the
upper triangular part of the $\eps_{ik}$'s, ($i <k$):
\br
\eps_{1l} &=& 0, \quad l=1, \cdots s; \quad
\eps_{1l} = ({k\o 2}\pa )^{l-s-1}(\eta  V^{-}),
\quad l=s+1, \cdots , n+1
\nonumber \\
\eps_{2m} &=& ({k\o 2}\pa )^{m-2} \eps_{22} +
({k\o 2}\pa )^{m-s-1}  (\eta W_n) +
\sum_{l=1}^{m-s-1} ({k\o 2}\pa
)^{m-s-l-1}[V^{+}({k\o 2}\pa )^{l-1}(\eta
V^{-})], \nonumber \\
&&\label{B.10}
\er
$m=2, \cdots , n+1$ . For generic $\eps_{ml}$,
($2 \leq m \leq n, \; m<l $) we have two
different formulas for $2\leq m \leq s$ and $ s<
m \leq n$,
\br
&&(a)\quad \quad 2\leq m \leq s,  \quad \quad l=m+1, \cdots
n+1 
\nonumber \\
&&\eps_{m,l} = \sum_{p=1}^{l-m} \left( \begin{array}{c} {p+m-4} \\ {m-3} \end{array}
\right)
 ({k\o 2}\pa
)^{p-1}\eps_{2,l-m-p+3} 
+\sum_{p=1}^{m-2} \left( \begin{array}{c} {l-m+p-2}
 \\ {p-1} \end{array}
\right)
({k\o 2}\pa )^{l-m}\eps_{m-p+1,m-p+1} 
\nonumber \\
&&\hskip 1.5cm + \sum_{p=1}^{m-2} \left( \begin{array}{c} {l-m+p-2} \\ {p-1} \end{array}
\right)
({k\o 2}\pa
)^{l-s-p}(\eta W_{n-m+p+1})
\label{B.11}\\
&&(b)\quad \quad s<m \leq n,  \quad \quad l=m+1, \cdots n+1
\nonumber \\
&&\hskip-1cm\eps_{m,l} = \sum_{p=1}^{l-m} \left( \begin{array}{c} {p+m-s-2} \\
 {m-s-1} \end{array}
\right)
 ({k\o 2}\pa
)^{p-1}\eps_{s,l-m-p+s+1}
+ \sum_{p=1}^{m-s} \left( \begin{array}{c} 
{l+p-m-2} \\ {p-1} \end{array}
\right)
({k\o 2}\pa )^{l-m}\eps _{m-p+1,m-p+1}\nonumber\\
\label{B.12}
\er
The solution of eqn. (\ref{B.11}) is given by
($2\leq m \leq s $):
\br
&&\hskip-1cm \eps_{ml} = \{ A_{l,m}(s) 
+ {1\o n}\left( \begin{array}{c} {n} \\ {n-s+2} \end{array}
\right)
[(n-s+1)\left( \begin{array}{c} {l-3} \\ {m-2} \end{array}
\right)
 -\left( \begin{array}{c} {l-3} \\ {m-3} \end{array}
\right)
 ]\} \times \nonumber\\
&&\times (-1)^{l-m} (-{k\o 2}\pa
)^{n-s-m+l+1} \eta  
+ \sum_{p=1}^{l-m-s+1} \left( \begin{array}{c} {l-s-p-1} \\ {m-2} \end{array}
\right)
 ({k\o 2}\pa
)^{l-m-s-p+1}[V^{+}({k\o 2}\pa )^{p-1} (\eta
V^{-})] +\nonumber \\
&&+ \sum_{p=0}^{n-s-1} \{ {1\o n}\left( \begin{array}{c} {l-2} \\ {m-2} \end{array}
\right)\left( \begin{array}{c} {p+s-1} \\ {p+1} \end{array}
\right)-
\left( \begin{array}{c} {l-3} \\ {m-2} \end{array}
\right)\left( \begin{array}{c} {p+s-2} \\ {p} \end{array}
\right)  -  B_{l,m}^{p}(s) \}\times \nonumber \\
&&\times (-1)^{l-m}
(-{k\o 2}\pa )^{p+l-m}(\eta W_{n-s-p+1}) 
+\sum_{p=1}^{m-1} \left( \begin{array}{c} {l-s-1} \\ {p-1} \end{array}
\right)
 ({k\o 2}\pa
)^{l-s-p}(\eta W_{n-m+p+1} ),
 \label{B.13}
\er
where 
\br
A_{lm}(s) = \sum_{p=0}^{s-3} \left( \begin{array}{c} {l-p-4} \\ {l-m-1} \end{array}
\right)\left( \begin{array}{c} {n-p-2} \\ {n-s+1} \end{array}
\right), \quad
B_{lm}^{p}(s) = \sum_{q=0}^{s-3} \left( \begin{array}{c} {l-q-4} \\ {l-m-1} \end{array}
\right)\left( \begin{array}{c} {p+s-q-3} \\ {p} \end{array}
\right)
\nonumber 
\er
The corresponding solution  of eqn. (\ref{B.12})
($s < m \leq n $) differs from (\ref{B.13})
only by the last term which in this case is
\br
&&\hskip-1.4cm \eps_{ml} = \{ A_{l,m}(s)+ \sum_{p=1}^{s-1} 
 \left( \begin{array}{c} {l-s-1} \\ {l-m-p} \end{array}
\right)
({k\o 2}\pa
)^{l-m-p}(\eta W_{n-s+p+1} )
\nonumber \\
&&\hskip-0.7cm + {1\o n}\left( \begin{array}{c} {n} \\ {n-s+2} \end{array}
\right)
[(n-s+1)\left( \begin{array}{c} {l-3} \\ {m-2} \end{array}
\right)
 -\left( \begin{array}{c} {l-3} \\ {m-3} \end{array}
\right)
 ]\} (-1)^{l-m} (-{k\o 2}\pa
)^{n-s-m+l+1} \eta  \nonumber \\
&&\hskip-0.7cm + \sum_{p=1}^{l-m-s+1} \left( \begin{array}{c} {l-s-p-1} \\ {m-2} \end{array}
\right)
 ({k\o 2}\pa
)^{l-m-s-p+1}[V^{+}({k\o 2}\pa )^{p-1} (\eta
V^{-})] \nonumber \\
&&\hskip-0.7cm + \sum_{p=0}^{n-s-1} \{ {1\o n}\left( \begin{array}{c} {l-2} \\ {m-2} \end{array}
\right)\left( \begin{array}{c} {p+s-1} \\ {p+1} \end{array}
\right)-
\left( \begin{array}{c} {l-3} \\ {m-2} \end{array}
\right)\left( \begin{array}{c} {p+s-2} \\ {p} \end{array}
\right)
 - B_{l,m}^{p}(s) \} \times\nonumber\\
&&\hskip-0.7cm \times (-1)^{l-m}
(-{k\o 2}\pa )^{p+l-m}(\eta W_{n-s-p+1})+\sum_{p=1}^{s-1} \left( \begin{array}{c} {l-s-1} \\ {l-m-p} \end{array}
\right)
 ({k\o 2}\pa )^{l-m-p}(\eta W_{n-s+p+1} ) ,
 \label{B.14}
\er
The $W_s$ transformation ($\eta \equiv \eta_s$) 
derived from eqn. (\ref{3.19}) ($3\leq s \leq
n-1$ ) can be written in a compact form as 
\br
\d_{\eta }V^{+} &=& \eps_{22}V^{+} - ({k\o 2}\pa
)\eps_{21}, \quad \d_{\eta }V^{-} =
\sum_{l=s+1}^{n} \eps_{1l} W_{n-l+2} -
\eps_{nn}V^{-} - ({k\o 2}\pa
)\eps_{1,n+1},
 \nonumber \\
\d_{\eta }W_{n} &=& \eps_{21}V^{-} -
\eps_{1,n+1}V^{+} + (\eps_{22} - \eps_{nn} )W_n +
\sum_{l=3}^{n} \eps_{2l}W_{n-l+2} -{k\o 2}\pa
\eps_{2,n+1}, 
\nonumber \\
\d_{\eta }T &=& \eps_{n,s-1}W_{n-s+3} +
\eps_{n,s}W_{n-s+2} - \eps_{n-1,n+1} -{k\o 2}\pa
\eps_{n,n+1},
\label{B.15}
\er
and
\br
(a) & & n-s+2 \leq q \leq n 
\nonumber \\ 
\d_{\eta }W_{q} &=& \eps_{n-q+2,1}V^{-} +
\sum_{l=2}^{n} (\eps_{n-q+2,l} - \d_{l,n-q+2}
\eps_{nn})W_{n-l+2} - \eps_{n-q+1,n+1} - {k\o 2}\pa \eps_{n-q+2,n+1}
\nonumber \\
(b) & & 2 \leq q < s \quad or \quad s \leq q <n-s+2
\nonumber \\
\d_{\eta }W_{q} &=& P_q(s) + (\eps_{n-q+2,n-q+2}
- \eps_{nn} ) W_q + \sum_{l=n-q+3}^{n}
\eps_{n-q+2,l}W_{n-l+2} \nonumber \\
&- & \eps_{n-q+1,n+1} -
{k\o 2}\pa \eps_{n-q+2,n+1}
\label{B.16}
\er
where
\be
P_q(s) = \left\{ \begin{array}{l}
{\sum_{l=s-q+1}^{s} \eps_{n-q+2,l} W_{n-l+2},
\quad for \quad 2\leq q <s }\\ {\sum_{l=2}^{s}
\eps_{n-q+2,l}W_{n-l+2} + \eps_{n-q+2,1}V^{-},
\quad for \quad s\leq q <n-s+2 }   \end{array}\right.
\label{B.17}
\ee
Taking into account the explicit form
(\ref{B.9}), (\ref{B.13}) and (\ref{B.14}) of
$\eps_{ml}$'s we realize the following
simplifications:
\be
\eps_{ll}-\eps_{nn} = 
\left( \begin{array}{c} {n-l+1} \\ {n-s+1} \end{array}
\right) [(-{k\o 2}\pa
)^{n-s+1}]\eta  - \sum_{p=0}^{n-s-1}\left( \begin{array}{c} {p+s-l} \\ {p} \end{array}
\right) 
 (-{k\o
2}\pa )^{p} (\eta W_{n-s-p+1})
\nonumber 
\ee
\br
&&\hskip-1cm C_{m,n}(s) = \eps_{m,n+1} + {k\o 2} \pa
\eps_{m+1,n+1}=\nonumber \\
 &&= \{ A_{n+2,m+1}(s) + {1\o n}\left( \begin{array}{c} {n} \\
 {n-s+2} \end{array}
\right)
[(n-s+1)\left( \begin{array}{c} {n-1} \\ {m-1} \end{array}
\right)
 -\left( \begin{array}{c} {n-1} \\ {m-2} \end{array}
\right)
 ]\} \times \nonumber\\
&&\times (-1)^{n-m+1} (-{k\o 2}\pa
)^{2n-s-m+2} \eta  
+ \sum_{r=1}^{n-m-s+2} \left( \begin{array}{c} {n-s -r+1} \\ {m-1} \end{array}
\right) ({k \o 2}\pa
)^{n-m-s-r+2} [V^{+}  ({k \o 2}\pa
)^{r-1}
(\eta V^{-}) ] \nonumber \\ 
&&+ \sum_{p=0}^{n-s-1} (-1)^{n-m+1} \{ {1\o
n}\left( \begin{array}{c} {n} \\ {m-1} \end{array}
\right)\left( \begin{array}{c} {p+s-1} \\ {p+1} \end{array}
\right)
 - \left( \begin{array}{c} {n-1} \\ {m-1} \end{array}
\right)\left( \begin{array}{c} {p+s-2} \\ {p} \end{array}
\right)\nonumber \\
  &&- B^{p}_{n+2,m+1}(s) \} 
(-{k\o 2}\pa
)^{n+p-m+1}( \eta   W_{n-s-p+1})+ D_{m,n}(s),
\label{B.18}
\er
where
\be
D_{m,n}(s) = \left\{ \begin{array}{c} { \hskip-.3cm
\sum_{p=1}^{m} \left( \begin{array}{c} {n-s+1} \\ {p-1} \end{array} \right)
({k\o 2} \pa )^{n-s-p+2}
(\eta W_{n-m+p}), \quad m \leq s } 
 \\ 
{\sum_{p=1}^{s-1}\left( \begin{array}{c} {n-s+1} \\ {m-s+p} \end{array} \right)
({k\o 2}\pa  )^{n-m-p+1}
(\eta W_{n-s+p+1}), \quad m > s } 
\end{array} \right.
\ee
Finally, we arrive at the $W_s$-transformations we
seek,
\br
&&\hskip-0.7cm \d_{{\eta }}V^{-} = {1\o n}\left( \begin{array}{c} {n} \\ {n-s+2} \end{array} \right)
 V^{-}(-{k\o 2}\pa 
)^{n-s+1} \eta  - ({k\o 2}\pa  )^{n-s+1}(\eta
V^{-} )\nonumber \\
&&\hskip0.3cm - {1\o n} V^{-} \sum_{p=0}^{n-s-1} 
\left( \begin{array}{c} {p+s-1} \\ {s+1} \end{array} \right)
(-{k\o 2}\pa  )^{p}(\eta W_{n-s-p+1} )
+\sum_{l=s+1}^{n} W_{n-l+2}({k\o 2}\pa 
)^{l-s-1} (\eta V^{-} ), \nonumber \\
&&\hskip-0.7cm \d_{{\eta }}V^{+} = {1\o n}V^{+} \{ (n-s+1) 
\left( \begin{array}{c} {n} \\ {n-s+2} \end{array}  \right)
(-{k\o 2}\pa )^{n-s+1} \eta + \sum_{l=0}^{n-s-1}
[\left( \begin{array}{c} {l+s-1} \\ {l+1} \end{array}  \right) \nonumber \\
 &&\hskip0.3cm  -n \left( \begin{array}{c} {l+s-2} \\ {l} \end{array}  \right)
  ] (-{k\o 2}\pa )^{l} (\eta
W_{n-s-l+1} )\}
+ \sum_{l=1}^{n-s+1} (-{k\o 2}\pa )^{l}\{
V^{+}[\left( \begin{array}{c} {n-l-1} \\ {s-2} \end{array}  \right)
 (-{k\o 2}\pa )^{n-l-s+1} \eta \nonumber \\
 &&\hskip0.3cm -
\sum_{m=1}^{n-l-s} \left( \begin{array}{c} {m+s-3} \\ {s-2} \end{array}  \right)
 (-{k\o 2}\pa )^{m-1}
(\eta W_{n-l-m-s+2} )] \},
\nonumber \\
&&\hskip-0.7cm \d_{{\eta }}T = -({k\o 2}\pa \eta )W_{n-s+2}
-(n-s+1) ({k\o 2}\pa )(\eta W_{n-s+2} ) -
{{(n-s+1)}\o 2} \left( \begin{array}{c} {n+1} \\ {n-s+3} \end{array}  \right) \times
\nonumber \\
&&\hskip0.3cm \times (-{k\o 2}\pa )^{n-s+3} \eta 
-\sum_{p=0}^{n-s-1} \{ {{n-1}\o 2} \left( \begin{array}{c} {p+s-1} 
\\ {p+1} \end{array} \right) 
 -(n-1)
 \left( \begin{array}{c} {p+s-2} \\ {p} \end{array} \right) \nonumber \\
&&\hskip0.3cm - (n-s-p)\left( \begin{array}{c} {p+s-2} \\ {p+1} \end{array} \right)
 -(p+1)\left( \begin{array}{c} {p+s-1} \\ {p+2} \end{array}\right)
  \} (-{k\o 2}\pa
)^{p+2}(\eta W_{n-s-p+1} )
\nonumber 
\er
and
$\d_{\eta} W_{n} $ is given by (\ref{B.15}) with 
\br
&&\hskip-0.8cm \eps_{21}= \sum_{l=1}^{n-s+1} (-{k\o 2}\pa
)^{l-1} \{ V^{+} [\left( \begin{array}{c} {n-l-1} \\ {s-2} \end{array} \right)
 (-{k\o 2}\pa )^{n-l-s+1}
\eta \nonumber \\
&&\hskip-0.2cm - \sum_{m=1}^{n-l-s}\left( \begin{array}{c} {m+s-3} \\ {s-2} \end{array} \right)
 (-{k\o 2}\pa
)^{m-1} (\eta W_{n-l-m-s+2})]\}
\nonumber \\
&&\hskip-0.8cm \eps_{2l}={{n-s+1}\o n} \left( \begin{array}{c} {n} \\ {n-s+2} \end{array} \right)
(-1)^{l-2} [
(-{k\o 2}\pa )^{n+l-s-1} \eta ] +
 \sum_{p=1}^{l-s-1} 
 ({k\o 2}\pa )^{l-s-p-1}
[V^{+} ({k\o 2}\pa )^{p-1} (\eta V^{-}) ]
\nonumber \\
&&\hskip-0.2cm + {1\o n} \sum_{p=0}^{n-s-1} [ \left( \begin{array}{c} {p+s-1} \\ 
{p+1} \end{array}\right)
 -n\left( \begin{array}{c} {p+s-2} \\ {p} \end{array}\right)
 ] (-1)^{p}
({k\o 2}\pa )^{p+l-2} (\eta W_{n-s-p+1} ) +
({k\o 2}\pa )^{l-s-1} (\eta W_n )
\nonumber
\er
Substituting $\eps_{n-q+2,1} $from (\ref{B.7}),
$\eps_{n-q+2,l}$ from (\ref{B.13}) and
(\ref{B.14}) and $\eps_{ll}- \eps_{nn}$ and
$C_{n-q+1,n}(s) $ from (\ref{B.18}) in the
transformation laws (\ref{B.16}) and (\ref{B.17})
we derive the explicit form of the $\eta_s$
transformation of the $W_q$ currents ($2 \leq q
<n$).


\begin{thebibliography}{99}
\bibitem{BPZ}
A. Belavin, A. Polyakov and A.B. Zamolodchikov, Nucl. Phys. {\bf B241}, (1984),
333
\bibitem{Z}
A.B. Zamolodchikov, Theor. Math. Phys. {\bf 65 },(1986), 1205
\bibitem{ZF1}
A. B. Zamolodchikov, V. A. Fateev, Sov. Phys. JETP {\bf 62} (1985) 215 
\bibitem{ZF2} 
V. A. Fateev, A. B. Zamolodchikov, Nucl. Phys. {\bf B280} [FS18] (1987) 644
\bibitem{Ber-Oo}
M. Bershadsky and H. Ooguri, Commun. Math. Phys. {\bf 126 } (1989) 49
\bibitem{Ge}
D. Gepner, Nucl. Phys. {\bf B290 },(1987),10
\bibitem{Pol1} 
A. Polyakov, Int. J. Mod. Phys. {\bf A 5} (1990) 833
\bibitem{Al-Shat}
A. Alekseev and S. Shatashvili, Nucl. Phys. {\bf B323 },(1989),719
\bibitem{Orai}
J. Balog, L. Feher, L. O'Raifeartaigh, P. Forgas, A. Wipf, Ann. of Phys. {\bf 203}
(1990) 76 
\bibitem{GS} 
J.-L. Gervais, M. V. Saveliev,  Phys. Lett. {\bf 286B} (1992) 271
\bibitem{Bila} 
A. Bilal,  Nucl. Phys. {\bf B422} (1994) 258
\bibitem{Gosz}
J.F.Gomes,G.M.Sotkov and A.H.Zimerman, $SL(2,R)_q$ symmetries of Non-Abelian 
Toda theories ,preprint IFT-P.017/97,CBPF-NF-018/98 ,hep-th /9803122
\bibitem{Bab}
O. Babelon, Phys. Lett {\bf B215 },(1988), 523
\bibitem{F-L}
V.A. Fateev and S.L. Lukyanov, Int. J. Mod. Phys {\bf A3} (1988) 507
\bibitem{F-L1}
V.A. Fateev and S.L. Lukyanov, Int. J. Mod. Phys {\bf A7} (1992) 853
\bibitem{Fad}
L.D. Faddeev and L. Takhtajan, ``Hamiltonian methods in the Theory of
Solitons'', (Springer, Berlin ),1987
\bibitem{Gomez-Siera}
L.Alvarez-Gaume, C. Gomez and G, Siera, Phys. Lett. {\bf B220 } (1989) 142
\bibitem{F-L2}
V.A. Fateev and S.L. Lukyanov, Int. J. Mod. Phys {\bf A7} (1992) 1325
\bibitem{Bab-Bon}
O. Babelon and L. Bonora, Phys. Lett. {\bf B253 } (1991) 365
\bibitem{Ger-Bil} 
A. Bilal, J.-L. Gervais, Nucl. Phys. {\bf B314} (1989) 646
\bibitem{GSSZ2}
J.F. Gomes, F.E.M. da Silveira, G.M. Sotkov and A.H. Zimerman, Preprint in
preparation
\bibitem{LAF}
L.A. Ferreira, J.L. Miramontes and J.S. Guillen, Nucl. Phys. {\bf B449 } (1995)
\bibitem{Witten1}
E. Witten,  Phys. Rev. Lett. {\bf 38} (1978) 121631
\bibitem{Ger} 
A. Gerasimov, A. Morozov, M. Olshanesky, A. Marshaskov, S. Shatashvili, Int. J. Mod.
Phys. {\bf A 5} (1990) 2495
\bibitem{SSZ}
 G.M. Sotkov, M. Stanishkov and C.J. Zhu, Nucl. Phys. {\bf B356 } (1991) 245
\bibitem{Bershadsky} 
M. Bershadsky, Comm. Math. Phys. {\bf 139} (1991) 71
\bibitem{Bern-Leclair}
D. Bernard and A. Leclair, Commun. Math. Phys. {\bf 142} (1991) 99
\bibitem{Skly}
E. Sklyanin, J. Sov. Math. {\bf 19 },(1982),1546
\bibitem{Bab-Bern}
 O. Babelon and D. Bernard, Commun. Math. Phys. {\bf 149 } (1992) 279
\bibitem{Bais}
F. A. Bais, H. Weldon,  Phys. Rev. {\bf D18} (1978) 561;  Phys. Lett. {\bf 79B}
(1978) 297
\bibitem{L-S} 
A. N. Leznov, M. V. Saveliev, Comm. Math. Phys. {\bf 74} (1980) 111;   Comm. Math. Phys.
{\bf 89} (1983) 59
\bibitem{Witten2} 
E. Witten, Phys. Rev. {\bf D14} (1991) 314
\bibitem{Dijkgraaf} 
R. Dijkgraaf, H. Verlinde, E. Verlinde, Nucl. Phys. {\bf 371} (1992) 269
\bibitem{Horne} 
J. H. Horne, G. T. Horowitz, Nucl. Phys. {\bf B368} (1992) 444
\bibitem{Ginsparg} 
P. Ginsparg, F. Quevedo, Nucl. Phys. {\bf 385} (1992) 527
\bibitem{Bil} 
A. Bilal,  Commun. Math. Phys. {\bf 170} (1995) 117
\bibitem{Ga-Ku}
K. Gawedzki, A. Kupiainen,  Nucl. Phys. {\bf B320} (1989) 625
\bibitem{Bardakci}
K. Bardakci, M. Crescimanno, E. Rabinovici,  Nucl. Phys. {\bf B344} (1990) 
344
\bibitem{Witten3}
E. Witten, Commun. Math. Phys. {\bf 144 } (1992) 189
\bibitem{LS}
A. N. Leznov, M. V. Saveliev, Group Theoretical Methods for 
Integration of
Nonlinear Dynamical Systems, Progress in Physics, Vol. 15 (1992), Birkhauser
Verlag, Berlin
\bibitem{Drinf}V. G. Drinfeld, Sov. Math. Dokl, {\bf 36} (1988) 212; M. Jimbo, Lett. Math. Phys. {\bf
11} (1986) 247
\end{thebibliography}
\end{document}